\documentclass[aps,prd,reprint,superscriptaddress,eqsecnum,nofootinbib]{revtex4-1}
\usepackage{bm}
\usepackage{amssymb}
\usepackage{amsmath}
\usepackage{hyperref}
\usepackage{graphicx,subfigure,wrapfig}
\usepackage{transparent}

\newcommand{\rmd}{{\rm d}}
\newcommand{\bx}{{\mathbf x}}
\newcommand{\bA}{{\mathbf A}}
\newcommand{\bB}{{\mathbf B}}

\begin{document}

\title{Caustics and wave propagation in curved spacetimes}

\date{\today}

\author{Abraham I. Harte}
\affiliation{Max-Planck-Institut f\"ur Gravitationsphysik, Albert-Einstein-Institut, 
	\\
	Am M\"{u}hlenberg 1, D-14476 Golm, Germany}
	\email{Electronic address: harte@aei.mpg.de}

\author{Theodore D. Drivas}
\affiliation{Department of Applied Mathematics and Statistics, The Johns Hopkins University,
Baltimore, Maryland 21218, USA}
\email{Electronic address: tdrivas2@jhu.edu}

\begin{abstract}
We investigate the effects of light cone caustics on the propagation of linear scalar fields in generic four-dimensional spacetimes. In particular, we analyze the singular structure of relevant Green functions. As expected from general theorems, Green functions associated with wave equations are globally singular along a large class of null geodesics. Despite this, the ``nature'' of the singularity on a given geodesic does not necessarily remain fixed. It can change character on encountering caustics of the light cone. These changes are studied by first deriving global Green functions for scalar fields propagating on smooth plane wave spacetimes. We then use Penrose limits to argue that there is a sense in which the ``leading order singular behavior'' of a (typically unknown) Green function associated with a generic spacetime can always be understood using a (known) Green function associated with an appropriate plane wave spacetime. This correspondence is used to derive a simple rule describing how Green functions change their singular structure near some reference null geodesic. Such changes depend only on the multiplicities of the conjugate points encountered along the reference geodesic. Using $\sigma(p,p')$ to denote a suitable generalization of Synge's world function, conjugate points with multiplicity 1 convert Green function singularities involving $\delta(\sigma)$ into singularities involving $\pm 1/\pi \sigma$ (and vice-versa). Conjugate points with multiplicity 2 may be viewed as having the effect of two successive passes through conjugate points with multiplicity 1. Separately, we provide an extensive review of plane wave geometry that may be of independent interest. Explicit forms for bitensors such as Synge's function, the van Vleck determinant, and the parallel and Jacobi propagators are derived almost everywhere for all non-singular four-dimensional plane waves. The asymptotic behaviors of various objects near caustics are also discussed.
\end{abstract}

\maketitle

\section{Introduction}

Disturbances in physical fields generically propagate throughout the causal future of that initial disturbance. The character of this propagation is well-understood for points that are sufficiently close to said disturbance. More specifically, there exist fairly straightforward procedures to construct Green functions\footnote{The term ``Green function'' as used here coincides with the standard definition of a fundamental solution. For the model equation \eqref{GenericWaveEqn}, a Green function is any solution to $L G(p,p') = -4 \pi I \delta(p,p')$, where $I$ denotes an appropriate identity operator (needed for non-scalar fields) and $\delta(p,p')$ the Dirac distribution. By contrast, some authors define Green functions somewhat more restrictively. See, e.g., \cite{BaerWaves}.} for linear (or linear\textit{ized}) wave equations in regions which are sufficiently small that, roughly speaking, characteristic rays starting at one point do not cross each other at any other point. What occurs outside of these regions -- where characteristics intersect (or ``almost intersect'') each other -- is considerably more complicated.

To be specific, consider linear hyperbolic equations whose principal part is the d'Alembertian associated with a spacetime metric $g_{ab}$. Suppressing possible indices on the field $\Phi$ and source $\rho$, let 
\begin{equation}
	L \Phi = (g^{ab} \nabla_a \nabla_b  + \ldots) \Phi = - 4 \pi \rho,
	\label{GenericWaveEqn}
\end{equation}
where $\nabla_a$ is the natural derivative operator associated with $g_{ab}$. The omitted part of $L$ in this equation may be any first order linear differential operator. Equations satisfied by Klein-Gordon fields, electromagnetic vector potentials, and linearized perturbations of Einstein's equation all fall into this class (for certain gauge choices). 

Considerable insight into \eqref{GenericWaveEqn} may be obtained by constructing an associated Green function. If the points $p$ and $p'$ are sufficiently close, the retarded Green function is known to have the form\footnote{Distributions like $\theta(p \geq p') \delta(\sigma)$ appearing here are not \textit{a priori}, well-defined. They are to be interpreted as, e.g., $\lim_{\epsilon \rightarrow 0^+} \theta(p \geq p') \delta(\sigma+\epsilon)$. See Sect. 4.1 of \cite{FriedlanderWaves}.\label{Foot:Vertex}}  \cite{PoissonRev, FriedlanderWaves} (again suppressing indices)
\begin{align}
	G_{\mathrm{ret}}(p,p') = \theta(p \geq p') \big[ \mathcal{U} (p,p') \delta ( \sigma(p,p') ) 
	\nonumber
	\\
	~ + \mathcal{V} (p,p') \Theta ( -\sigma(p,p') ) \big]
	\label{GretGeneral}
\end{align}
in four spacetime  dimensions. Here, $\theta(p \geq p')$ is defined to equal unity if $p$ is in the causal future of $p'$, and zero otherwise. $\Theta$ and $\delta$ are the one-dimensional Heaviside and Dirac distributions, respectively. $\sigma(p,p') = \sigma(p',p)$ denotes Synge's world function; a two-point scalar equal to one-half of the squared geodesic distance between its arguments \cite{PoissonRev, FriedlanderWaves, SyngeBook}. The bitensors $\mathcal{U} (p,p')$ and $\mathcal{V} (p,p')$ are more complicated to define, although explicit procedures to compute them are known \cite{PoissonRev, FriedlanderWaves}.

The interpretation of \eqref{GretGeneral} is very simple. It implies that disturbances at a point $p'$ are initially propagated ``sharply'' along future-directed null geodesics [where $\sigma(\cdot,p') = 0$] with an influence proportional to $\mathcal{U} (\cdot, p')$. At least for fields where the differential operator $L$ has no first-order component, $\mathcal{U} (\cdot, p')$ is closely related to the expansion of the congruence of geodesics emanating from $p'$. As might have been expected, the importance of the  $\delta(\sigma)$ term in the retarded Green function is related to the focusing of null geodesics in the sense expected from geometric optics. Apart from this, the second line of \eqref{GretGeneral} indicates that there may also be contributions to field disturbances -- known as a ``tail'' -- that propagate along all future-directed \textit{timelike} geodesics emanating from $p'$ [where $\sigma (\cdot, p') < 0$]. 

This description of wave propagation cannot usually be applied throughout an entire spacetime. The Hadamard form \eqref{GretGeneral} for the retarded Green function is guaranteed to be valid only in convex geodesic domains \cite{FriedlanderWaves}. Indeed, the standard definition of $\sigma$ breaks down when, e.g., pairs of points can be connected by more than one geodesic (or by none). It is the purpose of this paper to discuss \textit{global} properties of Green functions in curved spacetimes. In particular, we focus on changes in Green functions arising from the presence of light cone caustics.

Some insight into global wave propagation may be gained from general theorems on the propagation of singularities in wave equations (see, e.g., Corollary 5 on p. 121 of \cite{BarFredenhagen}).  Roughly speaking, these state that singularities are globally propagated along null geodesics. In particular, the fact that $G_\mathrm{ret}( \cdot, p' )$ is initially singular along all future-directed null geodesics emanating from $p'$ implies that it remains singular as these geodesics are extended arbitrarily far into the future of $p'$ [even though Eq. \eqref{GretGeneral} does not necessarily hold in the distant future]. Standard propagation-of-singularities theorems do not, however, describe the specific ``character'' of the singularity on a given null geodesic. Generically, the singular structure of $G_\mathrm{ret}( \cdot, p')$ can exhibit qualitative changes when passing each caustic associated with the future light cone of $p'$.

It is clear from \eqref{GretGeneral} that retarded Green functions initially contain a term involving $\delta(\sigma)$. Recent computations of retarded Green functions for linear scalar fields in Nariai \cite{CausticsNariai} and Schwarzschild \cite{CausticsSchw} spacetimes have demonstrated that there is a sense in which such terms are replaced by different singular distributions after each encounter with a caustic of the light cone. Following a null geodesic forwards in time from a source point $p'$, the singular structure of $G_\mathrm{ret}( \cdot ,p')$ appeared to ``oscillate'' in the repeating 4-fold pattern [modulo an appropriate extension of the (positive) prefactor $\mathcal{U}(p,p')$ appearing in \eqref{GretGeneral}]
\begin{equation}
	\delta(\sigma) \rightarrow \mathrm{pv} \left( \frac{1}{\pi \sigma} \right)\rightarrow - \delta(\sigma) \rightarrow - \mathrm{pv} \left( \frac{1}{\pi \sigma} \right) \rightarrow \ldots
	\label{SingStruct4}
\end{equation}
Here, ``pv'' denotes the Cauchy principal value.

Ori has heuristically argued \cite{Ori} that this phenomenon should be generic for waves propagating through ``astigmatic caustics'' where light rays are focused in only one transverse direction. Such caustics are associated with conjugate points of multiplicity 1.  Furthermore, Ori claims that the effect of ``stronger'' \textit{an}astigmatic caustics associated with multiplicity 2 conjugate points should have an effect equivalent to two passes through astigmatic caustics. If all caustics in a particular geometry are anastigmatic -- displaying perfect focusing -- this reasoning would imply that the associated Green functions display the 2-fold pattern of singular structures 
\begin{equation}
	\delta(\sigma) \rightarrow - \delta(\sigma) \rightarrow \ldots
	\label{SingStruct2}
\end{equation} 
Such patterns have indeed been observed in scalar Green functions for both the Einstein static universe and the Bertotti-Robinson spacetime \cite{MarcPrivate}. 

There is a vast literature on wave propagation through caustics in various contexts. A significant body of work has applied catastrophe theory to classify shapes of stable caustic surfaces in different contexts \cite{ArnoldCausticBook, PerlickCaustics, EhlersCaustics,OtherRussianCausticBook}. Additionally, the behavior of wave fields near caustics has been discussed in, e.g., \cite{OtherRussianCausticBook, Ludwig}. It is known from this work that there is a sense in which individual Fourier modes of a field experience a phase change of $\pi/2$ on passing through an astigmatic caustic [associated with the 4-fold pattern \eqref{SingStruct4}]. One might therefore expect four passes through astigmatic caustics to return a Green function to its ``original form.'' Stronger anastigmatic caustics associated with the pattern \eqref{SingStruct2} effectively add a phase of $2 (\pi/2) = \pi$ to each Fourier mode. Two passes through such caustics might therefore be expected to return a Green function to its original form. 

Despite this type of frequency-domain argument, the simple ``position-space'' patterns \eqref{SingStruct4} and \eqref{SingStruct2} do not appear to have been systematically derived before except in a few special cases. One problem is finding an appropriately-precise statement of the result. The usual definition of $\sigma$ breaks down once caustics arise, so even the meanings of the patterns \eqref{SingStruct4} and \eqref{SingStruct2} are not immediately clear in general spacetimes. Additionally, one can only hope that there is a sense in which such patterns hold ``near'' null geodesics where the singular portion of a Green function might be meaningfully disentangled from its remainder. It is not immediately clear how to precisely formulate a notion of this type.

The physical interpretation of the patterns \eqref{SingStruct4} and \eqref{SingStruct2} has also been somewhat mysterious. How, for example, can a ``sharp'' distribution like $\delta(\sigma)$ instantaneously jump into the much more ``spread out'' $\mathrm{pv}(1/\pi \sigma)$? Additionally, one may question how a causal Green function could extend into regions where $\sigma>0$ [as it does when the singular structure of a Green function involves $\mathrm{pv}(1/\pi \sigma)$]. Naively, this might appear to imply that disturbances in fields can propagate to points that are causally disconnected from that disturbance. 

This paper starts by investigating and resolving all of these issues in four-dimensional plane wave spacetimes. Retarded and advanced Green functions associated with massless scalar fields propagating in smooth plane wave spacetimes are derived explicitly. These geometries might be thought of as modelling gravitational waves emitted from some moderately distant astrophysical system. Certain characteristics of plane wave spacetimes are not, however, particularly realistic. More relevant are some of their mathematical properties:
\begin{itemize}
	\item Appropriately adjusting the amplitude and polarization profile of a plane wave geometry allows the construction of examples with any number and combination of astigmatic and anastigmatic caustics. This is accomplished in a spacetime with topology $\mathbb{R}^4$ and a metric whose coordinate components can be made globally smooth.
	
	\item Green functions associated with massless minimally-coupled scalar fields or Maxwell fields are known to have nonzero tails $\mathcal{V}(p,p')$ in almost every four-dimensional spacetime. Essentially the only nontrivial counterexamples are plane wave spacetimes \cite{FriedlanderWaves, Huygens}. 
	
	\item Although generic plane wave spacetimes fail to be (globally) geodesically convex, there is a natural definition for $\sigma(p,p')$ that holds almost everywhere.	
	
	\item The geodesic structure of plane wave spacetimes is understood essentially in its entirety. This allows $\sigma(p,p')$ and $\mathcal{U}(p,p')$ to be computed explicitly.
\end{itemize}
 
Most importantly, we choose to work in plane wave spacetimes due to the existence of a procedure known as the Penrose limit \cite{PenroseLimit, BlauPenrose, BlauNotes}. This provides a sense in which the geometry near any given null geodesic in an arbitrary spacetime looks like the geometry of an appropriate plane wave spacetime. The Penrose limit preserves various properties of the original spacetime \cite{BlauNotes, GerochHereditary, BlauHereditary}; in particular, the conjugate point structure of the chosen (or ``reference'') geodesic.

We use Penrose limits to argue that most of the ``leading order'' singular behavior of Green functions associated with wave propagation in \textit{arbitrary} spacetimes may be understood using knowledge of Green functions in appropriate plane wave spacetimes. This singular structure naturally splits into two components. One portion is associated with the appearance of conjugate points on the reference geodesic with which the Penrose limit is performed. The effects of such points are, in a sense, determined quasi-locally. They affect Green functions near the reference geodesic in a way that depends only on their multiplicities. Furthermore, the effects of conjugate points propagate into their future along the reference geodesic. In most cases, the singular structures that result from the appearance of conjugate points have either the 4- or 2-fold patterns \eqref{SingStruct4} or \eqref{SingStruct2}. There do, however, exist finely-tuned examples that fall into neither category because there are a mixture of astigmatic and anastigmatic caustics.

It is important to emphasize that despite this result, Green functions are not quasi-local objects. In general, it is not possible for all of a Green function's singular structure to be determined near some null geodesic using only knowledge of the geometry near that geodesic. ``Nonlocal singularities'' can be introduced near a reference geodesic when non-conjugate pairs of points on that geodesic are also connected by other null geodesics\footnote{A simple example of this phenomenon is provided by the spacetime of a straight cosmic string \cite{PerlickCaustics}. These geometries can be described by the metric $\rmd s^2 = -\rmd t^2 + \rmd z^2 + \rmd \rho^2 +(k \rho)^2 \rmd \phi^2$ with $t, z \in \mathbb{R}$, $\rho>0$, and $\phi \in [0, 2 \pi)$. They are locally flat, and therefore admit no conjugate points along any geodesic. Cosmic string spacetimes do, however, possess an angular defect (if $k \neq 1$) that forces some pairs of points on opposite sides of the string to be connected by more than one geodesic.}. We show that the effects of such intersections on a Green function are ``non-propagating.'' There is sense in which their associated singularities are confined to regions near the intersection points. Using the Penrose limit, singularities of this type correspond to isolated structures in plane wave Green functions occurring at locations which cannot be predicted from any given set of initial data. Plane wave spacetimes are not globally hyperbolic, so their Green functions cannot be uniquely specified in terms of any initial data set. This lack of uniqueness is, indeed, necessary if plane wave Green functions are to consistently capture certain characteristics of generic Green functions with intrinsically nonlocal components.

This paper is organized into three main parts. Sects. \ref{Sect:ppGeometry} and \ref{Sect:Geometry} define the plane wave geometry and provide an extensive discussion of its geometrical properties. While much of the material in these sections has been noted before \cite{BlauNotes, GlobalGeo, EhlersKundt, EhrlichEmch1, EhrlichEmch2, QED1, QED2}, some appears to be new (e.g., the behavior of various geometric objects near caustics). Only a few key results from Sect. \ref{Sect:Geometry} are needed to understand the majority of Sect. \ref{Sect:Green}, where explicit global Green functions are constructed for all smooth four-dimensional plane wave spacetimes. Sect. \ref{Sect:Penrose} finally shows that knowledge of plane wave Green functions is sufficient to understand the leading order singular structure of Green functions in arbitrary spacetimes. Appendix \ref{Sect:ABProperties} describes various properties of two matrices central to describing the geometry of plane wave spacetimes. Appendix \ref{Sect:Distributions} establishes that the Green function obtained for plane wave spacetimes is a well-defined distribution.

\subsection*{Notation}

In this paper, abstract indices are represented using letters taken from the beginning of the Latin alphabet: $a,b,\ldots$ Four-dimensional coordinate indices are represented using Greek characters, while indices referring only to the spatial coordinates $x^1,  x^2$ or $X^1,X^2$ introduced below are denoted by $i,j,\ldots$ Where appropriate, units are used in which $G=c=1$. Our sign conventions follow those of Wald \cite{Wald}: The metric signature is chosen to be $(-+++)$, the Riemann tensor is defined such that $2 \nabla_{[a} \nabla_{b]} \omega_c = R_{abc}{}^{d} \omega_d$ for any 1-form $\omega_a$, and the Ricci tensor satisfies $R_{ab} = R_{acb}{}^{c}$. Spacetime points on a manifold $M$ are generally denoted by $p, \, p'$, etc. Coordinates associated with (say) $p'$ are themselves primed. We often find occasion to abuse notation in various ways that should be understandable from context. For example, we often identify a function of spacetime points with the equivalent function acting on coordinates: e.g., $f(p) = f(u,v,x^1,x^2)$ in a global chart $(u,v,x^1,x^2): M \rightarrow \mathbb{R}^4$. We also make extensive use of elementary vector and matrix notation to denote the spatial coordinate components of various tensors: e.g., $\bx^\intercal = (x^1 \, \, \, \, x^2)$, $(\mathbf{A} \mathbf{B})_{ij} = A_{ik} B_{kj}$, etc. The majority of this paper is concerned with plane wave spacetimes. In Sect. \ref{Sect:Penrose}, more general spacetimes are considered as well. Quantities associated with these geometries are often distinguished by the presence of a check mark. A non-plane wave metric is often denoted by $\check{g}_{ab}$, for example.

\section{Plane wave spacetimes in general}
\label{Sect:ppGeometry}

\subsection{pp-waves}

A pp-wave is a spacetime which may be physically interpreted as a (not necessarily vacuum) gravitational wave with parallel rays orthogonal to a family of planar wavefronts. While definitions in the literature vary slightly, pp-waves are often prescribed as a manifold $M$ together with a metric $g_{ab}$ which everywhere admits a nonzero null vector field $\ell^a$ satisfying $\nabla_a \ell^b = 0$ (where $\nabla_a$ is the Levi-Civita connection associated with $g_{ab}$). $\ell^a$ is interpreted as the direction of wave propagation. Since it is covariantly constant, it must be Killing. This implies that the wave propagates without distortion. It is also clear that the integral curves of $\ell^a$ -- the characteristic rays of the gravitational wave -- form a null geodesic congruence that is non-expanding, shear-free, and twist-free. This implies that there is a sense in which such rays remain ``parallel'' to each other. They are also orthogonal to a family of planar 2-surfaces that may be interpreted as wavefronts.

A large class of pp-wave metrics in four dimensions can be written as \cite{EhlersKundt, HallBook, GenPPWave, GriffithsBook}
\begin{equation}
  \rmd s^2 = - 2 \rmd u \rmd v + H(u, \bx) \rmd u^2 + |\rmd \bx|^2,
  \label{ppMetricGen}
\end{equation}
where $|\rmd \bx|^2$ denotes $(\rmd x^1)^2 + (\rmd x^2)^2$. We assume for simplicity that the coordinates $(u,v,\bx) = (u,v,x^1,x^2)$ can take any values in $\mathbb{R}^4$. The unconstrained function $H(u,\bx) = H(u,x^1,x^2)$ fixes the waveform and its polarization. Note that $\partial/\partial v$ is both null and covariantly-constant. It may therefore be identified with the direction of wave propagation $\ell^a$:
\begin{equation}
	\ell^a = \left( \frac{\partial}{ \partial v } \right)^a.
	\label{EllDef}
\end{equation}
Surfaces spanned by the spatial coordinates $x^1$ and $x^2$ (at fixed $u,v$) are wavefronts transverse to the direction of propagation. They are spacelike surfaces with topology $\mathbb{R}^2$ and an induced metric that is everywhere flat: The wavefronts are 2-planes. Furthermore, note that $\ell_a = -\nabla_a u$. The $u$ coordinate may be viewed as labelling the phase of the gravitational wave.

All non-vanishing coordinate components of the Riemann tensor may be obtained from
\begin{equation}\label{Riemann}
  R_{uiuj} = - \frac{1}{2} \partial_i \partial_j H(u,\bx) ,
\end{equation}
where $\partial_i :=\partial/\partial x^i$ and $i=1,2$. The Ricci tensor can have at most one nonzero component:
\begin{equation}\label{Ricci}
  R_{uu} = - \frac{1}{2} \nabla^2 H(u,\bx).
\end{equation}
Here, $\nabla^2$ denotes the two-dimensional Euclidean Laplacian acting on the coordinates $(x^1,x^2)$. It follows that any pp-wave satisfying the vacuum Einstein equation $R_{ab} = 0$ has a waveform $H(u,\bx)$ that is harmonic in the spatial coordinates. The sum of any two harmonic functions is itself harmonic, so there is a sense in which vacuum pp-wave metrics propagating in the same direction remain vacuum under linear superposition.

Note that it follows from \eqref{ppMetricGen} and \eqref{Ricci} that the Ricci scalar $g^{ab} R_{ab}$ vanishes in all pp-wave spacetimes. More generally, \textit{all} locally-constructed curvature scalars vanish in these geometries. Spacetimes with this property -- of which the pp-waves are a special case -- are known to be members of the Kundt class \cite{VanishingCurvScalars} (The converse is not true: There do exist Kundt metrics with non-vanishing curvature scalars.). Geometries with vanishing curvature scalars are the gravitational analogs of ``null'' electromagnetic fields $F_{ab}$ satisfying $F_{ab} F^{ab} = \epsilon^{abcd} F_{ab} F_{cd} = \ldots = 0$. Some of the simplest nontrivial examples of null electromagnetic fields are plane waves propagating in flat spacetime. Similarly, some of the simplest nontrivial geometries with vanishing curvature scalars are plane wave spacetimes. These are a subclass of pp-wave spacetimes.

\subsection{Plane waves}
\label{Sect:Plane}

Plane wave spacetimes are special pp-waves where the curvature components $R_{\mu\nu\lambda\rho}$ depend only on the ``phase coordinate'' $u$. The wave amplitude and polarization can then be said to remain constant on each planar wavefront formed by varying $x^1, x^2$ while holding fixed $u$ and $v$. 

It follows from \eqref{Riemann} that plane waves arise if the profile function $H(u,\bx)$ is at most quadratic in the spatial variables $\bx$. A coordinate transformation may be used to eliminate any components of $H$ independent of or linear in $x^1$ and $x^2$. The metric of a general plane wave spacetime can therefore be written in the Brinkmann form
\begin{equation} 
  	\rmd s^2 = - 2 \rmd u \rmd v + H_{ij}(u) x^i x^j \rmd u^2 + | \rmd \bx|^2  .
	\label{PlaneWaveGen}
\end{equation}
Here, $H_{ij} (u)$ is an arbitrary symmetric $2 \times 2$ matrix that specifies the wave's amplitude and polarization profile. Except in Sect. \ref{Sect:Penrose}, the metric \eqref{PlaneWaveGen} is assumed to hold throughout this paper. We restrict attention to non-singular plane waves where $(u,v,\bx) \in \mathbb{R}^4$ and $H_{ij}(u)$ is a collection of smooth functions from $\mathbb{R}$ to $\mathbb{R}$.

It is convenient later to have a special notation for constant-phase surfaces associated with \eqref{PlaneWaveGen}. For any $u' \in \mathbb{R}$, let $S_{u'}$ denote the $u = u'$ hyperplane 
\begin{equation}
	S_{u'} := \{ p' \in M : u(p') = u' \}.
	\label{ConstU}
\end{equation}
It follows from \eqref{Riemann} that the Riemann tensor on one of these hypersurfaces is entirely determined by the components
\begin{equation}
	R_{uiuj} = - H_{ij}(u).
	\label{RiemannPlane}
\end{equation}
Furthermore,
\begin{equation}
	R_{uu} = - \mathrm{Tr} \, \mathbf{H}(u),
	\label{RicciPlane}
\end{equation}
where $\mathrm{Tr}$ denotes the ordinary trace of the $2 \times 2$ matrix $(\mathbf{H})_{ij} = H_{ij}$.

It follows that the vacuum Einstein equation $R_{ab} = 0$ is satisfied if and only if $H_{ij}$ is trace-free. The metric of any purely gravitational plane wave can therefore be put into the form
\begin{align}
	\rmd s^2 = - 2 \rmd u \rmd v + \big\{ h_{+}(u) \big[ (x^1)^2-(x^2)^2 \big] 
\nonumber
\\
	 ~ + 2 h_{\times}(u) x^1 x^2 \big\} \rmd u^2  + |\rmd \bx|^2 ,
\label{PlaneWaveRiccFlat}
\end{align}
where $h_{+}(u)$ and $h_{\times} (u)$ are arbitrary functions representing waveforms for the two polarization states of the gravitational wave. If these functions are proportional, the wave is said to be linearly polarized. A coordinate rotation can then be used to eliminate $h_{\times}(u)$ in favor of rescaling $h_{+}(u)$.

It is interesting to note that there is a sense in which metrics with the form \eqref{PlaneWaveRiccFlat} satisfy all generally covariant field equations that can be constructed purely from the metric and its derivatives \cite{PlaneWaveFieldEqs}. Ricci-flat plane wave spacetimes therefore provide a model for plane-symmetric gravitational radiation in general relativity as well as many alternative theories of gravity.

In this paper, we do not restrict the discussion only to vacuum plane waves. One interesting class of non-vacuum plane waves are those that are conformally-flat. Such geometries must have Riemann tensors that are ``pure trace.'' It follows from inspection of \eqref{PlaneWaveGen} and \eqref{RicciPlane} that the metric of a conformally-flat  plane wave can always be put into the form
\begin{equation}
	\rmd s^2 = - 2 \rmd u \rmd v - h^2(u) |\bx|^2 \rmd u^2 + |\rmd \bx|^2
\label{PlaneWaveConfFlat}
\end{equation}
for some function $h(u)$. There is at most one nonzero Ricci component in these coordinates:
\begin{equation}
	R_{uu} =  2 h^2(u) .
\label{RicciConfFlat}
\end{equation}
If Einstein's equation $R_{ab} - \frac{1}{2} g_{ab} R = 8\pi T_{ab}$ is imposed, the null energy condition holds for the stress-energy tensor $T_{ab}$ if and only if $h^2(u) \geq 0$. This condition also implies the weak, dominant and strong energy conditions.

Conformally-flat plane wave spacetimes satisfying $h^2(u) \geq 0$ may be interpreted as the gravitational fields associated with plane electromagnetic waves in Einstein-Maxwell theory. In general, the stress-energy tensor of an electromagnetic field $F_{ab}$ is
\begin{equation}
	T_{ab} = \frac{1}{4 \pi} (F_{ac} F_{b}{}^{c} - \frac{1}{4} g_{ab} F_{cd} F^{cd}).
\end{equation}
Inserting this into Einstein's equation and using \eqref{RicciConfFlat}, it is easily observed that the plane wave geometry \eqref{PlaneWaveConfFlat} may be associated with the electromagnetic plane wave
\begin{equation}
	F_{ab} = 2 h(u) \nabla_{[a} u \nabla_{b]} x^1 .
	\label{EMField}
\end{equation}
This electromagnetic field is a solution to the vacuum Maxwell equations. It also satisfies $F_{ab} F^{ab} = \epsilon_{abcd} F^{ab} F^{cd} = 0$. The electric and magnetic fields seen by any observer are therefore equal in magnitude and orthogonal:
\begin{equation}
	E_a E^a = B_a B^a , \qquad E_a B^a = 0.
\end{equation}
Note, however, that \eqref{EMField} is but one possible $F_{ab}$ that could be associated with a given $h(u)$. Other possibilities exist.

Although we make little use of it, it should be mentioned that plane waves are often described in the literature in terms of Rosen coordinates $(U,V,\mathbf{X})$ instead of the Brinkmann coordinates $(u,v,\bx)$ used in \eqref{PlaneWaveGen}. The metric then takes the form
\begin{equation}
	\rmd s^2 = - 2 \rmd U \rmd V + \mathcal{H}_{ij}(U) \rmd X^i \rmd X^j.
	\label{RosenMetric}
\end{equation}
The $2 \times 2$ matrix $\mathcal{H}_{ij}(U)$ appearing here depends non-algebraically and non-uniquely on $H_{ij}(u)$. Consider, in particular, the transformation
\begin{subequations}
\label{BrinkmannToRosen}
\begin{align}
	u &= U,
	\\
	v &= V + \frac{1}{2}  \dot{E}^{k}{}_{i} (U) E_{kj} (U) X^i X^j,
	\\
	x^i &= E^{i}{}_{j}(U) X^j,
\end{align}
\end{subequations}
where the matrix $E^{i}{}_{j}(U)$ is a nontrivial solution to the differential equation
\begin{equation}
	\ddot{\mathbf{E}}(U) = \mathbf{H} (U) \mathbf{E} (U).
	\label{EDef}
\end{equation}
We also require that
\begin{equation}
	\dot{\mathbf{E}}^\intercal \mathbf{E} = (\dot{\mathbf{E}}^\intercal \mathbf{E})^\intercal.
	\label{SymE}
\end{equation}
Applying \eqref{BrinkmannToRosen} to the Brinkmann line element \eqref{PlaneWaveGen} with these restrictions on $\mathbf{E}$, one finds the Rosen line element \eqref{RosenMetric} with
\begin{equation}
	\bm{\mathcal{H}}(U) = \mathbf{E}^\intercal (U) \mathbf{E}(U).
	\label{RosenH}
\end{equation}

Rosen coordinates have the advantage of being more closely related than Brinkmann coordinates to intuition for gravitational waves built up from linearizing Einstein's equation about Minkowski spacetime in transverse-traceless gauge. Some properties of a plane wave's geodesics and symmetries are also more easily expressed in terms of Rosen coordinates. Unfortunately, the metric \eqref{RosenMetric} does not generally cover the entire spacetime. Rosen coordinates generically develop singularities that are not present in Brinkmann coordinates. There is also a considerable degree of ``gauge freedom'' in $\mathcal{H}_{ij}$ [i.e., there are many allowed solutions to \eqref{EDef} for a given $H_{ij}$].

\section{Geometric properties of plane wave spacetimes}
\label{Sect:Geometry}

Before discussing wave propagation in some background spacetime, it is important to understand the geometry of that background. This section discusses the symmetries of plane wave spacetimes as well as their geodesic and causal structures. Bitensors such as Synge's function, the van Vleck determinant, and the parallel propagator are computed explicitly. Emphasis is placed on the focusing of geodesics and the asymptotic behavior of various bitensors near light cone caustics. These topics are all important for the understanding of Green functions associated with wave equations like \eqref{GretGeneral}. 

A recurring object in the geometry of plane wave spacetimes is the matrix differential equation \eqref{EDef}. This may be viewed as a generalized oscillator equation where the wave profile $\mathbf{H}(U)$ acts like (the negative of) a ``squared frequency matrix.'' It is useful to describe all solutions by a linear combination of two particular solutions $\bA(u,u')$ and $\bB(u,u')$. We choose to define these matrices to be solutions of
\begin{subequations}
	\label{JacobiSpatial}
\begin{align}
	\partial^2_u \bA(u,u') = \mathbf{H}(u) \bA(u,u') 
	\\
	\partial^2_u \bB(u,u') = \mathbf{H}(u) \bB(u,u')
\end{align}
\end{subequations}
satisfying the boundary conditions
\begin{align}
	\label{ABboundary}
	[\bA] = [\partial_u \bB]= \bm{\delta}, \qquad 	[ \bB ] = [ \partial_u \bA ] = 0.
\end{align}
Here, $\bm{\delta}$ denotes the $2 \times 2$ identity matrix and $[ \cdot]$ indicates the coincidence limit $u \rightarrow u'$. $\bA(u,u')$ and $\bB(u,u')$ are assumed to be matrices of functions that are smooth throughout $\mathbb{R} \times \mathbb{R}$. Some of their properties are discussed in Appendix \ref{Sect:ABProperties}.

We demonstrate below that $\bA(u,u')$ and $\bB(u,u')$ may be used not only to describe the transformation between Rosen and Brinkmann coordinates, but also to compute the spatial coordinate components of geodesics and Jacobi fields. Additionally, these matrices can be used to identify conjugate points, explicitly compute various bitensors, and construct Killing fields.

As an important example, consider a geodesic passing through two points $p$ and $p'$. It is shown in Sect. \ref{Sect:Conj} that these points are conjugate along the given geodesic if and only if -- abusing notation somewhat -- their Brinkmann wavefront coordinates $u = u(p)$ and $u' = u(p')$ satisfy
\begin{equation}
	\det \bB(u,u') = 0
	\label{ZeroDet}
\end{equation}
and $u \neq u'$. This means that there exists a nontrivial Jacobi field along the chosen geodesic which vanishes at both $p$ and $p'$. Defining the multiplicity of a conjugate pair as the number of nontrivial linearly independent Jacobi fields which vanish at these points (i.e., the number of focused directions), the multiplicity of $p$ and $p'$ is easily read off as the nullity of $\bB(u,u')$.  In the four spacetime dimensions considered here, the multiplicity cannot exceed two. If the pair $(p,p')$ is conjugate with multiplicity 1, the set of all null geodesics emanating from $p'$ and passing through the constant-$u$ surface $S_{u}$ is shown in Sect. \ref{Sect:CausticGeo} to form a one-dimensional curve on $S_u$ [recall \eqref{ConstU}]. This represents astigmatic focusing. Conjugate points with multiplicity 2 momentarily focus bundles of null geodesics to a single point. This represents anastigmatic focusing.

Note that \eqref{ZeroDet} does not depend on any details of the geodesic under consideration. It is purely a relation between pairs of $u$ coordinates. Geometrically, it may be interpreted as distinguishing certain pairs $(S_{u}, S_{u'})$ of hyperplanes associated with two different phases of the gravitational wave. We call these ``conjugate hyperplanes.'' They play a central role in the geometry of plane wave spacetimes.

It is convenient to denote the set of all solutions to $\det \bB(\cdot, u')=0$ by $T(u')$. We write the individual elements\footnote{Conjugate points always occur discretely in plane wave spacetimes. In more general Lorentzian metrics, it is possible for there to exist continuous sections of a geodesic that are conjugate to one particular point on that geodesic \cite{HelferConj, NewerConj}. This is, however, only possible along spacelike geodesics. For plane wave spacetimes, the appearance of conjugate points along spacelike and causal geodesics is governed by the same equation. $T(u')$ is therefore a countable set for every $u' \in \mathbb{R}$.} of $T(u')$ as $\tau_n(u') \in \mathbb{R} \setminus \{ u' \}$:
\begin{equation}
	T(u') = \bigcup_n \tau_n(u').
\label{CTau}
\end{equation}
Here, the $n$ are nonzero integers that order the elements of $T(u')$ (if any). By convention, we set $n>0$ if $\tau_n(u')>u'$ and $n<0$ otherwise. See Fig. \ref{Fig:Normal} and further discussion in Sects. \ref{Sect:Geodesics} and \ref{Sect:Conj} below.

\begin{figure}
	\centering
	\includegraphics[width= .85\linewidth]{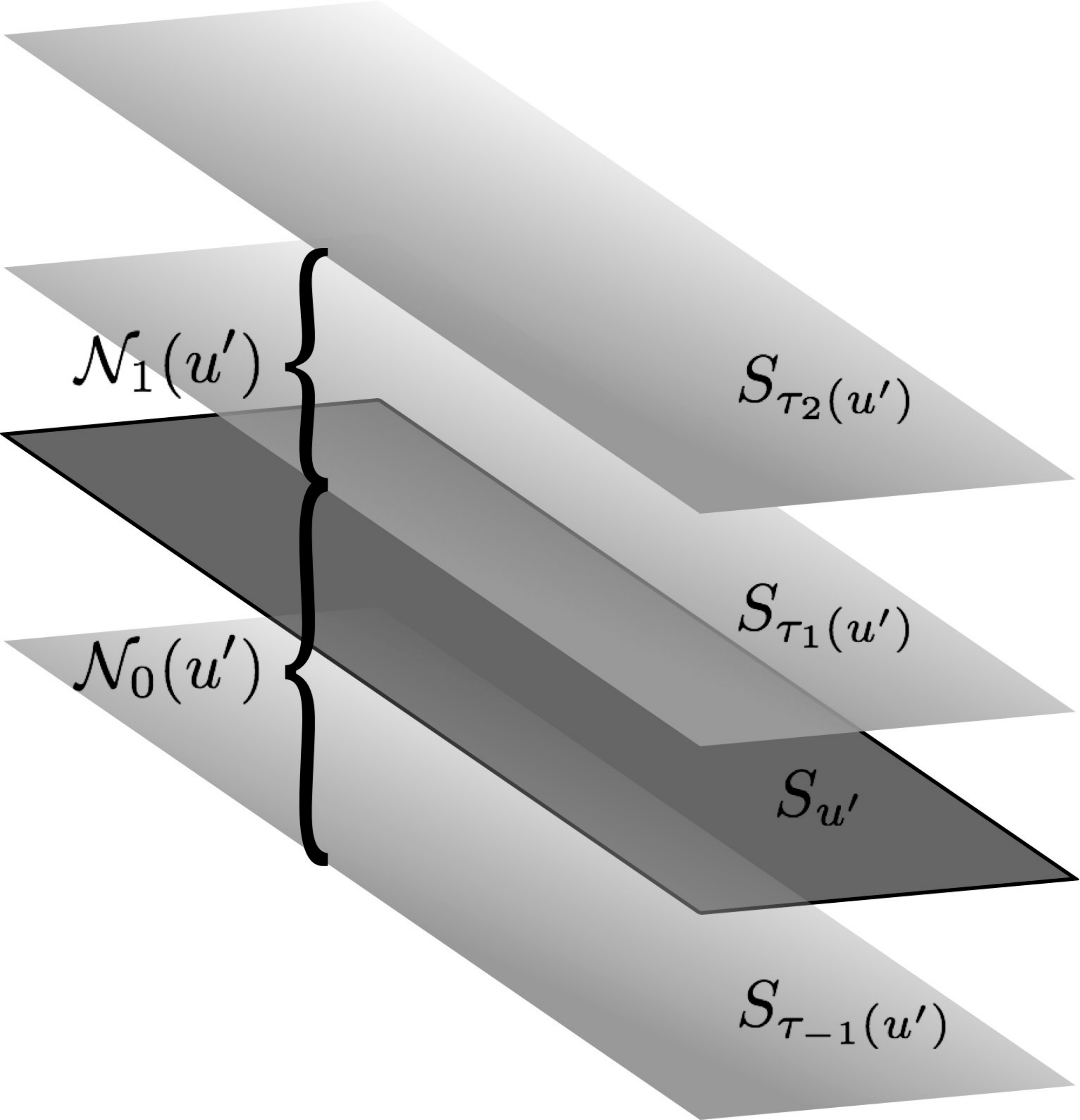}
	\caption{Fixing any $u' \in \mathbb{R}$, a plane wave spacetime naturally divides into a set of 3-surfaces $S_{\tau_n(u')}$ and open 4-volumes $\mathcal{N}_n(u')$ in between them. Every point in $S_{u'}$ is connected to every point in $\mathcal{N}_n(u')$ by exactly one geodesic. Such points are never conjugate. Points in $S_{u'}$ can be connected to points in $S_{\tau_n(u')}$ by either an infinite number of geodesics or by none. In the former case, both points are conjugate along every connecting geodesic. This justifies the description of $(S_{u'}, S_{\tau_n(u')})$ as a pair of ``conjugate hyperplanes.''}
	\label{Fig:Normal}
\end{figure}

Given some preferred phase coordinate $u'$, $T(u')$ naturally divides the spacetime into a collection of simply-connected open sets $\mathcal{N}_n(u')$ and their bounding hyperplanes $S_{\tau_n(u')}$. Define the region $\mathcal{N}_0(u')$ to be
\begin{align}
	\mathcal{N}_0 (u') := \{ p \in M : u(p) \in ( \tau_{-1} (u'), \tau_{+1} (u') ) \}
\label{NormalNeighborhood}
\end{align}
if $\tau_{\pm 1}(u')$ both exist in  $T(u')$. If these elements do not exist, the appropriate endpoint(s) of the interval in the definition of $\mathcal{N}_0(u')$ is to be replaced by $\pm \infty$. If, say, $\tau_{+1} (u')$ and $\tau_{+2}(u')$ exist, 
\begin{align}
	\mathcal{N}_{1}:= \{ p \in M: u(p) \in (\tau_1, \tau_2) \}.
\end{align}
Similarly,
\begin{equation}
	\mathcal{N}_{-1}:= \{p \in M: u(p) \in (\tau_{-2}, \tau_{-1}) \}
\end{equation}
if $\tau_{-1}(u')$ and $\tau_{-2}(u')$ both exist. In general, $\mathcal{N}_n(u')$ represents the region between the $n$th hyperplane conjugate to $S_{u'}$ and the ``next one.'' See Fig \ref{Fig:Normal}. 

The region $\mathcal{N}_0 (u')$ is a convex normal neighborhood of every point $p'$ for which $u' = u(p')$. It is, however, shown in Sect. \ref{Sect:Geodesics} below that all points in the open set
\begin{equation}
	\mathcal{N} (u') := \bigcup_n \mathcal{N}_n (u')
\label{GenNormalNeighborhood}
\end{equation}
are connected to $p'$ by exactly one geodesic. In this sense, $\mathcal{N}(u')$ is a ``generalized normal neighborhood'' of $p'$. Note, however, that this set is not path-connected unless $T(u')$ is the empty set (in which case $\mathcal{N}(u') = \mathcal{N}_0(u') =M$). The geodesic connecting $p'$ to a generic point in $\mathcal{N}(u')$ need not lie entirely in $\mathcal{N}(u')$. In general, it will pass through some of the $S_{\tau_n}(u')$. These hypersurfaces are not contained in $\mathcal{N}(u')$. Note that the portion of the spacetime not contained in $\mathcal{N}(u')$,
\begin{equation}
	M \setminus \mathcal{N}(u') = \bigcup_{n} S_{\tau_n(u')},
\end{equation}
has zero volume.

\subsection{Symmetries}
\label{Sect:Syms}

Plane wave spacetimes possess at least five linearly independent Killing vectors. One of these is clearly $\ell^a = (\partial/\partial v)^a$. The others have the form
\begin{equation}
  \left( x^i \dot{\Xi}_i(u)  \frac{\partial}{\partial v} + \Xi^i (u)  \frac{\partial}{\partial x^i} \right)^a
  \label{KillingVectsAuto}
\end{equation}
in terms of the Brinkmann coordinates $(u,v, \bx)$. Here, $\Xi^i(u) = \Xi_i(u)$ is any solution to
\begin{equation}
  \ddot{\bm{\Xi}}(u) = \mathbf{H} (u) \bm{\Xi}(u) .
  \label{KillingEq}
\end{equation}
This equation prescribes a total of four linearly independent Killing fields in addition to $\ell^a$. Even more Killing fields may be found in certain special cases. Note that \eqref{KillingEq} is very closely related to the modified oscillator equation \eqref{EDef}. In terms of the matrices $\bA$ and $\bB$ defined by \eqref{JacobiSpatial} and \eqref{ABboundary}, the general solution is
\begin{equation}
	\bm{\Xi}(u) = \bA(u,u') \bm{\Xi}(u') + \bB(u,u') \dot{\bm{\Xi}}(u'). 
	\label{KillingJacobi}
\end{equation}
The parameters $u'$, $\bm{\Xi}(u')$, and $\dot{\bm{\Xi}}(u')$ appearing in this equation may be varied arbitrarily.

Various types of non-Killing symmetries exist in generic plane wave spacetimes. For example, the vector field $\zeta^a := u \ell^a$ is always a (proper) affine collineation.\footnote{Affine collineations generate a family of diffeomorphisms that preserve all geodesics and their affine parameters. A vector field $\zeta^a$ is an affine collineation if and only if $\nabla_a \mathcal{L}_\zeta g_{bc} = 0$ \cite{HallBook}. A homothety $\psi^a$ is a special type of affine collineation satisfying $\mathcal{L}_\psi g_{ab} = (\mbox{constant}) \times g_{ab}$. Its associated diffeomorphisms preserve the metric up to changes in scale. A \textit{proper} homothety is a homothety that is not Killing. A proper affine collineation is an affine collineation that is not a homothety (and not Killing).} There is also a proper homothety $\psi^a$ given by
\begin{equation}
	\psi^a = \left( 2 v \frac{\partial}{\partial v} + x^i \frac{\partial}{\partial x^i} \right)^a.
	\label{Homothety}
\end{equation}
This satisfies $\mathcal{L}_\psi g_{ab} = 2 g_{ab}$. More extensive discussions of the symmetries of plane wave spacetimes may be found in \cite{ppWaveConfSyms, ppWaveKT}.

\subsection{Geodesics}
\label{Sect:Geodesics}

The geodesic structure of plane wave spacetimes is relatively straightforward to determine, yet still exhibits a number of nontrivial features. First recall that the vector field $\ell^a = (\partial/ \partial v)^a$ is Killing. The quantity $\ell_a \dot{z}^a$ is therefore conserved along any affinely-parameterized geodesic with tangent $\dot{z}^a$.

If $\ell_a \dot{z}^a =0$ for a particular geodesic, that geodesic remains in a single constant-$u$ hypersurface. In the Brinkmann coordinates where \eqref{PlaneWaveGen} holds, the coordinate components of such a geodesic satisfy 
\begin{equation}
	\frac{\rmd }{\rmd s} \dot{z}^\mu (s) = 0
\end{equation}
for all $\mu = u,v,x^1,x^2$. Any geodesic lying on a surface of constant phase $u$ therefore appears to be a (Euclidean) straight line in the coordinates $(v,\bx)$. It is also clear that there exists exactly one geodesic connecting any pair of points with the same $u$ coordinates. The hypersurfaces $S_u$ defined by \eqref{ConstU} are therefore totally-geodesic.

Geodesics are more complicated when $\ell_a \dot{z}^a \neq 0$. In these cases, the affine parameter of a geodesic may always be rescaled such that $\ell_a \dot{z}^a = -1$. Choosing the origin of this parameter appropriately then allows it to be identified with the coordinate $u$. Doing this, the spatial components $z^i(s) = z^i(u)$ of a geodesic are easily shown to satisfy
\begin{equation}
		\ddot{\mathbf{z}} (u) = \mathbf{H}(u) \mathbf{z}(u) .
		\label{Geodesic}
\end{equation}
This equation is identical to \eqref{KillingEq} and very similar to \eqref{EDef}. As in \eqref{KillingJacobi}, any possible $\mathbf{z}(u)$ can be written in terms of the matrices $\bA$ and $\bB$ introduced above:
\begin{equation}
	\mathbf{z} (u) = \bA(u,u') \mathbf{z}(u') + \bB(u,u') \dot{\mathbf{z}}(u').
\label{GeodesicJacobi}
\end{equation}

The $v$ component of any geodesic is most easily found using the conservation law associated with the homothety $\psi^a$ given by \eqref{Homothety}. In general, affine collineations -- of which homotheties (and Killing fields) are special cases -- are associated with conserved quantities of the form
\begin{equation}
	\dot{z}_a \psi^a - \frac{1}{2} s \dot{z}^a \dot{z}^b \mathcal{L}_\psi g_{ab}
\end{equation}
for any affinely-parameterized geodesic with tangent $\dot{z}^a(s)$ \cite{KatzinLevine}. Using this for a geodesic with initial coordinates $(u',v',\mathbf{z}')$ and initial spatial velocity $\dot{\mathbf{z}}'$,
\begin{align}
	z^v(u) &= v' + \varepsilon (u-u') 
	\nonumber
	\\
	& ~ + \frac{1}{2} \big[ \mathbf{z}(u) \cdot \dot{\mathbf{z}}(u)  - \mathbf{z}' \cdot \dot{\mathbf{z}}'  \big] .
	\label{vGeodesic}
\end{align}
Here, $s$ has again been identified with $u$ and the constant $\varepsilon$ is defined by 
\begin{equation}
	\varepsilon = - \frac{1}{2} \dot{z}_\mu(u) \dot{z}^\mu(u).
	\label{EpsilonDef}
\end{equation}
All geodesics not confined to the hyperplane $S_{u'}$ have spatial coordinates which evolve via \eqref{GeodesicJacobi}. The evolution of their $v$ coordinates is easily found by combining \eqref{GeodesicJacobi} and \eqref{vGeodesic}. 

Given any two distinct points on a particular geodesic where $\bB^{-1}(u,u')$ exists, \eqref{GeodesicJacobi} may be used to solve for the spatial velocity in terms of the starting and ending points $\bx = \mathbf{z}(u)$ and $\bx' = \mathbf{z}(u')$:
\begin{subequations}
	\label{GeodesicVelocities}
	\begin{align}
	&\dot{\mathbf{z}}' = \bB^{-1} (\bx - \bA \bx'),
	\\
	&\dot{\mathbf{z}} = \partial_u \bA \bx' + \partial_u \bB \bB^{-1} (\bx - \bA \bx' ) .
	\end{align}
\end{subequations}
Using \eqref{vGeodesic}, the $v$ coordinate of such a geodesic is given by
\begin{align}
	z^v(u) = v' + \varepsilon (u-u') + \frac{1}{2} \Big[ \bx^\intercal  \partial_u \bA \bx' 
\nonumber
\\
	~ + ( \bx^\intercal \partial_u \bB - \bx'^\intercal) \bB^{-1} (\bx - \bA \bx' ) \Big].
\label{vGeodesic2}
\end{align}
The constant $\varepsilon$ is unconstrained. It follows that two points $p$ and $p'$ are connected by exactly one geodesic whenever $\det \bB(u,u') \neq 0$. It was mentioned above that pairs of points are also connected by exactly one geodesic when $u=u'$. Recalling \eqref{GenNormalNeighborhood} and the surrounding discussion, there therefore exists exactly one geodesic connecting any point $p \in \mathcal{N}(u')$ to any point $p'$ with phase coordinate $u'=u(p')$.

In all other cases, $p$ and $p'$ lie on conjugate hyperplanes. The  rank of the $2 \times 2$ matrix $\bB(u,u')$ is then strictly less than two. If a particular pair of hyperplanes is fixed together with spatial coordinates $\bx'$ on one of them, it follows from \eqref{GeodesicJacobi} that the space of all possible $\bx$ that can be reached by geodesics with initial spatial coordinates $\bx'$ has a dimension less than two. This implies that almost all points on conjugate hyperplanes are geodesically disconnected (although they are always connected by continuous non-geodesic curves). 

\begin{figure}
	\centering
	\includegraphics[width= .9 \linewidth]{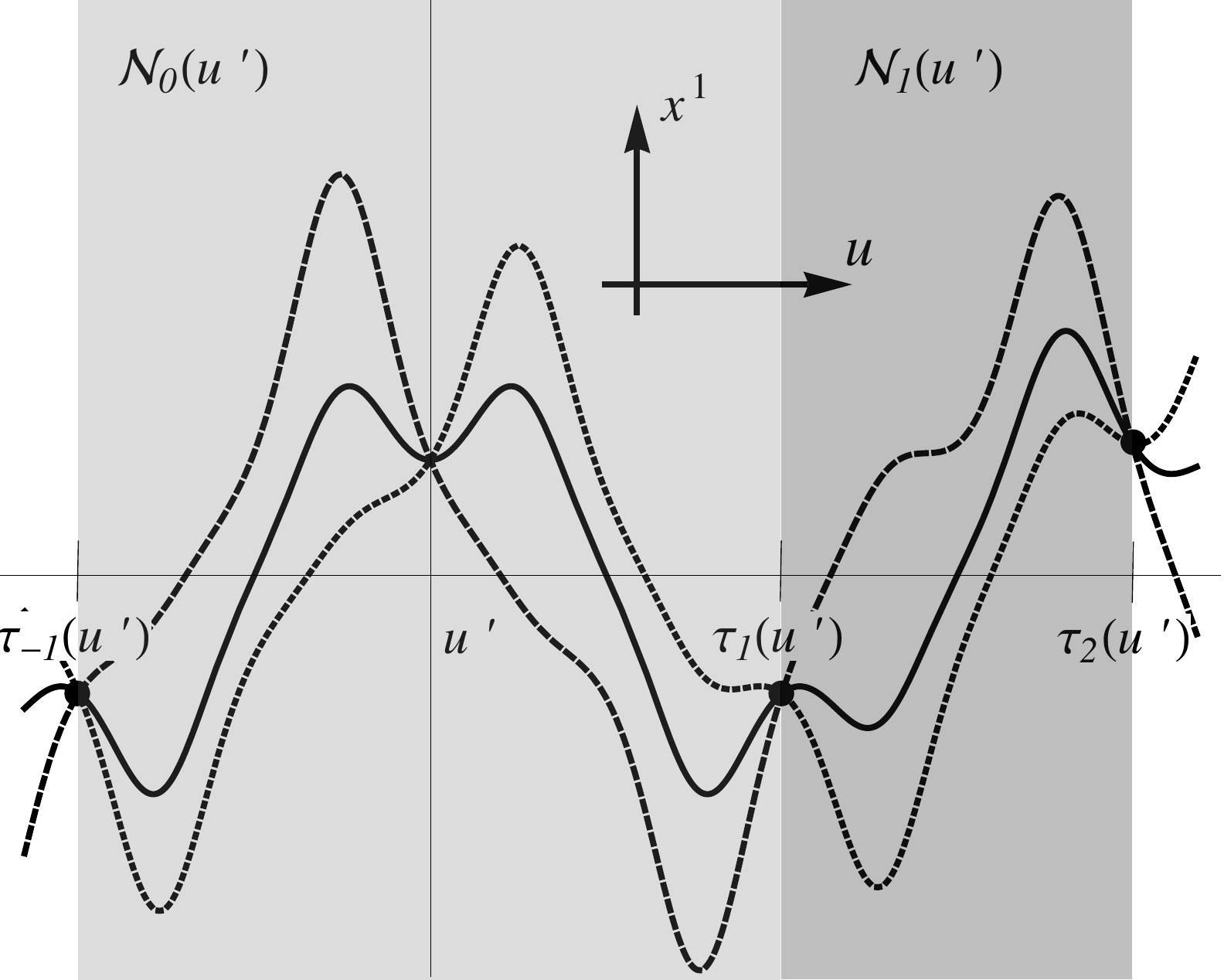}
	\caption{Schematic illustration of geodesic focusing in a plane wave spacetime. Three $u$ coordinates conjugate to $u'$ are indicated together with the projection of three geodesics onto the $x^1$-$u$ coordinate plane. The focusing here is assumed to act (at least) in the $x^1$-direction. 
		The regions $\mathcal{N}_0(u')$ and $\mathcal{N}_1(u')$ are also illustrated. See Figs. \ref{Fig:Mult2Focus} and \ref{Fig:Mult1Focus} for a different projection.}
		\label{Fig:GeodesicFocusing}
\end{figure}

Suppose, however, that two points lying on conjugate hyperplanes are known to be connected by one particular geodesic. If the initial data for this geodesic is modified by adding to its initial spatial velocity any nonzero vector in the null space of $\bB$, it follows from \eqref{GeodesicJacobi} that the spatial endpoints of this new geodesic will be the same as those of the original geodesic. Spacetime points that lie on conjugate hyperplanes and are connected by at least one geodesic are therefore connected by an infinite number of geodesics. See Fig. \ref{Fig:GeodesicFocusing}.

To summarize, all distinct pairs of points that do not lie on conjugate hyperplanes are connected by exactly one geodesic. Almost all points lying on conjugate hyperplanes fail to be connected by any geodesics. The remainder are connected by an infinite number of geodesics. More discussion may be found in Sect. \ref{Sect:CausticGeo} below.

\subsection{Conjugate points and Jacobi propagators}
\label{Sect:Conj}

In general, plane waves satisfying standard energy conditions focus geodesics. It is evident from the above discussion that, as claimed, pairs of points satisfying \eqref{ZeroDet} must be conjugate along any geodesic connecting them. We now show that \textit{all} conjugate points are of this form. Furthermore, we establish that the matrices $\bA$ and $\bB$ coincide with the spatial components of Jacobi propagators $A^{a}{}_{a'}$ and $B^{a}{}_{a'}$ (when the associated geodesic satisfies $\ell_a \dot{z}^a \neq 0$ and the affine parameter is identified with $u$). The full Jacobi propagators are known to be useful for computing ``generalized Killing fields'' which have found application in understanding the motion of extended matter distributions \cite{Dix70a, Dix74, HarteSyms, HarteScalar, HarteEM, HarteGrav}. Although they will not be used later, we provide explicit forms for the full Jacobi propagators in terms of their spatial components $\bA$ and $\bB$.   

By definition, two points $p$ and $p'$ lying on a particular geodesic are said to be conjugate if there exists a nontrivial Jacobi field on that geodesic which vanishes at both $p$ and $p'$ (see, e.g., \cite{Wald}). The multiplicity of a conjugate pair is defined to be the number of linearly independent Jacobi fields with this property. In general, the presence of conjugate points indicates that a family of geodesics starting at one point later intersect (or come arbitrarily close to intersecting). The multiplicity of a conjugate pair indicates the number of transverse directions that are so focused. 

Consider an affinely parameterized geodesic $z(s)$ as above. By definition, Jacobi fields $\xi^{a}(s)$ on this curve satisfy
\begin{equation}
	\frac{\mathrm{D}^{2} \xi^{a}}{\rmd s^{2}}  - R_{bcd}{}^{a} \xi^{b} \dot{z}^{c} \dot{z}^{d} 	=0 ,
	\label{Jacobi}
\end{equation}
where $\mathrm{D}/\rmd s$ denotes a covariant derivative in the direction $\dot{z}^a$. The Jacobi equation is linear, so its general solution has the form
\begin{equation}
	\xi^{a}(s) = A^{a}{}_{b'}(s,s') \xi^{b'}(s') + B^{a}{}_{b'}(s,s') \frac{\mathrm{D}  }{\rmd s'} \xi^{b'}(s') ,
	\label{JacobiPropDef}
\end{equation}
for some bitensors $A^{a}{}_{b'}(s,s')$ and $B^{a}{}_{b'}(s,s')$ that depend only on the spacetime metric and the chosen geodesic. These are called Jacobi propagators. 
They are solutions to
\begin{subequations}
\label{JacobiPropEqs}
\begin{eqnarray}
	0 &=&	\frac{\mathrm{D}^{2}}{\rmd s^{2}} A^{a}{}_{a'}  - R_{bcd}{}^{a} A^{b}{}_{a'} \dot{z}^{c}  \dot{z}^d ,
	\\
	&=& \frac{\mathrm{D}^{2}}{\rmd s^{2}} B^{a}{}_{a'}  - R_{bcd}{}^{a} B^{b}{}_{a'} \dot{z}^{c}  \dot{z}^d,
\end{eqnarray}
\end{subequations}
with the boundary conditions
\begin{subequations}
\label{ABBoundary}
\begin{align}
	\left[ \frac{\mathrm{D} A^{a}{}_{a'} }{\rmd s }  \right] = [ B^{a}{}_{a'} ] = 0,
	\\
	[A^{a}{}_{a'} ] = \left[   \frac{\mathrm{D} B^{a}{}_{a'} }{ \rmd s} \right] = \delta^{a}_{a'}.
\end{align}
\end{subequations}
If $s$ and $s'$ are not too widely separated, $A^{a}{}_{a'}(s,s')$ and $B^{a}{}_{a'}(s,s')$ may be written  explicitly in terms of Synge's function $\sigma(p,p')$ and its first two derivatives \cite{Dix70a}. Our main interest lies in geometric properties of plane waves outside of the normal neighborhood where the traditional derivation of such formulae breaks down. We therefore consign ourselves for now to somewhat less explicit comments on the Jacobi propagators that hold globally. It is shown in Sect. \ref{Sect:Bitensors} below that Synge's function may, in fact, be usefully extended beyond the normal neighborhood. It is, however, more convenient to express $\sigma$ in terms of the Jacobi propagators rather than writing the Jacobi propagators in terms of $\sigma$.

First note that for geodesics confined to a single constant-$u$ hypersurface, the Jacobi propagators are simply
\begin{equation}
A^{\mu}{}_{\mu'} = \delta^\mu_{\mu'}, \qquad B^{\mu}{}_{\mu'} = (s-s') \delta^{\mu}_{\mu'}
\end{equation}
in the Brinkmann coordinates where the metric takes the form \eqref{PlaneWaveGen}. It is therefore clear that there can be no conjugate points along any geodesic satisfying $\ell_a \dot{z}^a = 0$.

Now consider a geodesic where $\ell_a \dot{z}^a \neq 0$. As before, we may identify its affine parameter with $u$. Doing so, it is apparent from inspection of \eqref{Jacobi} that $\dot{z}^{a}(u)$ and $(u-u') \dot{z}^{a}(u)$ are both Jacobi fields. This means that
\begin{subequations}
\label{JacobiEig1}
\begin{eqnarray}
	A^{a}{}_{a'}(u,u') \dot{z}^{a'}(u') &=& \dot{z}^{a}(u),
	\\
	B^{a}{}_{a'}(u,u') \dot{z}^{a'}(u') &=& (u-u') \dot{z}^{a}(u).
\end{eqnarray}
\end{subequations}
Also note that the restriction of any affine collineation (such as a Killing vector) to a particular geodesic is a Jacobi field on that geodesic. This means, for example, that the covariantly-constant null vector $\ell^{a}$ associated with the direction of propagation of a generic plane wave spacetime can be used to generate Jacobi fields. So can $(u-u') \ell^{a}$. Hence,
\begin{subequations}
\label{JacobiEig2}
\begin{eqnarray}
	A^{a}{}_{a'}(u,u') \ell^{a'}(z(u')) &=& \ell^{a}(z(u)),
	\\
 	B^{a}{}_{a'}(u,u') \ell^{a'}(z(u')) &=& (u-u') \ell^{a}(z(u)).
\end{eqnarray}
\end{subequations}

Furthermore, application of \eqref{Riemann} and \eqref{JacobiPropEqs} shows that
\begin{subequations}
\label{lzDotAB}
\begin{eqnarray}
	0 &= & \partial^2_u ( \ell_a A^{a}{}_{a'} ) = \partial^2_u ( \ell_a B^{a}{}_{a'} ),
	\\
	&=& \partial^2_u ( \dot{z}_a A^{a}{}_{a'} ) = \partial^2_u ( \dot{z}_a B^{a}{}_{a'} ).
\end{eqnarray}
\end{subequations}
Using the coincidence limits \eqref{ABBoundary} of $A^{a}{}_{a'}$ and $B^{a}{}_{a'}$, the appropriate solutions to these differential equations are seen to be
\begin{subequations}
\label{lzDotAB2}
\begin{align}
	\ell_a A^{a}{}_{a'} (u,u') &= \ell_{a'}, 
	\\
	\ell_a B^{a}{}_{a'} (u,u') &= (u-u') \ell_{a'},
	\\
	\dot{z}_a A^{a}{}_{a'} (u,u') &= \dot{z}_{a'},
	\\
	\dot{z}_a B^{a}{}_{a'} (u,u') &= (u-u') \dot{z}_{a'}.
\end{align}
\end{subequations}
Also note that the components $A^{i}{}_{i'}$ and $B^{i}{}_{i'}$ in Brinkmann coordinates coincide with the matrices $\bA$ and $\bB$ defined by \eqref{JacobiSpatial} and \eqref{ABboundary} above.

If two points $z(u)$ and $z(u')$ (with $u \neq u'$) are conjugate on a particular geodesic, there must exist nonzero vectors $\lambda^{a'}$ at $z(u')$ such that
\begin{equation}
	 B^{a}{}_{a'}(u,u') \lambda^{a'} = 0.
 	\label{JacobiConjCond}
\end{equation}
Contracting this with $\ell_a$ and $\dot{z}_a$ while using \eqref{lzDotAB2} shows that
\begin{equation}
	\ell_{a'} \lambda^{a'} = \dot{z}_{a'} \lambda^{a'} = 0.
	\label{vOrtho}
\end{equation}
Applying \eqref{lzDotAB2} again then shows that \eqref{JacobiConjCond} can always be replaced by the weaker condition
\begin{equation}
	B^{i}{}_{\mu'} (u, u') \lambda^{\mu'} = 0.
	\label{BSimpInt}
\end{equation}
Using \eqref{JacobiEig2} and \eqref{vOrtho} further demonstrates that distinct points $z(u)$ and $z(u')$ in plane wave spacetimes are conjugate if and only if $u$ and $u'$ satisfy \eqref{ZeroDet}. As claimed at the beginning of Sect. \ref{Sect:Geometry}, all conjugate points may be identified by finding the zeros of $\det \bB(u,u')$. 

For completeness, we now write down all coordinate components of the Jacobi propagators using the eigenvector equations \eqref{JacobiEig1}, \eqref{JacobiEig2}, and \eqref{lzDotAB2}. Applying the relations involving $\ell^a$,
\begin{subequations}
\begin{align}
	A^{\mu}{}_{v'} = \delta^\mu_v , \qquad B^{\mu}{}_{v'} = (u-u') \delta^\mu_v,
\\
	A^{u}{}_{\mu'} = \delta^{u'}_{\mu'} , \qquad B^{u}{}_{\mu'} = (u-u') \delta^{u'}_{\mu'}	.
\end{align}
\end{subequations}
Using \eqref{vGeodesic},
\begin{subequations}
\begin{align}
	A^{i}{}_{u'} &= \big( \dot{\mathbf{z}} - \bA \dot{\mathbf{z}}' \big)^i,
	\\
	A^{v}{}_{i'} &= \big( \dot{\mathbf{z}}^\intercal \bA - \dot{\mathbf{z}}' \big)_{i'},
	\\
	A^{v}{}_{u'} &= \frac{1}{2} \big[ (\bx^\intercal \mathbf{H} \bx - \bx'^\intercal \mathbf{H}' \bx') 
	\nonumber
	\\
	&\qquad \quad + |\dot{\mathbf{z}}'|^2 + |\dot{\mathbf{z}}|^2  -2 \dot{\mathbf{z}}^\intercal \bA \dot{\mathbf{z}}'\big],
\end{align}
\end{subequations}
and
\begin{subequations}
\begin{align}
	B^{i}{}_{u'} &= \big[ (u-u') \dot{\mathbf{z}} - \bB \dot{\mathbf{z}}' \big]^i,
	\\
	B^{v}{}_{i'} &= \big[ \dot{\mathbf{z}}^\intercal \bB - (u-u') \dot{\mathbf{z}}' \big]_{i'},
	\\
	B^{v}{}_{u'} &= \frac{1}{2} (u-u') \big[(\bx^\intercal \mathbf{H} \bx - \bx'^\intercal \mathbf{H}' \bx') 
	\nonumber
	\\
	& \qquad ~ + |\dot{\mathbf{z}}'|^2 + |\dot{\mathbf{z}}|^2 - 2 \dot{\mathbf{z}}^\intercal \bB \dot{\mathbf{z}}'  \big].
\end{align}
\end{subequations}

Although the Jacobi propagators are defined along a particular geodesic, they are easily reinterpreted as bitensors \textit{on spacetime} for all pairs of points $p$ and $p'$ that do not lie on conjugate hyperplanes. This may be done explicitly by using \eqref{GeodesicVelocities} to replace $\dot{\mathbf{z}}$ and $\dot{\mathbf{z}}'$ with $\bx$, $\bx'$, $\bA$, and $\bB$. It is then straightforward to build vector fields on spacetime equal to Jacobi fields along all geodesics emanating from some preferred origin. Fixing that origin, the resulting fields form a 20-dimensional vector space. They can be interpreted as ``generalized affine collineations'' associated with the chosen origin \cite{HarteSyms}. A certain ten-dimensional subset generalize the Killing fields (and include any real Killing fields that may exist).

\subsection{Bitensors}
\label{Sect:Bitensors}

Two-point tensors (or bitensors) are useful for, among other things, the evaluation of Green functions in curved spacetimes. Foremost among these is Synge's world function $\sigma(p,p') = \sigma(p',p)$, which is equal to one-half of the squared geodesic distance between its arguments. This is typically defined only in those regions where $p$ and $p'$ can be connected by exactly one geodesic. More generally, it is possible to define a closely related two-point scalar $\sigma_z (s,s')$ associated with a particular geodesic $z(s)$:
\begin{equation}
\sigma_z(s,s') := \frac{1}{2} (s'-s) \int_{s}^{s'} g_{ab}(z(t)) \dot{z}^a (t) 	\dot{z}^b(t) \rmd t.
\label{SigDef}
\end{equation}
If the geodesic in this equation is the only geodesic connecting two points $z(s)$ and $z(s')$, the ordinary world function is related via
\begin{equation}
\sigma(z(s),z(s')) = \sigma_z(s,s') .
\label{TwoSigmas}
\end{equation} 
Sect. \ref{Sect:Geodesics} establishes that distinct points in plane wave spacetimes are connected by exactly one geodesic as long as they do not lie on conjugate hyperplanes. Eq. \eqref{TwoSigmas} may therefore be used to define Synge's function for all pairs of points that do not lie on conjugate hyperplanes.

The definition \eqref{SigDef} for $\sigma_z$ is straightforward to evaluate explicitly in the general plane wave metric \eqref{PlaneWaveGen} along any geodesic satisfying $\ell_{a} \dot{z}^{a} = -1$ (with, once again, $s$ identified with $u$). Integrating by parts and using \eqref{Geodesic} gives
\begin{equation}
\sigma_z (u,u') = \frac{1}{2} (u-u') \left. ( \mathbf{z} \cdot \dot{\mathbf{z}} - 2 z^{v} ) \right|_{u'}^{u}.
\label{Sigma1}
\end{equation}
Removing the $\dot{\mathbf{z}}$ appearing here using \eqref{GeodesicVelocities} and substituting the result into \eqref{TwoSigmas},
\begin{align}
\sigma(p,p') = \frac{1}{2} (u-u') \Big[ - 2 (v - v') + \bx^\intercal \partial_{u} \bA  \bx'   \nonumber
\\
~ + ( \bx^\intercal \partial_{u} \bB - \bx'^\intercal ) \bB^{-1}  ( \bx - \bA \bx' ) \Big] .
\label{SigmaGen}
\end{align}
$\partial_u \bA$ may be eliminated from this equation using \eqref{Abel} and the symmetry of $\partial_u \bB \bB^{-1}$ established in Appendix \ref{Sect:ABProperties}:
\begin{align}
\sigma(p,p') = \frac{1}{2} (u-u') \Big[ - 2 (v - v') + \bx^\intercal \partial_u \bB \bB^{-1} \bx 
\nonumber
\\
~ + \bx'^\intercal \bB^{-1} \bA \bx' - 2   \bx'^\intercal \bB^{-1} \bx \Big].
\label{SigmaGen2}
\end{align}
Both of these relations are valid as long as $u \notin T(u')$ and $u \neq u'$. If $u=u'$, Synge's function reduces to the Euclidean expression
\begin{equation}
\sigma(p,p') = \frac{1}{2}  |\bx -\bx'|^2.
\end{equation}
Note that unless $\bx$ and $\bx'$ are chosen in a very particular way, $\sigma$ diverges as $u \rightarrow \tau_n(u') \in T(u')$. This is a manifestation of the aforementioned fact that most pairs of points on conjugate hyperplanes cannot be connected by any geodesics. 

The (scalarized) van Vleck determinant $\Delta(p,p')$ is often defined\footnote{It is also common to define the van Vleck determinant as the solution to a certain ``transport equation'' along the geodesic connecting its arguments \cite{FriedlanderWaves, PoissonRev}. Unfortunately, $\Delta(p,p')$ becomes unbounded near conjugate hyperplanes. It is not clear how to unambiguously extend the solution of a differential equation through these singularities, so we adopt the definition \eqref{vanVleckDef} instead.} in terms of $\sigma(p,p')$ via
\begin{equation}
	\Delta(p,p') = - \frac{ \det [ - \nabla_\mu \nabla_{\mu'} \sigma(p,p') ] }{ \sqrt{-g(p)} \sqrt{-g(p')} } ,
\label{vanVleckDef}
\end{equation}
where $g(p) := \det g_{\mu\nu}(p)$. This is a biscalar with coincidence limit $[\Delta] = 1$. It is closely related to the expansion of the congruence of geodesics emanating from $p'$ \cite{PoissonRev}. Inserting \eqref{PlaneWaveGen} and \eqref{SigmaGen2} into \eqref{vanVleckDef} shows that
\begin{align}
	\Delta(p,p') = \det ( \partial_{i} \partial_{i'} \sigma ) = \frac{ (u-u')^{2} }{ \det \bB(u,u') } 
\label{vanVleck}
\end{align}
wherever $\sigma$ is defined. It is evident from \eqref{ZeroDet} that $\Delta(\cdot, p')$ is unbounded near any hyperplane conjugate to $S_{u'} \ni p'$. This is related to the well-known fact that the expansion of a congruence of geodesics emitted from a particular source point diverges when approaching a point that is conjugate to that source \cite{Wald}.

Recall that $\bB(u,u')$ has been assumed to remain everywhere finite [which follows from assuming that $\mathbf{H}(u)$ is sufficiently well-behaved]. Using this together with \eqref{vanVleckDef} shows that $\Delta(\cdot, p')$ can never pass through zero. It may switch signs, however. It is shown in Sect. \ref{Sect:CausticGeo} that this occurs only when passing through conjugate hyperplanes with multiplicity $1$. 

The parallel propagator $g^{a}{}_{a'}(p,p')$ is another important bitensor. Although this is not required to construct the scalar Green functions discussed in most of this paper, it does appear in Green functions associated with electromagnetic fields and metric perturbations \cite{PoissonRev}. We therefore include it for completeness. $g^{a}{}_{a'}(p,p')$ satisfies
\begin{equation}
		\frac{\mathrm{D} }{\rmd s} g^{a}{}_{a'}(z(s),z(s')) = 0
\label{ParPropDef}
\end{equation}
along any geodesic $z(s)$ connecting its two arguments. It also has the coincidence limit $[g^{a}{}_{a'} ] = \delta^a_{a'}$.  As the name implies, $g^{a}{}_{a'}(p,p')$ parallel transports vectors from $p'$ to $p$ (or covectors from $p$ to $p'$) when there is a unique geodesic connecting these points.

If $p$ and $p'$ are connected by a single geodesic with tangent $\dot{z}^a(u)$, it is clear that $\ell^a$ and $\dot{z}^a$ are both left- and right-eigenvectors of $g^{a}{}_{a'}(p,p')$:
\begin{subequations}
\begin{align}
	\ell_a g^{a}{}_{a'} = \ell_{a'}, \qquad g^{a}{}_{a'} \ell^{a'} = \ell^a,
	\\
	\dot{z}_a g^{a}{}_{a'} = \dot{z}_{a'}, \qquad g^{a}{}_{a'} \dot{z}^{a'} = \dot{z}^a	.
\end{align}
\end{subequations}
Applying the first of these equations demonstrates that
\begin{equation}
	g^{u}{}_{\mu'} = \delta^{u'}_{\mu'}, \qquad g^{\mu}{}_{v'} = \delta^{\mu}_{v}.
\end{equation}
A direct calculation using \eqref{ParPropDef} also shows that
\begin{equation}
	g^{i}{}_{i'} = \delta^{i}_{i'}.
\end{equation}
The remaining components of the parallel propagator are
\begin{subequations}
	\begin{align}
		g^{i}{}_{u'} = (\dot{\mathbf{z}} - \dot{\mathbf{z}}')^i, \qquad g^{v}{}_{i'} = (\dot{\mathbf{z}} - \dot{\mathbf{z}}')_{i'},
		\\
		g^{v}{}_{u'} = \frac{1}{2} \big[ ( \bx^\intercal \mathbf{H} \bx - \bx'^\intercal \mathbf{H}' \bx') + | \dot{\mathbf{z}} - \dot{\mathbf{z}}' |^2 \big] . 
	\end{align}
\end{subequations}
Applying \eqref{GeodesicVelocities} removes any reference to geodesic velocities in these equations:
\begin{subequations}
	\begin{align}
		g^{i}{}_{u'} = g^{v}{}_{i} = \big[ (\partial_u \bB - \bm{\delta}) \bB^{-1} (\bx - \bA \bx') \nonumber
	\\
		~  + \partial_u \bA \bx'  \big]_i,
	\\
		g^{v}{}_{u'} = \frac{1}{2} \big[ ( \bx^\intercal \mathbf{H} \bx - \bx'^\intercal \mathbf{H}' \bx') + |g^{i}{}_{u'}|^2 \big].
	\end{align}
\end{subequations}

\subsection{Effects of caustics}
\label{Sect:CausticGeo}

Essentially all difficulties related to deriving Green functions in plane wave spacetimes arise from the poor behavior of various bitensors near conjugate hyperplanes. This is, in turn, a consequence of geodesic non-uniqueness in these regions. We now discuss the structure of geodesics and the asymptotic forms of $\sigma(p,p')$ and $\Delta(p,p')$ near conjugate hyperplanes. Both of these topics depend on knowledge of the spatial Jacobi propagators $\bA(u,u')$ and $\bB(u,u')$ near conjugate hyperplanes.

\subsubsection{$\bA$ and $\bB$ near conjugate hyperplanes}

It is simplest to discuss the behavior of the spatial Jacobi propagators near conjugate hyperplanes associated with degenerate (multiplicity 2) conjugate points. Consider a particular $\tau_n(u') \in T(u')$ where
\begin{equation}
	\hat{\bB}_n(u') := \bB (\tau_n(u'),u') = 0.
\end{equation}
It follows from \eqref{Abel} that $\hat{\bA}_n(u') := \bA(\tau_n(u'),u')$ and $\widehat{\partial_u \bB}_n (u') := \partial_u \bB(u,u')|_{u=\tau_n(u')}$ are both invertible [even though $\hat{\bB}_n(u')$ is not]. Furthermore,
\begin{equation}
	\widehat{\partial_u \bB}_n = (\hat{\bA}_n^{-1})^\intercal \neq 0.
	\label{BDotDeg}
\end{equation}

Using these expressions to expand $\bB(u,u')$ when $u$ is near $\tau_n$,
\begin{equation}
	\bB(u,u') \sim (u-\tau_n) (\hat{\bA}_n^{-1})^\intercal .
\end{equation}
The ``$\sim$'' symbol is used here to denote a relation that holds asymptotically as $u \rightarrow \tau_n$. To leading order, the determinant of $\bB(u,u')$ in this limit is
\begin{equation}
	\det \bB(u,u') \sim \frac{(u-\tau_n)^2}{\det \hat{\bA}_n} .
	\label{DetBDeg}
\end{equation}
Its inverse is clearly
\begin{equation}
	\bB^{-1}(u,u') \sim (u-\tau_n)^{-1} \hat{\bA}_n^\intercal .
	\label{InvBDeg}
\end{equation}

Now consider a different $\tau_n \in T(u')$ associated with a \textit{non}-degenerate (multiplicity 1) conjugate point. $\hat{\bB}_n$ is then nonzero and has matrix rank $1$. Using the matrix determinant lemma \cite{NumericalMethods} together with \eqref{Abel}, the Jacobi propagators at $(\tau_n(u'),u')$ are easily seen to satisfy
\begin{equation}
	\det \hat{\bA}_n \det \widehat{\partial_u \bB}_n = 1 + \mathrm{Tr} \, ( \widehat{\partial_u \bA}_n^\intercal \hat{\bB}_n ).
	\label{DetCond}
\end{equation}
We shall only consider cases where
\begin{equation}
	\mathrm{Tr} \, ( \widehat{\partial_u \bA}_n^\intercal \hat{\bB}_n ) \neq -1.
	\label{TrCond}
\end{equation}
This is a technical condition which we use to ensure that $\widehat{\partial_u \bB}_n$ is invertible. There do exist plane waves where \eqref{TrCond} is violated, but these examples must be very finely tuned.

Applying the matrix determinant lemma gives an approximation for the determinant of $\bB(u,u')$ when $u$ is near $\tau_n$. To lowest nontrivial order,
\begin{align}
	\det \bB(u,u') \sim (u-\tau_n)  \mathrm{Tr} \, [ \hat{\bB}_n (\widehat{\partial_u \bB}_n)^{-1} ]
	\nonumber
	\\
	~ \times \det \widehat{\partial_u \bB}_n.
	\label{DetBSimp}
\end{align}
The inverse of $\bB(u,u')$ in the limit $u \rightarrow \tau_n$ follows from the Sherman-Morrison formula \cite{NumericalMethods}:
\begin{align}
	\bB^{-1} (u,u') \sim ~ \! & (u-\tau_n)^{-1}  (\widehat{\partial_u \bB}_n)^{-1} 
	\nonumber
	\\
	& ~ \times \bigg( \bm{\delta}   - \frac{ \hat{\bB}_n (\widehat{\partial_u \bB}_n)^{-1} }{ \mathrm{Tr} \, [ \hat{\bB}_n (\widehat{\partial_u \bB}_n)^{-1}] } \bigg) .
	\label{InvBSimp}
\end{align}

Assumption \eqref{TrCond} implies that $\widehat{\partial_u \bB_n}$ has matrix rank $2$. $\hat{\bB}_n$ has rank $1$ for the non-degenerate case considered here, so $\hat{\bB}_n (\widehat{\partial_u \bB}_n)^{-1}$ must also have rank $1$. It is shown in Appendix \ref{Sect:ABProperties} that $\hat{\bB}_n (\widehat{\partial_u \bB}_n)^{-1}$ is also symmetric. As a result, it possesses one nonzero eigenvalue and a non-vanishing trace. The matrix in parentheses in \eqref{InvBSimp} is therefore a two-dimensional projection operator. It is symmetric with eigenvalues $0$ and $1$. There therefore exists a unit vector $\hat{\mathbf{q}}_n$ satisfying
\begin{equation}
	\bigg( \bm{\delta}   - \frac{ \hat{\bB}_n (\widehat{\partial_u \bB}_n)^{-1} }{ \mathrm{Tr} \, [ \hat{\bB}_n (\widehat{\partial_u \bB}_n)^{-1}] } \bigg) \hat{\mathbf{q}}_n = \mathbf{\hat{q}}_n.
	\label{nDef}
\end{equation}
This is unique up to a sign. In terms of $\hat{\mathbf{q}}_n$,
\begin{align}
	\bB^{-1}(u,u') \sim (u-\tau_n)^{-1} (\widehat{\partial_u \bB}_n)^{-1} (\hat{\mathbf{q}}_n \otimes \hat{\mathbf{q}}_n).
	\label{BInvDiag}
\end{align}
Eq. \eqref{nDef} may be used to see that $\hat{\mathbf{q}}_n$ is in the (left-) null space of $\hat{\bB}_n$:
\begin{equation}
	\hat{\mathbf{q}}_n^\intercal \hat{\bB}_n = 0. 
	\label{nDotB}
\end{equation}

Using \eqref{Abel}, the $u$-derivative of $\bB$ on a non-degenerate conjugate hyperplane is given by
\begin{equation}
	\widehat{\partial_u \bB}_n =  (\hat{\bA}^{-1}_n)^\intercal ( \bm{\delta} + \widehat{\partial_u \bA}_n^\intercal \hat{\bB}_n ) . 
\label{BDotSimp}
\end{equation}
This may be substituted into \eqref{DetBSimp} and \eqref{InvBSimp} in order to provide explicit approximations for the inverse and determinant of $\bB(u,u')$ near a non-degenerate conjugate hyperplane in terms of $\hat{\bA}_n$, $\widehat{\partial_u \bA}_n$, and $\hat{\bB}_n$. Eqs. \eqref{DetBDeg} and \eqref{InvBDeg} serve the same purpose for degenerate conjugate points. 

$\bB$ may be interpreted as a ``focusing matrix.'' Near degenerate conjugate points, \eqref{InvBDeg} implies that $\bB^{-1}$ diverges in both spatial directions. In the case of a non-degenerate conjugate point, \eqref{BInvDiag} illustrates how $\bB^{-1}$ diverges in only ``one direction.'' This distinction is closely related to the behavior of geodesics near conjugate points with different multiplicities.

\subsubsection{Geodesics on conjugate hyperplanes}
\label{Sect:GeoConj}

It is shown in Sect. \ref{Sect:Geodesics} that pairs of points on conjugate hyperplanes are connected either by an infinite number of geodesics or by none. This is a consequence of the fact that all geodesics -- timelike, spacelike, or null -- emanating from a single point and reaching a conjugate hyperplane are focused down to a one- or two-dimensional region on that three-dimensional surface. The null geodesics are focused to either a point or a line. We now demonstrate that the latter case occurs on non-degenerate conjugate hyperplanes, and is an example of astigmatic focusing. Scenarios where all null geodesics momentarily focus to a single point occur only on degenerate conjugate hyperplanes. This is anastigmatic focusing.

\begin{figure}[t!]
	\centering
	\includegraphics[width=1\linewidth]{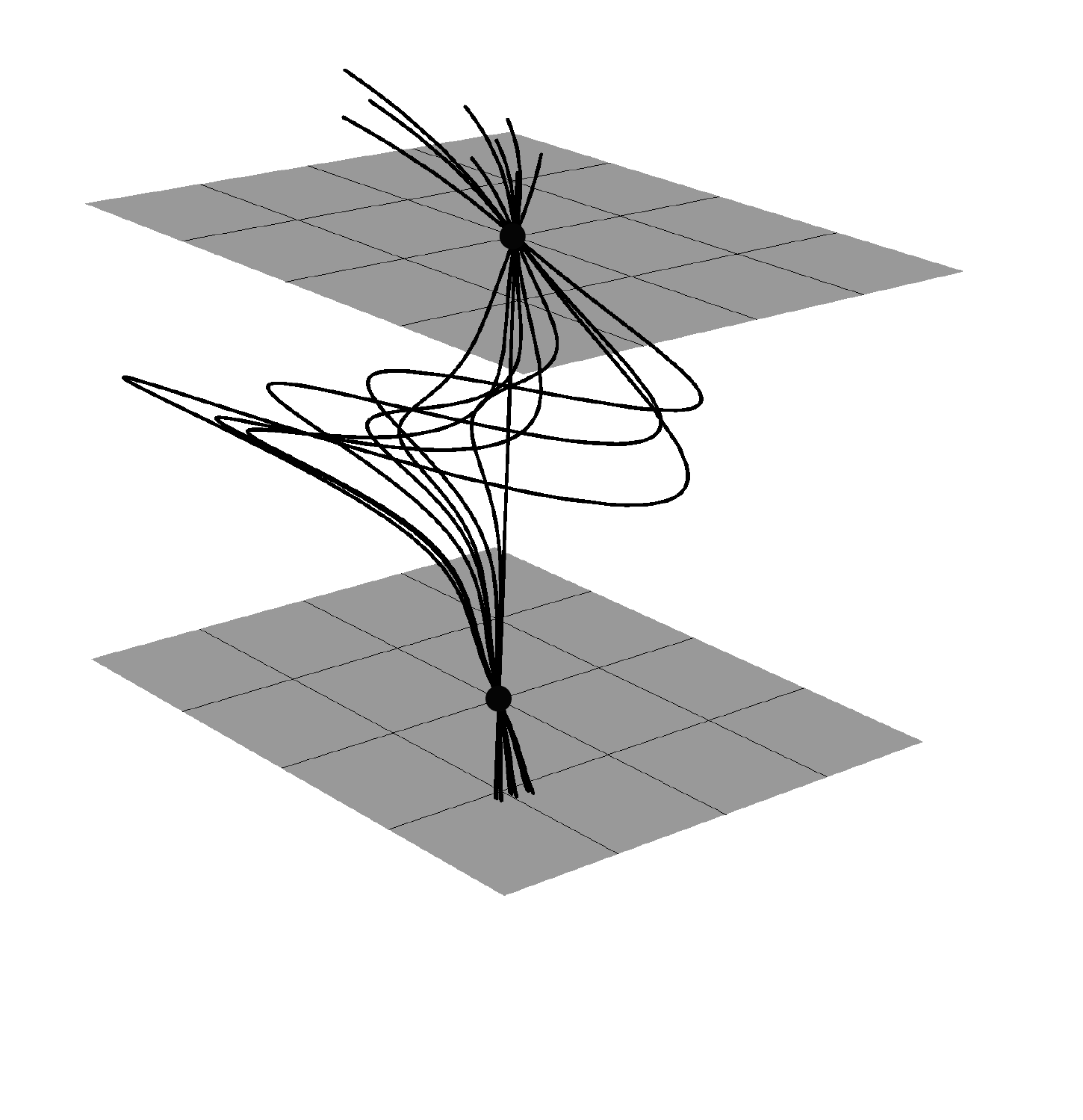}
	\vskip -.35 cm
	\caption{A collection of null geodesics emanating from a point on $S_{u'}$ (the lower plane) and focusing back to a single point on a conjugate hyperplane $S_{\tau_n(u')}$ with multiplicity 2 (the upper plane). One spatial coordinate has been suppressed.}
	\label{Fig:Mult2Focus}
\end{figure}

Consider a point $p'$ and a constant-$u$ hyperplane $S_{\tau_n(u')}$ that is conjugate to $S_{u'}$. Suppose that the conjugate points associated with this pair are degenerate, so $\hat{\bB}_n = \bB(\tau_n,u') = 0$. It then follows from \eqref{GeodesicJacobi} that all geodesics (of any type) emanating from a particular point $p'$ with spatial coordinates $\bx'$ are focused to
\begin{equation}
	\bx = \hat{\bA}_n(u') \bx'
\label{xFocusDeg}
\end{equation}
as they pass through $S_{\tau_n(u')}$. Use of \eqref{GeodesicJacobi}, \eqref{vGeodesic}, and \eqref{BDotDeg} then shows that the $v$ coordinates of all geodesics are given by
\begin{equation}
	v' + \varepsilon (\tau_n-u') + \frac{1}{2} \bx'^\intercal \hat{\bA}_n^\intercal \widehat{\partial_u \bA}_n \bx'
\end{equation}
on $S_{\tau_n}$. The only free parameter here is $\varepsilon$, which is defined by \eqref{EpsilonDef}. The set of all geodesics emanating from $p'$ are therefore focused down to a line on $S_{\tau_n}$. Null geodesics are all characterized by $\varepsilon=0$, and are therefore focused down to a single point. See Fig. \ref{Fig:Mult2Focus}. Almost the entire null cone of a point $p'$ is focused to a point on every degenerate hyperplane conjugate to $S_{u'}$. The lone exception is the null geodesic generated by $\ell^a$. This lies entirely in $S_{u'}$, so it never passes through any conjugate hyperplanes. 

The situation is slightly more complicated for non-degenerate conjugate points. In these cases,  $\hat{\bB}_n$ can be written as the outer product of two nonzero vectors:
\begin{equation}
	\hat{\bB}_n = \hat{\mathbf{p}}_n \otimes \hat{\mathbf{m}}_n.
\label{BRank1}
\end{equation}
There is no loss of generality in supposing that $| \hat{\mathbf{p}}_n |^2 =1$. Substitution of \eqref{BRank1} into \eqref{GeodesicJacobi} shows that the spatial components of all geodesics starting at $\bx'$ lie on the line
\begin{equation}
	\bx = \hat{\bA}_n \bx'+ t \hat{\mathbf{p}}_n
\label{xFocusSimp}
\end{equation}
as they pass through $S_{\tau_n}$. The parameter $t = \hat{\mathbf{m}}_n \cdot \dot{\mathbf{z}}'$ appearing in this equation can be any real number. Combining \eqref{BRank1} with \eqref{nDotB} shows that
\begin{equation}
\hat{\mathbf{p}}_n \cdot \hat{\mathbf{q}}_n = 0,
\label{nmOrtho}
\end{equation}
where $\hat{\mathbf{q}}_n$ is defined by \eqref{nDef}. In this sense, a plane wave may be thought of as focusing geodesics in the direction $\hat{\mathbf{q}}_n$.

\begin{figure}[t!]
	\centering
	\includegraphics[width=.95\linewidth]{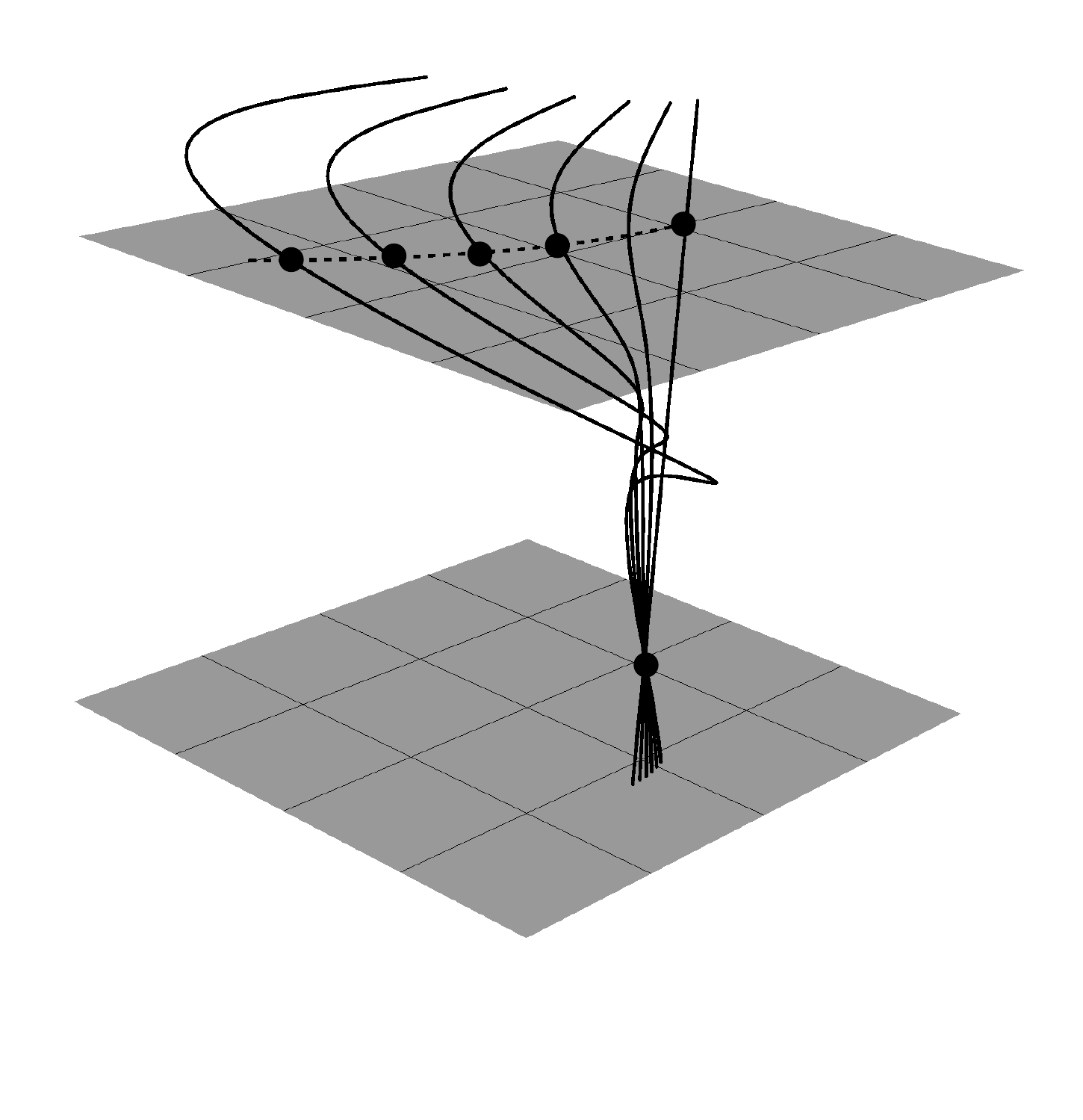}	
	\caption{A collection of null geodesics emanating from a point on $S_{u'}$ (the lower plane) and focusing to a parabola on a conjugate hyperplane $S_{\tau_n(u')}$ with multiplicity 1 (the upper plane). One spatial coordinate aligned with $\hat{\mathbf{p}}_n(u')$ is displayed. The other [aligned with $\hat{\mathbf{q}}_n(u')$] is suppressed.}
	\label{Fig:Mult1Focus}
\end{figure}

Eqs. \eqref{nDef} and \eqref{nmOrtho} imply that $\hat{\mathbf{p}}_n$ is an eigenvector of $\hat{\bB}_n ( \widehat{\partial_u \bB}_n )^{-1}$. In particular,
\begin{equation}
	\hat{\bB}^\intercal_n \hat{\mathbf{p}}_n = \mathrm{Tr} \, [\hat{\bB}_n ( \widehat{\partial_u \bB}_n )^{-1} ] ( \widehat{\partial_u \bB}_n )^\intercal \hat{\mathbf{p}}_n.
\end{equation}
Using this together with \eqref{GeodesicJacobi}, \eqref{vGeodesic}, \eqref{xFocusSimp}, and \eqref{Abel}, the $v$ coordinates of geodesics starting at a single point and passing through a non-degenerate conjugate hyperplane $S_{\tau_n}$ satisfy
\begin{align}
	v' + \varepsilon (\tau_n - u') + \frac{1}{2} \bx'^\intercal \hat{\bA}_n^\intercal \widehat{ \partial_u \bA}_n \bx' +
	t \hat{\mathbf{p}}_n^\intercal \widehat{\partial_u \bA}_n \bx' 
	\nonumber
	\\
	~ + \frac{1}{2} t^2 \mathrm{Tr}\, [\hat{\bB}_n ( \widehat{\partial_u \bB}_n )^{-1} ]^{-1} .
\end{align}
There are two free parameters here: $\varepsilon$ and $t$. For null geodesics, $\varepsilon$ vanishes. The intersection of $S_{\tau_n}(u')$ with the light cone of a point $p'$ is therefore a one-dimensional curve. It is a parabola in the coordinates $( v,\bx)$. See Fig. \ref{Fig:Mult1Focus}.

\subsubsection{Bitensors near conjugate hyperplanes}

The bitensors discussed in Sect. \ref{Sect:Bitensors} are not defined if their arguments lie on conjugate hyperplanes. Despite this, expansions for $\bB(u,u')$ obtained above may be used to understand how $\sigma(p,p')$ and $\Delta(p,p')$ behave near conjugate hyperplanes.

First consider $\sigma(p,p')$ if the $u$ coordinate of $p$ is close (but not equal) to some $\tau_n(u') \in T(u')$ associated with degenerate conjugate points. In this region, the unbounded matrix $\bB^{-1}(u,u')$ almost always dominates in \eqref{SigmaGen}. Using that equation together with \eqref{BDotDeg} and \eqref{InvBDeg},
\begin{equation}
	\sigma(p,p') \sim - \frac{1}{2} \left( \frac{\tau_n-u'}{\tau_n-u} \right) |\bx - \hat{\bA}_n \mathbf{x'} |^2 .
	\label{SigmaDeg}
\end{equation}
As before, $\hat{\bA}_n := \bA(\tau_n,u')$. It follows that $\sigma$ diverges as one approaches a degenerate conjugate hyperplane except if the approach is at the special spatial coordinates $\bx = \hat{\bA}_n \mathbf{x'}$. Recalling \eqref{xFocusDeg}, these are the coordinates to which all geodesics emanating from $p'$ are focused to on $S_{\tau_n}$. 

The behavior of the van Vleck determinant near a degenerate conjugate hyperplane is easily found using \eqref{vanVleck} and \eqref{DetBDeg}:
\begin{equation}
	\Delta(p,p') \sim \left( \frac{\tau_n-u'}{\tau_n-u} \right)^2 \det \hat{\bA}_n .
	\label{VanVleckDeg}
\end{equation}
$\det \hat{\bA}_n$ cannot vanish, so $\Delta$ always diverges like $(\tau_n-u)^{-2}$  as $u \rightarrow \tau_n$. Also note that the van Vleck determinant retains its sign before and after the singularity.

Similar equations may be derived if $\tau_n$ is associated with a non-degenerate conjugate point. First note that in this case, \eqref{Abel}, \eqref{nDotB}, and the symmetry of $\bB (\partial_u \bB)^{-1}$ imply that
\begin{equation}
	(\hat{\mathbf{q}}_n \otimes \hat{\mathbf{q}}_n ) \hat{\bA}_n = (\hat{\mathbf{q}}_n \otimes \hat{\mathbf{q}}_n ) [(\widehat{\partial_u \bB}_n)^{-1}]^\intercal,
\end{equation}
where $\hat{\mathbf{q}}_n$ is defined by \eqref{nDef}. Using this identity together with \eqref{SigmaGen} and \eqref{BInvDiag} establishes that
\begin{align}
	\sigma(p,p') \sim - \frac{1}{2} \left(\frac{ \tau_n-u' }{\tau_n - u} \right) [ \hat{\mathbf{q}}_n \cdot ( \bx - \hat{\bA}_n \bx' ) ]^2 .
	\label{SigmaSimp}
\end{align}
It is clear from this equation and \eqref{nmOrtho} that $\sigma$ diverges as $u \rightarrow \tau_n$ unless $\bx$ satisfies \eqref{xFocusSimp}. This is analogous to what occured in the case of degenerate point: $\sigma$ diverges as one attempts to approach pairs of points that are not connected by any geodesics.

The behavior of the van Vleck determinant near a non-degenerate conjugate point is easily determined using \eqref{DetBSimp} and \eqref{vanVleck}:
\begin{align}
	\Delta(p,p') \sim  -  \frac{ (\tau_n - u)^{-1} (\tau_n-u')^2 }{ \mathrm{Tr} [ \hat{\bB}_n (\widehat{\partial_u \bB}_n)^{-1} ] \det (\widehat{\partial_u \bB}_n) }.
	\label{VanVleckSimp}
\end{align}
Note that this diverges more slowly as $u \rightarrow \tau_n$ than in the case of degenerate conjugate hyperplanes (as expected due to the weaker focusing). It is also evident that $\Delta$ switches sign after passing through a non-degenerate conjugate hyperplane. 

\subsection{Causal structure}
\label{Sect:Causality}

Plane wave spacetimes are not globally hyperbolic \cite{PenrosePlane}. This is easily confirmed by considering two points $p$ and $p'$ that are conjugate along some causal geodesic with initial spatial velocity $\dot{\mathbf{z}}'$. As argued in Sect. \ref{Sect:Geodesics}, such points are connected by an infinite number of geodesics. Indeed, they are connected by an infinite number of \textit{causal} geodesics. $p$ and $p'$ are conjugate on all of them.

This may be seen by considering a causal geodesic connecting two points $p$ and $p'$. Suppose that $u(p)=\tau_n(u')$. If the affine parameter of the connecting geodesic is identified with $u$, consider a new geodesic (with the same affine parameter) where the initial data is shifted such that
\begin{align}
	\dot{\mathbf{z}}' &\rightarrow \dot{\mathbf{z}}' + t \bm{\lambda},
\\
	\varepsilon &\rightarrow \varepsilon + \frac{t}{2} \left( \frac{ \mathbf{z}' \cdot \bm{\lambda} - \mathbf{z} \cdot (\widehat{\partial_u \bB}_n \bm{\lambda} ) }{ \tau_n-u' } \right).
\end{align}
Here, $\bm{\lambda}$ is any vector satisfying $\hat{\bB}_n \bm{\lambda} = 0$ and $t \in \mathbb{R}$. $\varepsilon$ denotes the constant defined by \eqref{EpsilonDef}. It is easily verified that the resulting geodesic still passes through $p$ and $p'$. Indeed, varying $t$ produces a 1-parameter family of geodesics passing through these points. It is clear that $t$ may be increased without bound in at least one direction while retaining the causal nature of the geodesics (implied by $\varepsilon \geq 0$). There therefore exist causal geodesics connecting $p$ and $p'$ with arbitrarily large initial velocities.

Recall that $\bB(u,u')$ has maximal rank when $u \notin T(u')$. It then follows from \eqref{GeodesicJacobi} and the unboundedness of $| \dot{\mathbf{z}}'|$ that there exist causal geodesics connecting $p$ and $p'$ that reach arbitrarily large values of $|\bx|$ between these points. If $p$ is in the future of $p'$, $(\mathrm{causal \, past \, of} \, p) \cap (\mathrm{causal \, future \, of} \, p')$ is therefore unbounded. Global hyperbolicity requires that all such sets be compact, so plane waves with conjugate points cannot be globally hyperbolic.

One consequence of this is that causally-connected points can fail to be connected by any causal geodesics. Certain pairs of points connected by spacelike geodesics (and not by any other types of geodesic) may also be connected by accelerated curves that are everywhere causal. Avez and Seifert have shown that this cannot happen in globally hyperbolic spacetimes \cite{Avez, Seifert}.

To see that this does indeed occur in plane wave spacetimes, consider two points $p$ and $p'$ that do not lie on conjugate hyperplanes. Suppose that $u>u'$, and that there exists exactly one $\tau_n(u') \in T(u')$ that lies between $u$ and $u'$. The discussion in Sect. \ref{Sect:Geodesics} implies that $p$ and $p'$ are connected by a unique geodesic. That geodesic is spacelike whenever 
\begin{equation}
\sigma(p,p') > 0.
\label{SpacelikeSep}
\end{equation}	

Choose a third point $p''$, where $u'' \neq \tau_n$ lies between $u$ and $u'$. Consider a curve constructed by stitching together the unique geodesic connecting $p'$ to $p''$ with the unique geodesic connecting $p''$ to $p$. We now show that it is possible to choose $p''$ such that, despite \eqref{SpacelikeSep}, both of these geodesics are causal:
\begin{equation}
\sigma(p',p'') \leq 0, \; \qquad \sigma(p'',p) \leq 0.
\label{PiecewiseCausal}
\end{equation}

Suppose for definiteness that $u'' = \tau_n - \epsilon$ for some $\epsilon>0$.  If the $\bx''$ are not spatial coordinates to which geodesics starting at $\bx'$ must focus to as $u \rightarrow \tau_n$, the expansions \eqref{SigmaDeg} and \eqref{SigmaSimp} show that $\sigma(p',p'')$ can be made arbitrarily negative by choosing $\epsilon$ to be sufficiently small. Essentially any geodesic from $p'$ to $p''$ can therefore be made timelike by placing $u''$ sufficiently close to (but less than) $\tau_n$. One then needs to choose $v''$ and $\bx''$ such that $\sigma(p'',p) \leq 0$. $v''$ is entirely free, while $\bx''$ is only constrained not to equal \eqref{xFocusDeg} or \eqref{xFocusSimp} (with $\bx \rightarrow \bx''$). These parameters can always be adjusted to ensure that the geodesic from $p''$ to $p$ is causal. It follows that all pairs of points separated by exactly one conjugate hyperplane are causally connected. This is true despite that some such pairs are not connected by any causal geodesics.

This argument may be extended to points separated by multiple conjugate hyperplanes using curves constructed by stitching together increasing numbers of geodesic segments. The result is the same: \textit{Any two points separated by at least one conjugate hyperplane are in causal contact}. More discussion of causality in plane wave -- and more generally, pp-wave -- spacetimes may be found in \cite{Causalpp}.

\subsection{Examples}
\label{Sect:Examples}

We now illustrate the concepts just discussed by considering three examples of plane wave spacetimes.

\subsubsection{Symmetric electromagnetic plane wave}

The simplest nontrivial plane wave spacetime is a symmetric conformally-flat geometry whose associated stress-energy tensor satisfies the weak energy condition. Following the discussion in Sect. \ref{Sect:Plane}, the profile $\mathbf{H}(u)$ of such a wave is given by $\mathbf{H}(u) = - h^2 \bm{\delta}$ for some constant $h^2 > 0$. Rescaling the $u$ and $v$ coordinates, there is no loss of generality in setting $h = 1$:
\begin{equation}
	\mathbf{H} = - \bm{\delta}.
\end{equation}
Hence,
\begin{equation}
	\rmd s^2 = - 2 \rmd u \rmd v - |\bx|^2 \rmd u^2 + |\rmd \bx |^2.
	\label{BrinkmannConfFlatEx}
\end{equation}

Recalling \eqref{EMField}, this metric may be interpreted as the geometry associated with the electromagnetic field
\begin{equation}
	F_{ab} = 2 \nabla_{[a} u \nabla_{b]} x^1.
\end{equation}
A timelike geodesic observer at the spatial origin $\bx = 0$ and with the unit 4-velocity
\begin{equation}
	\dot{z}^a = \frac{1}{\sqrt{2}} \left( \frac{\partial}{\partial u}  + \frac{\partial}{\partial v} \right)^a 
\end{equation}
would view $F_{ab}$ as being composed of crossed electric and magnetic fields with constant (and equal) magnitude lying in the $x^1$-$x^2$ plane:
\begin{subequations}
\begin{align}
	E_a := F_{ab} \dot{z}^b &= - \frac{1}{\sqrt{2}} (\rmd x^1)_a ,
	\\
	B^a := - \frac{1}{2} \epsilon^{abcd} \dot{z}_b F_{cd} &= - \frac{1}{\sqrt{2}} \left( \frac{\partial}{\partial x^2} \right)^a.
\end{align}
\end{subequations}
Also note that $F_{ab}$ is covariantly constant everywhere.

Regardless of interpretation, it follows from \eqref{JacobiSpatial} and \eqref{ABboundary} that the spatial components of the Jacobi propagators are 
\begin{subequations}
\label{BEx1}	
\begin{align}
	\bA(u,u') = \bm{\delta} \cos (u-u') ,
\\
	\bB(u,u') = \bm{\delta} \sin (u-u') .	
\end{align}
\end{subequations}
It is evident from \eqref{ZeroDet} that the conjugate hyperplanes are equally spaced and occur at the $u$ coordinates
\begin{equation}
	\tau_n(u') = u' + n \pi ,
	\label{TauEx}
\end{equation}
where $n$ is any nonzero integer. In these spacetimes, there are an infinite number of conjugate points along any (inextendible) geodesic satisfying $\ell_a \dot{z}^a \neq 0$. All of these conjugate points have multiplicity $2$. Regardless of initial velocity, all geodesics with initial spatial coordinates $\mathbf{z}(u')$ on $S_{u'}$ have spatial coordinates $(-1)^n \mathbf{z}(u')$ on $S_{\tau_n(u')}$. 

The van Vleck determinant is easily computed using \eqref{vanVleck} and \eqref{BEx1}: 
\begin{equation}
	\Delta(p,p') = \left[ \frac{  ( u-u' ) }{ \sin (u-u') } \right]^2.
	\label{VVConfFlat}
\end{equation}
As expected from the discussion in Sect. \ref{Sect:CausticGeo}, $\Delta(p,p')$ is positive everywhere it is defined and diverges like $(\tau_n - u)^{-2}$ if $u \rightarrow \tau_n (u')$. 

For reference, Synge's function may be computed using \eqref{SigmaGen} and \eqref{BEx1}:
\begin{align}
	\sigma = \frac{1}{2} (u-u') \Big[ -2 (v-v') + \cot (u-u') 
	\nonumber
	\\
	~ \times \left( | \bx|^2 + | \bx'|^2 - 2 \bx \cdot \bx' \sec (u-u') \right)\Big].
	\label{SigConfFlat}
\end{align}

In terms of the Rosen coordinates $(U,V,\mathbf{X})$ discussed at the end of Sect. \ref{Sect:Plane}, the metric of a homogeneous conformally-flat plane wave may be written in the form \eqref{RosenMetric} with, e.g.,
\begin{equation}
	\bm{\mathcal{H}}(U) = \bA^\intercal(U,U') \bA(U,U') = \bm{\delta} \cos^2 (U-U').
	\label{HRosenConfFlat}
\end{equation}
Here, $U'$ is interpreted as an arbitrary parameter. It is evident that (no matter the choice of $U'$), there exist values of $U$ where $\bm{\mathcal{H}}(U) = 0$. The Rosen metric is singular at these points even though the Brinkmann metric \eqref{BrinkmannConfFlatEx} is everywhere well-defined.

\subsubsection{Symmetric gravitational plane wave}

The simplest example of a plane wave admitting non-degenerate conjugate points is the linearly polarized and symmetric gravitational wave described by 
\begin{equation}
\mathbf{H}(u) = \left(
	\begin{matrix}
	1 & 0 \\ 
	0 & -1
	\end{matrix}
	\right) .
	\label{HGravWave}
\end{equation}
This has vanishing trace, so the resulting spacetime satisfies the vacuum Einstein equation. The geometry represents a ``pure'' gravitational wave.

It is easily verified that
\begin{equation}
	\bA(u,u') = \left(
	\begin{matrix}
		\cosh (u-u') & 0 \\ 
		0 & \cos (u-u')
	\end{matrix}
	\right),
\label{AGravWave}
\end{equation}
and 
\begin{equation}
	\bB(u,u') = \left(
	\begin{matrix}
		\sinh (u-u') & 0 \\ 
		0 & \sin (u-u')
	\end{matrix}
	\right).
\label{BGravWave}
\end{equation}
The conjugate hyperplanes are again given by \eqref{TauEx}. Unlike in the previous example, however, the associated conjugate points all have multiplicity 1: Focusing occurs only in the $x^2$-direction.

The van Vleck determinant for this spacetime is
\begin{equation}
	\Delta(p,p') = \frac{ (u-u')^2 }{ \sin  (u-u') \sinh  (u-u')}.
	\label{VVRiccFlat}
\end{equation}
Following the general trends derived in Sect. \ref{Sect:CausticGeo}, $\Delta(p,p')$ diverges like $(\tau_n-u)^{-1}$ if $u \rightarrow \tau_n (u')$; a slower growth than occurs in the degenerate example \eqref{VVConfFlat}. It is also clear that the van Vleck determinant switches sign on passing through each conjugate hyperplane in this example.

A plane wave spacetime with $\mathbf{H}(u)$ given by \eqref{HGravWave} may be written in Rosen coordinates \eqref{RosenMetric} using, e.g.,
\begin{equation}
	\bm{\mathcal{H}}(U) = \left(	
	\begin{matrix}
		\cosh^2 (U-U') & 0 \\ 
		0 & \cos^2 (U-U')
	\end{matrix}
	\right).
\label{HRosenGravWave}
\end{equation}
Once again, the Rosen coordinates become singular while the Brinkmann coordinates do not.

\subsubsection{A more realistic example}

Although very simple, neither \eqref{HRosenConfFlat} nor \eqref{HRosenGravWave} look very much like ``ordinary'' oscillating waves. Consider instead a linearly polarized vacuum plane wave with the profile
\begin{align}
	\mathbf{H}(u)  = \frac{h}{2}  \left(
	\begin{matrix}
	1 & 0 \\ 
	0 & -1
	\end{matrix}
	\right)  \cos u.
	\label{HGravWaveReal}
\end{align}
Here, $0 < h <1$ is a constant. In this case, $\bA(u,u')$ and $\bB(u,u')$ are linear combinations of Mathieu functions.

Properties of these functions are not particularly well-known, so it is instructive to consider perturbative solutions when $h \ll 1$. One such solution of \eqref{EDef} is
\begin{align}
	\mathbf{E}(U) & = \bm{\delta} - \frac{h}{2} 
	 \left(
	 \begin{matrix}
	 1 & 0 \\ 
	 0 & -1
	 \end{matrix}
	 \right) 
	 \cos U + O(h^2).
\end{align}
Substituting this into \eqref{RosenMetric} and \eqref{RosenH}, the metric may be written in Rosen coordinates as
\begin{align}
	\rmd s^2 & = -2 \rmd U \rmd V + |\rmd \mathbf{X}|^2 
	\nonumber
	\\
	& ~ - h [ (\rmd X^1)^2 - (\rmd X^2)^2 ] \cos U + O(h^2).
\end{align}
This can be recognized as the line element of a polarized monochromatic gravitational plane wave as one would expect from linearized general relativity in transverse-traceless gauge. 

Continuing to assume that $h$ is small, the spatial Jacobi propagators are approximately given by
\begin{align}
	& \bA (u,u') = \bm{\delta} - \frac{ h }{2} \left(
	\begin{matrix}
	1 & 0 \\ 
	0 & -1
	\end{matrix}
	\right) 
	\nonumber
	\\
	 & ~ \times \big[ (\cos u - \cos u')  + (u-u') \sin u' \big] 
	 + O(h^2)
\end{align}
and
\begin{align}
	\bB(u,u')& = (u-u') \bm{\delta} +  h \left(
	\begin{matrix}
	1 & 0 \\ 
	0 & -1
	\end{matrix}
	\right)   \big[ (\sin u - \sin u')
	\nonumber
	\\
	& ~  -  \frac{1}{2} (u-u') (\cos u + \cos u') \big] + O(h^2). 
\end{align}
Hence,
\begin{equation}
	\det \bB(u,u') = (u-u')^2 + O(h^2)
\end{equation}
and
\begin{equation}
	\Delta(p,p') = 1 + O(h^2).
\end{equation}
It follows that there are no conjugate points in this approximation. 

\begin{figure}
	\centering
	\includegraphics[width= .9 \linewidth]{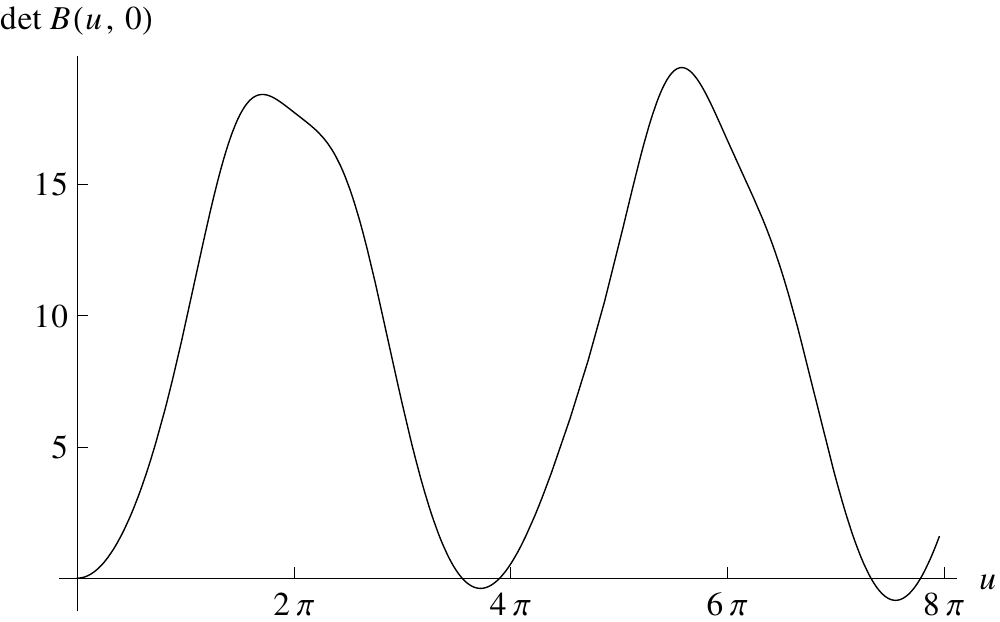}
	\caption{$\det \bB(u,0)$ for a plane wave spacetime with $\mathbf{H}(u)$ given by \eqref{HGravWaveReal} and $h = 2/3$. The zeros correspond to locations of conjugate hyperplanes. They occur in closely-spaced pairs separated by approximately $3.8 \pi$. Note that $\mathbf{H}(u)$ has the shorter period $2 \pi$.}
		\label{Fig:Mathieu}
	\end{figure}

Exact solutions for $\bA$ and $\bB$ in terms of Mathieu functions display much more interesting behavior. Conjugate points occur generically. Indeed, $\det \bB(u,u')$ is an approximately sinusoidal function of $u$: 
\begin{equation}
	\det \bB(u,u') \approx (\mathrm{const.}) \times [ 1 - \cos \nu(h) (u-u') ].
\end{equation}
This heuristic approximation rapidly improves as $h \rightarrow 0$. See Fig. \ref{Fig:Mathieu} for a case where it starts to break down ($h = 2/3$). The period of the oscillations in $\det \bB$ is determined by the Mathieu characteristic exponent $\nu(h)$, and is always greater than $2 \pi$. It is roughly given by
\begin{equation}
	\frac{2 \pi}{ \nu(h) } \approx \frac{ 2.82 \pi }{  h}
\end{equation}
if $h$ is not too large (the relative error in this estimate for the period is approximately $10 \%$ if $h = 2/3$). As illustrated in Fig. \ref{Fig:Mathieu}, conjugate points generically (but not universally) occur in closely-spaced pairs separated by roughly $2 \pi/\nu(h)$. Such points have multiplicity $1$. It is possible for there to exist conjugate points of multiplicity $2$ -- which do not occur in pairs -- although this requires finely-tuned values of $h$.

\section{Green functions in plane wave spacetimes}
\label{Sect:Green}

Consider a massless scalar field $\Phi$ propagating (without gravitational backreaction) in a plane wave spacetime. Allowing for a scalar charge density $\rho$, such a field satisfies the wave equation
\begin{align}
	-  4 \pi \rho =& \, \nabla^{a} \nabla_{a} \Phi  
	\nonumber
	\\
	=& \, [- 2 \partial_u \partial_v - H_{ij}(u) x^i x^j \partial_v^2 + \nabla^2] \Phi.
	\label{WaveEqScalar}
\end{align}
Significant insight into the solutions of this equation may be obtained by computing an associated Green function $G(p,p')$. Green functions are defined here to be any distributional solutions to the wave equation with (zeroth-order) sources localized to a single spacetime point:
\begin{equation}
	\nabla^{a} \nabla_{a} G(p,p') = - 4 \pi \delta(p,p').
	\label{BoxGAbstr}
\end{equation}
There are, of course, many solutions to this equation. Any one of them may be used to obtain some solution
\begin{equation}
	\Phi_\rho(p) := \int \rho(p') G(p,p') \rmd V'
\end{equation}
to \eqref{WaveEqScalar} (at least if $\rho$ satisfies certain constraints). More general solutions can be built by adding to $\Phi_\rho$ an appropriate homogeneous solution $\Phi_0$ satisfying $\nabla^a \nabla_a \Phi_0 =0$. Alternatively, appropriate Green functions together with initial data may be used to convert the wave equation into a Kirchhoff-type integral equation \cite{PoissonRev, FriedlanderWaves}.

If $p$ and $p'$ are sufficiently close, one particular solution to \eqref{BoxGAbstr} is the ``retarded\footnote{\label{Foot:AdvRet}Notions of ``advanced'' and ``retarded'' used here are quasi-local. They refer only to the causal properties of a solution when its arguments are sufficiently close together. No claims are made regarding the behavior of, e.g., $G_\mathrm{ret}$ at infinity [where expressions like \eqref{GretGeneral} are not valid]. Additionally, it should be noted that we derive \textit{particular} solutions that look retarded or advanced when their arguments remain close. They are not unique. See Sect. \ref{Sect:NonUnique}.} solution'' \eqref{GretGeneral} \cite{PoissonRev, FriedlanderWaves}. The bitensors $\mathcal{U}(p,p')$ and $\mathcal{V}(p,p')$ appearing in that formula are known for the four-dimensional plane wave spacetimes considered here \cite{FriedlanderWaves,Huygens}. In such cases, the tail of the Green function $\mathcal{V}(p,p')$ vanishes and the direct portion $\mathcal{U}(p,p')$ is determined by the van Vleck determinant described in Sect. \ref{Sect:Bitensors}:
\begin{equation}
	\mathcal{U}(p,p') = \sqrt{\Delta(p,p')}.
\end{equation}
It follows that
\begin{equation}
	G_{\mathrm{ret}}(p,p') = \theta(p \geq p') \sqrt{\Delta(p,p')} \delta \big( \sigma(p,p') \big) .
\label{Hadamard}
\end{equation}
The advanced solution $G_{\mathrm{adv}}(p,p')$ is the same with the obvious replacement $\theta(p \geq p') \rightarrow \theta( p' \geq p)$. We mainly focus on the ``symmetric Green function''
\begin{equation}
	G_{\mathrm{S}}(p,p') := \frac{1}{2} [ G_\mathrm{ret}(p,p') + G_\mathrm{adv}(p,p')],
\end{equation}
from which the advanced and retarded solutions are easily extracted. If $p$ and $p'$ are sufficiently close, it is clear from \eqref{Hadamard} that
\begin{equation}
	G_\mathrm{S}(p,p') = \frac{1}{2} \sqrt{\Delta(p,p')} \delta \big( \sigma(p,p') \big) .
	\label{HadamardSym}
\end{equation}

At least for short distances, \eqref{Hadamard} implies that disturbances in $\Phi$ are propagated only \textit{on} -- and not inside -- the light cones of those disturbances. Signals from sources that turn on and off sharply are themselves sharp. This ``Huygens' principle'' is a very special property of massless scalar fields in four-dimensional plane wave spacetimes. In almost all other cases, retarded Green functions have support inside the light cone [i.e., $\mathcal{V}(p,p') \neq 0$ in \eqref{GretGeneral}] \cite{FriedlanderWaves, Huygens}. Even for massless scalar fields in plane wave spacetimes, Huygens' principle is not necessarily valid globally.  It is shown below that the appropriate extension of \eqref{Hadamard} fails to be everywhere sharp if there exist non-degenerate (i.e., multiplicity 1) conjugate hyperplanes.

It is evident from the discussion in Sect. \ref{Sect:CausticGeo} that the form \eqref{HadamardSym} for the symmetric Green function becomes problematic if $p$ and $p'$ are too widely separated. If a plane wave spacetime admits conjugate hyperplanes, there exist pairs of points for which the bitensors $\sigma(p,p')$ and $\Delta(p,p')$ appearing in that formula fail to be defined. One can therefore expect \eqref{Hadamard} to be valid only for $p$ in a neighborhood of $p'$ that does not intersect any hyperplanes conjugate to $S_{u'}$. The largest such neighborhood is the set $\mathcal{N}_0(u')$ defined by \eqref{NormalNeighborhood} and illustrated in Fig. \ref{Fig:Normal}. Eqs. \eqref{Hadamard} and \eqref{HadamardSym} are indeed valid solutions to \eqref{BoxGAbstr} throughout
\begin{equation}
 \{ p, p' \in M : p \in \mathcal{N}_0(u(p')) \}.
\label{NormalNeighb0}
\end{equation}

Our strategy for constructing a global Green function $G_\mathrm{S}(\cdot,p')$ first demands that \eqref{HadamardSym} hold throughout the ``zeroth normal neighborhood'' $\mathcal{N}_0(u')$. Sect. \ref{Sect:GreenInNormal} then derives similar formulae in all of the remaining $\mathcal{N}_n(u')$ [where $\sigma(\cdot, p')$ and $\Delta(\cdot, p')$ remain well-defined]. The result involves two free parameters for each $n$, and is a valid solution to \eqref{BoxGAbstr} throughout the generalized normal neighborhood $\mathcal{N}(u')$. Sect. \ref{Sect:GreenOnConj} demonstrates how to extend this solution through the conjugate hyperplanes that separate the disjoint components $\mathcal{N}_n(u')$ of $\mathcal{N}(u')$. Enforcing the wave equation \textit{on} conjugate hyperplanes relates the various free parameters to each other in a simple way. This fixes the singularity structure of $G_\mathrm{S}(\cdot,p')$ along almost all null geodesics passing through $p'$.

It is important to emphasize that our construction produces only one of many possible Green functions. We essentially state that a solution to \eqref{BoxGAbstr} is known in some region, and extend this using ``initial data'' on the boundary of that region. Here, the relevant boundaries are hypersurfaces of constant $u$. Even in flat spacetime, initial data imposed in this way does not yield a unique solution to a wave equation. Unlike flat spacetime, however, plane wave geometries do not admit appropriate Cauchy surfaces that can be used instead of constant-$u$ hypersurfaces. As explained in Sect. \ref{Sect:Causality}, plane waves are not globally hyperbolic. That the Green function we construct fails to be unique is shown in Sect. \ref{Sect:Penrose} to provide an important freedom if the leading order singularity structure of Green functions in generic spacetimes is to be determined by plane wave Green functions. 

\subsection{Green functions in the generalized normal neighborhood}
\label{Sect:GreenInNormal}

Outside of the normal neighborhood \eqref{NormalNeighb0}, the symmetric Green function $G_\mathrm{S}(p,p')$ must be a solution to the homogeneous wave equation
\begin{equation}
	\nabla^a \nabla_a G_\mathrm{S} (p,p') = 0.
\label{BoxGHom}
\end{equation}
Although the Hadamard form \eqref{HadamardSym} breaks down outside of \eqref{NormalNeighb0}, one might still consider a ``Hadamard-like'' ansatz
\begin{equation}
	G_{\mathrm{S}}(p,p') = \frac{1}{2} \sqrt{|\Delta(p,p')|} g_n\big( \sigma(p,p') \big)
	\label{GAnsatz}
\end{equation}
for all $p \in \mathcal{N}_n(u')$ and $p' \in M$. Here, $g_n(\sigma)$ is some as-yet undetermined distribution (for $n \neq 0$). Recall that the bitensors $\sigma(p,p')$ and $\Delta(p,p')$ appearing in \eqref{GAnsatz} are well-defined throughout the region where that equation is valid. Also note that, as shown in Sect. \ref{Sect:Bitensors}, $\Delta(p,p')$ is finite and nonzero everywhere it is defined. This biscalar may be negative, however, which necessitates the absolute value appearing in \eqref{GAnsatz}. 

Substituting \eqref{GAnsatz} into \eqref{BoxGHom} yields the ordinary differential equation
\begin{equation}
	\sigma \frac{\rmd^2 g_n}{\rmd \sigma^2} + 2 \frac{ \rmd g_n}{\rmd \sigma}  = 0.
\end{equation}
The general distributional solution of this is
\begin{equation}
	g_n(\sigma) = \alpha_n \delta(\sigma) + \mathrm{pv} \, (\beta_n /\sigma) + \gamma_n,
\label{FGeneral}
\end{equation}
where $\alpha_n$, $\beta_n$, and $\gamma_n$ are arbitrary constants and ``$\mathrm{pv}$'' denotes the Cauchy principal value. The term involving $\gamma_n$  is not interesting, so we discard it at this point\footnote{The term involving $\gamma_n$ adds to the Green function something which depends only on $u$ and $u'$. All distributions in these variables are solutions to the homogeneous equation \eqref{BoxGHom}, and may therefore be freely added or removed from a particular Green function.}.

It follows that for any $p \in \mathcal{N}_n(u')$ (with $n$ possibly vanishing), 
\begin{align}
	G_\mathrm{S}(p,p') = \frac{1}{2} \sqrt{|\Delta(p,p')|} \Big[\alpha_n \delta\big( \sigma(p,p') \big) 
	\nonumber
	\\
	~ + \mathrm{pv} \big( \beta_n / \sigma(p,p') \big) \Big].
\label{GAnsatz2}
\end{align}
Comparison with \eqref{HadamardSym} shows that
\begin{equation}
	\alpha_{0} = 1, \qquad \beta_{0} = 0.
\label{InitData}
\end{equation}

Eq. \eqref{GAnsatz2} provides a class of possible forms for $G_{\mathrm{S}}(p,p')$ for all $p$ in the generalized normal neighborhood $\mathcal{N}(u')$. The coefficients $\alpha_n, \beta_n$ are undetermined at this point (for $n \neq 0$), which reflects the fact that the wave equation \eqref{BoxGAbstr} has been not been solved everywhere. In particular, it has not been solved on the conjugate hyperplanes $S_{\tau_n}$ separating the disconnected components of $\mathcal{N}$. Demanding that the wave equation be solved everywhere provides algebraic matching conditions that relate $(\alpha_n, \beta_n)$ to $(\alpha_{n+1}, \beta_{n+1})$ or $(\alpha_{n-1}, \beta_{n-1})$. Using \eqref{InitData} as ``initial data,'' these matching conditions provide a unique prescription for all $\alpha_n, \beta_n$.

\subsection{Green functions on conjugate hyperplanes}
\label{Sect:GreenOnConj}

Demanding that the wave equation \eqref{BoxGAbstr} be satisfied on a boundary $\partial \mathcal{N}_n$ first requires defining what could possibly be meant by $G_{\mathrm{S}}(p,p')$ in such regions. Roughly speaking, one would like to define objects that behave like, e.g.,
\begin{subequations}
		\label{FakeDistribution}
\begin{align}
	\sqrt{|\Delta|} \delta(\sigma) \Theta(u - \tau_n), 
	\\
	\mathrm{pv} \left( \frac{\sqrt{|\Delta|}}{\sigma} \right) \Theta(u - \tau_n).
\end{align}
\end{subequations}
$\Delta(p,p')$ and $\sigma(p,p')$ are both ill-behaved as  $u \rightarrow \tau_n$, so it is not obvious that there is any way to define distributions of this type. Nevertheless, appropriate distributions may be guessed that are well-defined everywhere and ``look like'' \eqref{FakeDistribution} for all $u$ away from conjugate hyperplanes (where the meaning of those expressions is unambiguous).
	
To be more precise, we must now treat Green functions properly as distributions. They are linear functionals acting on an appropriate space of test functions.\footnote{Not every linear functional on test functions is a distribution. There must additionally be a certain sense in which the functional is continuous with respect to sequences of test functions. Equivalently, it must be possible to bound the action of a distribution on an arbitrary test function using certain semi-norm estimates. See, e.g., Appendix \ref{Sect:Distributions} or \cite{FriedlanderDistributions}.} Specifically, $G_{\mathrm{S}}(p,p')$ takes as input a source point $p' \in M$ as well as a test function $\varphi(p) : M \rightarrow \mathbb{R}$ that is in the space $C_0^\infty(M)$ of smooth scalar functions with compact support:
\begin{equation}
	G_{\mathrm{S}} : C_0^\infty(M) \times M \rightarrow \mathbb{R}.
\end{equation}
Given any test function $\varphi \in C_0^\infty(M)$, the action of the Green function at $p'$ is denoted by
\begin{equation}
	\langle G_{\mathrm{S}}(p,p') , \varphi(p) \rangle.
\end{equation}
It is also common to write this as
\begin{equation}
	\int G_{\mathrm{S}}(p,p') \varphi(p) \rmd V,
\end{equation}
which is the notation we have already been using.

Differential equations like \eqref{BoxGAbstr} are really a type of shorthand notation. For every $\varphi \in C_0^\infty(M)$ and every $p' \in M$,
\begin{equation}
	\langle G_{\mathrm{S}}(p,p'), \nabla^{a} \nabla_{a} \varphi(p) \rangle = - 4 \pi \varphi(p').
	\label{BoxGWeak}
\end{equation}
Arguments given above already imply that this equation is satisfied by \eqref{GAnsatz2} if the support of $\varphi$ lies entirely in $\mathcal{N}(u')$. Equivalently, \eqref{GAnsatz2} is valid as long as $\varphi$ does not pass through any hyperplanes conjugate to $S_{u'}$. 

We now proceed by providing an ansatz for $\langle G_{\mathrm{S}} (p,p'), \varphi(p) \rangle$ that applies for test functions with supports that do not lie entirely in $\mathcal{N}(u')$. By linearity, it suffices to consider test functions $\varphi_n$ (with $n \neq 0$) whose supports intersect at most one conjugate hyperplane; specifically $S_{\tau_n(u')}$. Denote by $\mathcal{T}_n(u')$ a connected open neighborhood of the hyperplane $S_{\tau_n(u')}$ that does not intersect any other conjugate hyperplanes (or, for technical reasons, $S_{u'}$). One could choose, for example,
\begin{equation}
	\mathcal{T}_2 = \mathcal{N}_1 \cup \mathcal{N}_2 \cup S_{\tau_2} 
\end{equation}
if $\tau_2$ exists. Regardless, use the notation $\varphi_n$ to denote test functions in $\mathcal{T}_n$:
\begin{equation}
	\varphi_n \in C^{\infty}_0 \big( \mathcal{T}_n(u') \big).
	\label{PhiN}
\end{equation}
The action of $G_{\mathrm{S}}(\cdot, p')$ on a general test function $\varphi \in C^\infty_0(M)$ may be obtained by summing its action on various $\varphi_n$ and, perhaps, on a test function with support only in $\mathcal{N}$. 

We now introduce two new functionals $\mathcal{G}^{\sharp}_{n^\pm}(p,p')$ and $\mathcal{G}^{\flat}_{n^\pm}(p,p')$ that act on arbitrary $\varphi_n \in C^{\infty}_0 ( \mathcal{T}_n(u') )$:
\begin{subequations}
	\label{DistEquiv}
	\begin{align}
		\mathcal{G}^\sharp_{n^\pm} := \lim_{\epsilon \rightarrow 0^+} \sqrt{|\Delta|} \delta(\sigma) \Theta\big( \pm (u-\tau_n) - \epsilon \big),
	\\
		\mathcal{G}^\flat_{n^\pm} := \lim_{\epsilon \rightarrow 0^+} \mathrm{pv} \left( \frac{\sqrt{|\Delta|}}{ \sigma } \right) \Theta\big( \pm (u-\tau_n)- \epsilon \big).
	\end{align}
\end{subequations}
The $\sharp$ notation on $\mathcal{G}^\sharp_{n^\pm}$ indicates that this functional is related to the ``sharp'' propagation of signals associated with $\delta$-functions. The $1/\sigma$-like behavior of $\mathcal{G}^\flat_{n^\pm}$ is, by comparison, rather ``flat.'' The $n^\pm$ subscripts on $\mathcal{G}^\sharp_{n^\pm}(p,p')$ and $\mathcal{G}^\flat_{n^\pm}(p,p')$ denote support either in the future (+) or past (-) of the $n$th hyperplane $S_{\tau_n(u')}$ conjugate to $S_{u'} \ni p'$. 

The explicit coordinate representations of $\mathcal{G}^\sharp_{n^\pm}(p,p')$ and $\mathcal{G}^\flat_{n^\pm}(p,p')$ are
\begin{align}
	\langle \mathcal{G}^\sharp_{n^\pm} , \varphi_n \rangle = \pm \lim_{\epsilon \rightarrow 0^+} \int_{\tau_n \pm \epsilon}^{\pm \infty} \! \rmd u \int_{\mathbb{R}^2} \! \rmd^2 \bx
\nonumber
\\
	~ \times \left( \frac{\sqrt{|\Delta|}}{|u-u'|} \right)  \varphi_n (u, v'+\chi, \bx),
\label{IdeltaDef}
\end{align}
and
\begin{align}
	\langle \mathcal{G}^{\flat}_{n^\pm} , \varphi_n \rangle = \mp \lim_{\epsilon \rightarrow 0^+} \int_{\tau_n \pm \epsilon}^{\pm \infty} \! \! \rmd u \int_{\mathbb{R}^2} \! \rmd^2 \bx \int_0^\infty \! \rmd \Sigma 
\nonumber
\\
	 ~ \times \Sigma^{-1} \left( \frac{ \sqrt{|\Delta|} }{ u-u' } \right) \big[ \varphi_n (u, v'+\chi+ \Sigma, \bx) 
	\nonumber
	\\
	~ - \varphi_n (u, v'+\chi - \Sigma, \bx) \big] .
\label{IpvDef}
\end{align}
Here, the function $\chi(u,u';\bx,\bx')$ is defined to be the value of $v-v'$ which ensures that $p$ is connected to $p'$ via a null geodesic:
\begin{equation}
	\sigma \big(u, v'+\chi(u,u';\bx,\bx'), \bx; u', v', \bx' \big) = 0.
\label{VDefSig}
\end{equation}
Referring to \eqref{SigmaGen2}, $\chi(u,u'; \bx, \bx')$ is given by
\begin{align}
	\chi = \frac{1}{2} \big[\bx^\intercal \partial_u \bB \bB^{-1} \bx + \bx'^\intercal \bB^{-1} \bA \bx' 
\nonumber
\\
	\qquad \qquad \qquad ~ - 2 \bx'^\intercal \bB^{-1} \bx \big]
\label{VDef}
\end{align}
in terms of the matrices $\bA(u,u')$ and $\bB(u,u')$ defined by \eqref{JacobiSpatial} and \eqref{ABboundary}. It is shown in Appendix \ref{Sect:Distributions} that $\mathcal{G}^\sharp_{n^\pm}$ and $\mathcal{G}^\flat_{n^\pm}$ are well-defined distributions: All integrals in \eqref{IdeltaDef} and \eqref{IpvDef} converge and appropriate semi-norm estimates may be derived. 

Given the form \eqref{GAnsatz2} for $G_\mathrm{S}$ as it would act on test functions confined to $\mathcal{N}_n$, \eqref{DistEquiv} can be used to guess a natural extension valid for all test functions $\varphi_n \in C^{\infty}_0 ( \mathcal{T}_n(u') )$. Suppose that
\begin{align}
	\langle G_{\mathrm{S}}, \varphi_n \rangle = \frac{1}{2} \bigg( \alpha_{n-1} \langle \mathcal{G}^\sharp_{n^-} , \varphi_n \rangle + \alpha_n \langle \mathcal{G}^\sharp_{n^+} , \varphi_n \rangle
\nonumber
\\
	~ + \beta_{n-1} \langle \mathcal{G}^{\flat}_{n^-} , \varphi_n \rangle + \beta_n \langle \mathcal{G}^{\flat}_{n^+} , \varphi_n \rangle \bigg)
\label{GAnsatz3}
\end{align}
if $n > 0$. The same expression holds with the replacements
\begin{equation}
	( \alpha_{n-1} , \beta_{n-1} , \alpha_n, \beta_n ) \rightarrow 	( \alpha_n , \beta_n , \alpha_{n+1}, \beta_{n+1} )
\end{equation}
if $n<0$. It is clear from \eqref{GAnsatz2} and \eqref{DistEquiv} that the form \eqref{GAnsatz3} for $G_\mathrm{S}$ satisfies the wave equation \eqref{BoxGWeak} if $\varphi_n$ has no support on $S_{\tau_n(u')}$. For more general test functions, $\langle G_{\mathrm{S}}, \nabla^a \nabla_a \varphi_n \rangle \neq 0$ unless the $\alpha_n$ and $\beta_n$ are related in a particular way. We now compute $\langle \mathcal{G}_{n^\pm}^\sharp, \nabla^a \nabla_a \varphi_n \rangle$ and $\langle \mathcal{G}_{n^\pm}^\flat, \nabla^a \nabla_a \varphi_n \rangle$ in order to derive these relations.

Letting
\begin{align}
	\bar{\varphi}  := \varphi_n(u, v' + \chi(u,u';\bx, \bx'), \bx),
	\\
	\bar{\varphi}' := \partial_v \varphi_n(u, v' + \chi(u,u';\bx, \bx'), \bx),
\end{align}	
note that
\begin{align}
	&\frac{ \sqrt{|\Delta|} }{ |u-u'| } \nabla^a \nabla_a \varphi_n (u,v'+\chi,\bx) = - 2 \partial_u \left( \frac{ \sqrt{|\Delta|} }{ |u-u'| } \bar{\varphi}' \right)
	\nonumber
	\\
	&~ + \nabla^2 \left( \frac{ \sqrt{|\Delta|} }{ |u-u'| } \bar{ \varphi } \right) - 2 \partial_i \left( \frac{ \sqrt{|\Delta|} }{ |u-u'| }  \bar{\varphi}' \partial_i \chi \right).
\end{align}
This is easily verified using direct computation together with \eqref{JacobiSpatial}, \eqref{vanVleck}, \eqref{VDef}, \eqref{BASym} and the symmetry of the matrices $\partial_u \bA \bA^{-1}$ and $\partial_u \bB \bB^{-1}$ established in Appendix \ref{Sect:ABProperties}. Substitution into \eqref{IdeltaDef} shows that
\begin{align}
	\langle \mathcal{G}^\sharp_{n^\pm} , \nabla^{a} \nabla_{a} \varphi_n \rangle = \pm 2 \lim_{u \rightarrow \tau_n^\pm } \int_{\mathbb{R}^2} \! \rmd^2 \bx  \left( \frac{ \sqrt{|\Delta|} }{ |u-u'| } \right)
	\nonumber
	\\
	~ \times  \partial_{v} \varphi_n(u,v'+\chi(u,u';\bx,\bx'),\bx).
	\label{IDelta}
\end{align}
This measures the degree to which $\mathcal{G}^\sharp_{n^\pm}$ fails to satisfy the wave equation. 

Using \eqref{IpvDef}, the equivalent expression for the $1/\sigma$-type distribution $\mathcal{G}^\flat_{n^\pm}$ is
\begin{align}
	\langle \mathcal{G}^\flat_{n^\pm} , \nabla^{a} \nabla_{a} \varphi_n \rangle = \mp 2 \lim_{u \rightarrow \tau_n^\pm } \int_{\mathbb{R}^2} \! \rmd^2 \bx \int_0^\infty \rmd \Sigma  
	\nonumber
	\\
	~ \times  \Sigma^{-1} \left( \frac{ \sqrt{|\Delta|} }{ |u-u'|  } \right) \big[ \partial_{v} \varphi_n(u,v'+\chi + \Sigma,\bx) 
	\nonumber
	\\
	~ - \partial_{v} \varphi_n(u,v'+\chi - \Sigma,\bx) \big].
	\label{Ipv}
\end{align}
Evaluating $\langle G_\mathrm{S}, \nabla^a \nabla_a \varphi_n \rangle$ requires simplifying these two expressions and then applying \eqref{GAnsatz3}. The result depends on the multiplicity of the conjugate points associated with $\tau_n$, and requires the expansions derived in Sect. \ref{Sect:CausticGeo}.

\subsubsection{Degenerate conjugate points}

First consider the case where the conjugate hyperplanes associated with some $\tau_n(u') \in T(u')$ are related to degenerate (multiplicity 2) conjugate points. Applying \eqref{SigmaGen2} and \eqref{SigmaDeg} to \eqref{VDef} then shows that for all $p \in \mathcal{T}_n(u')$,
\begin{align}
	\chi(u,u' ; \bx, \bx')  = -\frac{1}{2} \frac{| \bx - \hat{\bA}_n \bx'|^2}{\tau_n-u} 
	\nonumber
	\\
	~ + \chi_n(u, u' ; \bx, \bx')  .
	\label{VExpandDeg}
\end{align}
Here, $\hat{\bA}_n(u') = \bA( \tau_n(u'), u' )$ and $\chi_n(u,u'; \bx, \bx')$ is a function that is well-behaved in all of its arguments.

Substituting \eqref{VExpandDeg} into \eqref{IDelta} and using \eqref{VanVleckDeg} together with the change of variables
\begin{equation}
\bx \rightarrow \bar{\bx} := \frac{ \bx - \hat{\bA}_n \bx' }{ \sqrt{|\tau_n - u|} }
\label{xBarDef}
\end{equation}
yields
\begin{align}
	\langle \mathcal{G}^\sharp_{n^\pm} , \nabla^a \nabla_a & \varphi_n \rangle = \pm 2 \sqrt{ |\det \hat{\bA}_n | } \int_{\mathbb{R}^2} \rmd^2 \bar{ \bx } 
	\nonumber
	\\
	& \partial_v \varphi_n (\tau_n, v' \pm \frac{1}{2} |\bar{\bx}|^2 + \chi_n, \hat{\bA}_n \bx' ) .
	\label{IDeg}
\end{align}
Transforming into polar coordinates in the usual way, the integral on the right-hand side of this equation may be evaluated explicitly:
\begin{align}
	\langle \mathcal{G}^\sharp_{n^\pm} & , \nabla^a \nabla_a \varphi_n \rangle = - 4\pi \sqrt{ | \det \hat{\bA}_n | } 
	\nonumber
	\\
	& ~ \times \varphi_n (\tau_n,v'+\chi_n (\tau_n, u'; \hat{\bA}_n \bx', \bx'),\hat{\bA}_n \bx').
	\label{BoxGSharpDeg}
\end{align}
The discussion in Sect. \ref{Sect:CausticGeo} may be used to show that the argument of $\varphi_n$ appearing here is the point to which all null geodesics starting at $p'$ focus to on $S_{\tau_n(u')}$.

Using similar arguments together with \eqref{Ipv}, the wave operator acting on the $1/\sigma$ part of the Green function produces
\begin{align}
	\langle \mathcal{G}^\flat_{n^\pm}, \nabla^a \nabla_a \varphi_n \rangle = 4 \pi \sqrt{| \det \hat{\bA}_n | } \int_0^\infty \rmd \Sigma 
	\nonumber
	\\
	~ \times \Sigma^{-1} \big[ \varphi_n (\tau_n, v' + \chi_n + \Sigma, \hat{\bA}_n \bx') 
	\nonumber
	\\
	~ -\varphi_n (\tau_n, v' + \chi_n - \Sigma, \hat{\bA}_n \bx') \big].
	\label{BoxGFlatDeg}
\end{align}
This depends on $\varphi_n$ at all points on $S_{\tau_n(u')}$ that are connected to $p'$ by geodesics of any type. It is not proportional to $\langle \mathcal{G}^\sharp_{n^\pm}, \nabla^a \nabla_a \varphi_n \rangle$ as given by \eqref{BoxGSharpDeg}.

Substituting \eqref{GAnsatz3}, \eqref{BoxGSharpDeg}, and \eqref{BoxGFlatDeg} into \eqref{BoxGWeak} shows that the wave equation can be satisfied on a degenerate conjugate hyperplane $S_{\tau_n(u')}$ if and only if
\begin{equation}
	\alpha_n = - \alpha_{n-1} , \qquad \beta_n = - \beta_{n-1}
	\label{MatchingDegPos}
\end{equation}
when $n>0$ or
\begin{equation}
	\alpha_n = - \alpha_{n+1}, \qquad \beta_n = - \beta_{n+1} 
	\label{MatchingDegNeg}
\end{equation}
when $n<0$. If these relations are satisfied for a particular $n$, $\nabla^a \nabla_a G_\mathrm{S}(p,p') = 0$ throughout $\mathcal{T}_n(u')$. They imply that Green functions tend to switch sign on passing through degenerate conjugate hyperplanes. 

\subsubsection{Non-degenerate conjugate points}

The non-degenerate (multiplicity 1) case is treated similarly. Choose a particular $\tau_n(u') \in T(u')$ associated with non-degenerate conjugate points. Eq. \eqref{SigmaSimp} then implies that if $u$ is sufficiently close to $\tau_n(u')$, $\chi(u,u';\bx,\bx')$ has the form
\begin{align}
	\chi(u,u';\bx, \bx') = - \frac{1}{2} \frac{ [ \hat{ \mathbf{q} }_n \cdot ( \bx - \hat{\bA}_n \bx' ) ] ^2 }{ \tau_n - u } 
	\nonumber
	\\
	~ + \chi_n(u, u'; \bx,\bx'),
	\label{VExpandSimp}
\end{align}
where $\chi_n(u,u';\bx,\bx')$ is smooth. Recall that the unit vector $\hat{\mathbf{q}}_n$ is associated with $\tau_n(u')$ as a solution to the eigenvector problem \eqref{nDef}.

The form of $\chi$ near a simple conjugate hyperplane suggests the coordinate transformation $\bx \rightarrow \tilde{\bx}$, where
\begin{subequations}
	\label{XTilde}
\begin{align}
	\tilde{x}^1 := \frac{\hat{ \mathbf{q} }_n \cdot ( \bx - \hat{\bA}_n \bx' )}{ \sqrt{ | \tau_n - u| } },
	\\
	\tilde{x}^2 := \hat{ \mathbf{p} }_n \cdot ( \bx - \hat{\bA}_n \bx' ).
\end{align}
\end{subequations}
Here, $\hat{ \mathbf{p} }_n$ is a unit vector satisfying $\hat{\mathbf{p}}_n \cdot \hat{\mathbf{q}}_n = 0$. The signs of $\hat{\mathbf{p}}_n$ and $\hat{\mathbf{q}}_n$ are to be chosen such that the ordered pair of coordinates $(\tilde{x}^1,\tilde{x}^2)$ has the same orientation as $(x^1, x^2)$. Using \eqref{VanVleckSimp} and \eqref{IDelta},
\begin{align}
	\langle \mathcal{G}^\sharp_{n^\pm} , \nabla^a \nabla_a \varphi_n \rangle = \frac{ \pm 2 }{ \sqrt{ | 
	\mathrm{Tr} [ \hat{\bB}_n (\widehat{ \partial_u \bB }_n )^{-1} ] \det (\widehat{\partial_u \bB}_n ) |} }
	\nonumber
	\\
	~ \times \int_{\mathbb{R}^2} \! \! \rmd^2 \tilde{\bx} \partial_v \varphi_n (\tau_n , v' \pm \frac{1}{2} ( \tilde{x}^1 )^2 + \chi_n, \hat{\bA}_n \bx' + \hat{\mathbf{p}}_n \tilde{x}^2 ).
\end{align}
Similar simplifications of \eqref{Ipv} result in
\begin{align}
	\langle \mathcal{G}^\flat_{n^\pm}, \nabla^a \nabla_a \varphi_n \rangle = \frac{ 2 \pi }{ \sqrt{ | 
			\mathrm{Tr} [ \hat{\bB}_n (\widehat{ \partial_u \bB }_n )^{-1} ] \det (\widehat{\partial_u \bB}_n) |} }
			\nonumber
			\\
			~ \times \int_{\mathbb{R}^2} \! \! \rmd^2 \tilde{\bx} \partial_v \varphi (\tau_n, v' \mp \frac{1}{2} (\tilde{x}^1)^2 + \chi_n, \hat{\bA}_n \bx' + \hat{\mathbf{p}}_n \tilde{x}^2 ).
\end{align}

Substituting these two equations into \eqref{GAnsatz3}, one sees that $G_{\mathrm{S}}(p,p')$ satisfies the wave equation throughout $\mathcal{T}_n(u')$ when
\begin{equation}
	\alpha_n = - \pi \beta_{n-1} , \qquad \beta_n = \frac{ \alpha_{n-1} }{ \pi },
	\label{MatchingSimpPos}
\end{equation}
if $n>0$ or
\begin{equation}
	\alpha_n = \pi \beta_{n+1} , \qquad \beta_n = - \frac{ \alpha_{n+1} }{ \pi }
	\label{MatchingSimpNeg}
\end{equation}
if $n<0$. Unlike in the degenerate case, these relations show that the qualitative character of a plane wave Green function changes on passing through a non-degenerate conjugate hyperplane. It switches from having a $\delta(\sigma)$-type singularity to a $1/\sigma$-type singularity (or vice-versa).

\subsection{A global solution}
\label{Sect:GlobalSoln}

We now have a recipe for constructing a global Green function associated with the massless scalar wave equation \eqref{WaveEqScalar}. Fixing $p'$, suppose that $\tau_{1}(u')$ and $\tau_{-1}(u')$ both exist. The symmetric Green function can then be written as
\begin{widetext}
\begin{align}
	G_\mathrm{S} = \frac{1}{2} \lim_{ \epsilon \rightarrow 0^{+} } \sqrt{|\Delta|} \bigg\{ \lim_{\bar{\epsilon} \rightarrow 0^+} \delta(\sigma + \bar{\epsilon} ) \Theta (u - \tau_{-1} -\epsilon) \Theta (\tau_1 - \epsilon -u) + \bigg[ \alpha_{1} \delta(\sigma) + \mathrm{pv} \left( \frac{\beta_{1}}{\sigma} \right) \bigg] \Theta( u-\tau_1 - \epsilon ) \Theta (\tau_{2} - \epsilon - u )
\nonumber
\\
	~ + \bigg[ \alpha_{-1} \delta(\sigma) + \mathrm{pv} \left( \frac{\beta_{-1}}{\sigma} \right) \bigg] \Theta( u-\tau_{-2} - \epsilon ) \Theta (\tau_{-1} - \epsilon - u ) + \ldots \bigg\}.
\label{GenG}
\end{align}
\end{widetext}
If $\tau_{2}(u')$ or $\tau_{-2}(u')$ does not exist, it is to be replaced by $\pm \infty$ here. Note that the three groups of step functions displayed in this equation confine various terms to $\mathcal{N}_0(u')$, $\mathcal{N}_1(u')$, and $\mathcal{N}_{-1}(u')$ [recall Fig. \ref{Fig:Normal} and the discussion surrounding \eqref{NormalNeighborhood} for definitions of these regions]. Terms in $\mathcal{N}_n(u')$ (with $|n|>1$) are also understood to be present if $\tau_n(u')$ exists. Roughly speaking, the limit $\epsilon \rightarrow 0$ ensures that $\Delta(p,p')$ and $\sigma(p,p')$ are only evaluated in regions where they are well-defined. The limit $\bar{\epsilon} \rightarrow 0$ present in the first term of \eqref{GenG} takes into account footnote \ref{Foot:Vertex}. It is necessary because $\delta(\sigma)$ is ill-defined in the coincidence limit $p \rightarrow p'$ (where $\nabla_a \sigma \rightarrow 0$).

The coefficients $\alpha_n$ and $\beta_n$ appearing in \eqref{GenG} are determined by the multiplicities of the various $\tau_n(u') \in T(u')$. $\alpha_0 = 1$ and $\beta_0 = 0$ are used as initial conditions for the matching equations \eqref{MatchingDegPos}-\eqref{MatchingDegNeg} and \eqref{MatchingSimpPos}-\eqref{MatchingSimpNeg} that fix $\alpha_n$ and $\beta_n$
when $n \neq 0$.  Given some particular $n$, either $\alpha_n = \pm 1$ and $\beta_n = 0$ or $\beta_n = \pm 1/\pi$ and $\alpha_n=0$.

Consider an ``observer'' moving on some (not necessarily causal) curve starting at $p'$. After passing through a hyperplane $S_{\tau_n(u')}$ conjugate to $S_{u'}$, the matching conditions \eqref{MatchingDegPos}-\eqref{MatchingDegNeg} and \eqref{MatchingSimpPos}-\eqref{MatchingSimpNeg} imply that such an observer would see $G_\mathrm{S}(\cdot, p')$ change according to the rules (modulo an overall factor of $\sqrt{|\Delta|}$):
\begin{itemize}
	\item If the conjugate pair $(S_{\tau_n(u')}, S_{u'})$ is associated with non-degenerate (multiplicity 1) conjugate points and $S_{\tau_n(u')}$ is traversed in a direction of increasing $u$, either
	\begin{equation}
	\pm \delta(\sigma) \rightarrow \pm \mathrm{pv} \left( \frac{1}{\pi \sigma} \right)
	\label{Rule1}
	\end{equation}
	or
	\begin{equation}
	\pm \mathrm{pv}\left( \frac{1}{\pi \sigma} \right) \rightarrow \mp \delta(\sigma).
	\end{equation}
	Signs on the right-hand sides of both of these replacement rules are reversed if traversing $S_{\tau_n(u')}$ in a direction of decreasing $u$.
	
	\item When the conjugate pair $(S_{\tau_n(u')}, S_{u'})$ has multiplicity $2$, the form of the Green function switches sign:
	\begin{align}
	\pm \delta(\sigma) &\rightarrow \mp \delta(\sigma),
	\\
	\pm \mathrm{pv} \left( \frac{1}{\pi \sigma} \right) &\rightarrow \mp \mathrm{pv} \left( \frac{1}{\pi \sigma} \right).
	\label{Rule4}
	\end{align}
	This is equivalent to the effect of two passes through distinct conjugate hyperplanes with multiplicity $1$.
\end{itemize}

When these rules are satisfied, expression \eqref{GenG} for $G_\mathrm{S}$ is everywhere a solution to \eqref{WaveEqScalar}. Retarded and advanced Green functions may easily be constructed from $G_\mathrm{S}$. For example,
\begin{widetext}
\begin{align}
G_\mathrm{ret} = \lim_{\epsilon \rightarrow 0^+} \sqrt{|\Delta|} \bigg\{ \lim_{\bar{\epsilon} \rightarrow 0^+} \delta( \sigma + \bar{\epsilon}) \Theta(u-u') \Theta(\tau_1 - \epsilon - u)
+ \sum_{n \geq 1} \bigg[ \alpha_n \delta(\sigma) + \mathrm{pv} \left( \frac{ \beta_n }{  \sigma } \right)  \bigg] \Theta (u-\tau_n-\epsilon) \Theta(\tau_{n+1} - u - \epsilon)
\bigg\}.
\label{GenGret}
\end{align}
\end{widetext}
This is a global solution to \eqref{WaveEqScalar}. It looks like a retarded Green function for $p$ near $p'$, but it is not the only solution with this property. See footnote \ref{Foot:AdvRet} and Sect. \ref{Sect:NonUnique}.

\subsubsection{Examples}

The simplest nontrivial examples of Green functions in specific plane wave spacetimes occur when all conjugate points are degenerate. In these cases, one finds from \eqref{MatchingDegPos} and \eqref{MatchingDegNeg} that $\alpha_n = (-1)^n$ and $\beta_n =0$. The retarded Green function is therefore given by
\begin{align}
	G_{\mathrm{ret}}(p,p')  =  (-1)^n \sqrt{ | \Delta | } \delta (\sigma) 
	\label{GretDeg}
\end{align}
when $p \in \mathcal{N}_n (u')$ and $u(p)>u(p')$. The form of this Green function changes sign on each pass through a conjugate hyperplane. The singular structure of $G_\mathrm{ret}$ (or $G_\mathrm{S}$ or $G_\mathrm{adv}$) follows the 2-fold pattern \eqref{SingStruct2} when all conjugate points are degenerate.

Conjugate points associated with conformally-flat plane waves are always degenerate, so their retarded Green functions are given by \eqref{GretDeg}. In the symmetric case where the wave amplitude $h(u)$ in \eqref{PlaneWaveConfFlat} remains constant, it is shown in Sect. \ref{Sect:Examples} that there exist an infinite number of degenerate conjugate hyperplanes (for any $u'$) at locations given by \eqref{TauEx}. Using \eqref{VVConfFlat}, the retarded Green function for such a spacetime is
\begin{equation}
	G_\mathrm{ret}(p,p') = \Theta(u-u') \left[  \frac{ (u-u') }{ \sin  (u-u') } \right] \delta(\sigma)
\end{equation}
if $h=1$ and $u-u' \neq n \pi$ (for all nonzero integers $n$). Eq. \eqref{SigConfFlat} provides an explicit coordinate expression for $\sigma$ in this case.

If all conjugate points in a particular plane wave spacetime are \textit{non}-degenerate, the scalar Green function has the repeating 4-fold singularity structure \eqref{SingStruct4} rather than the 2-fold structure \eqref{SingStruct2} found in the purely degenerate case. Applying \eqref{MatchingSimpPos} and \eqref{MatchingSimpNeg} to \eqref{GenGret} for some $p \in \mathcal{N}_n(u')$, $u(p)>u(p')$,
\begin{align}
	G_\mathrm{ret}(p,p') &=  \sqrt{|\Delta|}
\nonumber
\\
	&~ \times
	\begin{cases}
		(-1)^{ \frac{n}{2} } \delta(\sigma) & \text{if $n$ even} ,
	\\
		(-1)^{ \frac{n-1}{2} } \mathrm{pv} (1/\pi \sigma) & \text{if $n$ odd} .
	\end{cases}
	\label{Green4Fold}
\end{align}

This is, in a sense, the ``physically generic'' form for retarded Green functions in plane wave spacetimes. It is not correct if there exist degenerate conjugate hyperplanes\footnote{Degenerate and non-degenerate conjugate hyperplanes may exist in the same spacetime. Examples of this may be found by fine-tuning the parameter $h$ appearing in \eqref{HGravWaveReal}. The singular structure of the Green function in such cases deviates from the simple patterns \eqref{SingStruct4} and \eqref{SingStruct2}. Rules \eqref{Rule1}-\eqref{Rule4} must then be applied on a case-by-case basis.}, although such structures tend to be ``fragile.'' Consider, for example, a plane wave that initially possesses a degenerate conjugate hyperplane. If such a spacetime is perturbed by slightly changing $H_{ij}(u)$, the original degenerate hyperplane tends -- but is not guaranteed -- to split into two closely-spaced non-degenerate conjugate hyperplanes. Passing through one non-degenerate hyperplane might therefore be viewed as physically equivalent to quickly passing through two non-degenerate conjugate hyperplanes. Indeed, we have found that two passes through non-degenerate hyperplanes has the same effect on a scalar Green function as one pass through a degenerate conjugate hyperplane.

As a simple example of the 4-fold singularity structure exhibited in \eqref{Green4Fold}, consider a symmetric gravitational plane wave where the wave profile $H_{ij}(u)$ is given by \eqref{HGravWave}. Such spacetimes have an infinite number of non-degenerate conjugate hyperplanes at the locations \eqref{TauEx}. The retarded Green function in this case is explicitly
\begin{align}
	G_\mathrm{ret}(p,p') =  \left( \frac{  u-u'   }{ \sqrt{ | \sin (u-u') | \sinh (u-u')   }  } \right)
	\nonumber
	\\
	~ \times \begin{cases}
		(-1)^{ \frac{n}{2} } \delta(\sigma) & \text{if $n$ even} ,
		\\
		(-1)^{ \frac{n-1}{2} } \mathrm{pv} (1/\pi \sigma) & \text{if $n$ odd} .
	\end{cases}
	\label{GGravWave}
\end{align}
It is assumed here that $u>u'$. Also note that $n$ is given by
\begin{equation}
	n =	\left\lfloor (u - u' ) /\pi \right\rfloor ,
\end{equation}
where $\lfloor \cdot \rfloor$ denotes the floor function. A coordinate expression for the $\sigma$ appearing here may be found by substituting \eqref{AGravWave} and \eqref{BGravWave} into \eqref{SigmaGen}.

\subsubsection{Some comments}

Before moving on, recall that two questions are posed in the introduction regarding the qualitative way in which a Green function may change its singular structure. First, how can a very localized distribution like $\delta(\sigma)$ ``smoothly transition'' into something as apparently spread out as $\mathrm{pv}(1/\sigma)$? In plane wave Green functions, this change occurs on $u= \mathrm{const.}$ hyperplanes. Furthermore, the discussion in Sect. \ref{Sect:CausticGeo} implies that $|\sigma| \rightarrow \infty$ on almost all (but not quite all) approaches to such surfaces. This means that like $\delta(\sigma)$, $\mathrm{pv}(1/\sigma)$ vanishes almost everywhere when approaching a conjugate hyperplane where it might transition into a $\delta$-function.

It is also asked in the introduction how a retarded Green function can involve a term proportional to $\mathrm{pv}(1/\sigma)$ when $\sigma(p,p')>0$ traditionally implies that the points $p$ and $p'$ are not in causal contact. In plane wave spacetimes, $\sigma(p,p') >0$ implies that the (only) geodesic connecting $p$ and $p'$ is spacelike. Despite this, the discussion of Sect. \ref{Sect:Causality} implies that such points are still in causal contact as long as there exists at least one hyperplane conjugate to $S_{u(p')}$ that cuts through the geodesic segment connecting $p$ and $p'$. It is only in this case that our Green function can have support in regions where $\sigma (p,p') >0$. All of the support of $G_\mathrm{ret}(\cdot,p')$ is therefore in causal contact with $p'$.

A somewhat weaker version of this argument holds in any spacetime (including those that are not plane waves). Consider a null geodesic satisfying $z(s') = p'$ and $z(s) = p$. It follows from theorem 9.3.8 of \cite{Wald} that if there exists at least one point conjugate to $p'$ on the geodesic segment between $p'$ and $p$, these two points may be connected by timelike curves. When this condition holds, it follows that $p$ is in the chronological past or future of $p'$. Furthermore, there exists an open neighborhood of every point in the chronological past or future of $p'$ that remains entirely in this set. It follows that an open neighborhood of $p$ lies entirely in causal contact with $p'$ if $p$ is connected to $p'$ by a null geodesic segment with at least one point conjugate to $p'$. 

For plane wave spacetimes, this argument guarantees that two points $p$ and $p'$ satisfying $\sigma(p,p')>0$ and separated by at least one hyperplane conjugate to $S_{u(p')}$ are in causal contact at least if $\sigma$ is sufficiently small. It is a special property of plane wave spacetimes that this result continues to hold even when $\sigma$ is large.

In Sect. \ref{Sect:Penrose}, we show that some features of Green functions in generic spacetimes very near null geodesics are captured by appropriate plane wave Green functions. After a conjugate point, there is a sense in which a generic Green function may again be nonzero when $\sigma>0$. Here, $\sigma$ is interpreted as the world function of an associated plane wave spacetime. It acts like a coordinate for an infinitesimal region around the reference null geodesic. The argument above guarantees that terms like $\mathrm{pv}[1/\sigma(\cdot,p')]$ appearing in (say) retarded Green functions on generic spacetimes remain in causal contact with $p'$ near the reference geodesic.

\subsubsection{Non-uniqueness of plane wave Green functions}
\label{Sect:NonUnique}

Recall that we have constructed a ``retarded Green function'' $G_\mathrm{ret}(p,p')$ by demanding that it solve \eqref{BoxGAbstr} everywhere and that it be equal to \eqref{Hadamard} for all $p \in \mathcal{N}_0(u')$. Other distributions also satisfy these constraints. One may consider, e.g.,
\begin{equation}
	G_{\mathrm{ret}}(p,p') + \Gamma(p,p'),
\end{equation}
where $\Gamma(p,p')$ is some solution to $\nabla^a \nabla_a \Gamma(p,p') = 0$ that vanishes when $u<\tau_1(u')$. Any object of this form may reasonably be interpreted as a ``retarded Green function.'' Indeed, one might only require that $\nabla^a \nabla_a \Gamma(p,p') = 0$ and that $\Gamma(p,p')$ vanish when $u < u' + (\mbox{something positive})$.

Nontrivial distributions $\Gamma(p,p')$ always exist. Consider, for example, anything which depends purely on $u$ and $u'$ and that vanishes when, say, $u<\tau_1(u')$. As another possibility, suppose that $\Gamma(p,p')$ is, for fixed $p'$, concentrated on a constant-$u$ hypersurface $S_{t(u')}$. One such example is
\begin{equation}
	\Gamma(p,p') = \delta(t(u')-u) \gamma(\bx, \bx'),
\end{equation}
where $\gamma(\bx,\bx')$ is harmonic at least in its first argument: $\nabla^2 \gamma(\bx,\bx') = 0$. 

\section{Green functions in general spacetimes}
\label{Sect:Penrose}

Up to this point, we have focused on the propagation of (test) scalar fields $\Phi$ on plane wave backgrounds. As outlined in the introduction, plane wave spacetimes have a number of mathematically attractive features. They are not, however, physically realistic on large scales. Plane wave geometries are not asymptotically flat, nor even globally hyperbolic. Despite this, one might hope that there is a sense in which our results remain ``essentially correct'' for physically realistic plane waves where the metric is adequately approximated by \eqref{PlaneWaveGen} only in some finite region\footnote{A somewhat more realistic model of a simple gravitational wave is a pp-wave where the profile function $H(u,\bx)$ appearing in \eqref{ppMetricGen} is quadratic in $\bx$ in some finite region and subquadratic as $|\bx| \rightarrow \infty$. Geometries of this type are discussed in, e.g., \cite{GenPPWave0,GenPPWave}. Their causal properties do not display the pathologies of ``pure'' plane wave spacetimes.}. We now argue for a significantly stronger result: The singular behavior of Green functions in \textit{generic} spacetimes is, to leading order, equivalent to the singular behavior of Green functions in appropriate plane wave spacetimes. This is similar to a statement proposed in \cite{QED1}. 

The correspondence with plane wave spacetimes is motivated by two observations. First, general theorems regarding the propagation of singularities imply that the singular supports of generic Green functions lie on null geodesics \cite{BarFredenhagen}. Second, there is a sense in which the geometry ``near'' a null geodesic in any spacetime is equivalent -- via what is known as a Penrose limit -- to the geometry of an appropriate plane wave spacetime \cite{PenroseLimit, BlauPenrose, BlauNotes}. Furthermore, one might suppose that the behavior of a generic Green function near its singular support (i.e., near a null geodesic) could be at least partially understood using the geometry of that region. It would then appear to follow that some aspects of the singular structure of a Green function in a generic spacetime $(\check{M} ,\check{g}_{ab})$ near an affinely-parameterized null geodesic $\check{z}(u)$ might be understood by computing a Green function associated with a plane wave spacetime $(M,g_{ab})$ obtained from $(\check{M},\check{g}_{ab})$ and $\check{z}(u)$ using a Penrose limit. 

Before establishing that this line of reasoning is correct, we first provide a review of Penrose limits in Sect. \ref{Sect:PenroseDef}. An appropriate notion of a Green function's ``leading order singular behavior'' is then defined in Sect. \ref{Sect:GenericGreen}. Near a given null geodesic, it is argued that this structure is reproduced by a Green function associated with an appropriate plane wave spacetime. The results of Sect. \ref{Sect:Green} are then applied to determine the singular structure of Green functions for scalar fields propagating in arbitrary four-dimensional spacetimes. Lastly, a similar argument is provided in Sect. \ref{Sect:TensorGeneralize} for the behavior of Green functions associated with wave equations involving fields of nonzero tensor rank.

\subsection{Penrose limits}
\label{Sect:PenroseDef}

As formulated in \cite{BlauPenrose}, the Penrose limit takes as input a null geodesic $\check{z}(u)$ in a spacetime $(\check{M}, \check{g}_{ab})$, and uses this to construct a null generalization of a Fermi normal coordinate system\footnote{A check mark is omitted on the symbol $u$ because this coordinate is not rescaled in \eqref{PenroseScale} below.} $(u,\check{v},\check{\bx})$. Assume for simplicity that the reference geodesic $\check{z}(u)$ is defined for all $u \in \mathbb{R}$ and that $\check{g}_{ab}$ is smooth along this curve. Next, construct a tetrad $\{ e^a_\pm (u), e^a_i(u) \}$ on $\check{z}(u)$ that is parallel-propagated along $\check{z}(u)$ with respect to $\check{g}_{ab}$. Let the first element of this tetrad be the null tangent to the reference geodesic:
\begin{equation}
	e^a_{+}(u) = \frac{\rmd \check{z}(u)}{\rmd u}.
\end{equation}
Let the second element $e^a_{-}(u)$ of the tetrad also be null, and suppose that it satisfies $\check{g}_{ab} e^a_{+} e^b_{-} = -1$.  The final two elements $e^{a}_{i}(u)$ of the tetrad are to be spacelike and orthonormal. They are orthogonal to the two null vectors $e^a_{+}(u)$ and $e^a_{-}(u)$:
\begin{equation}
	\check{g}_{ab} e^a_{+} e^b_{i}  = \check{g}_{ab} e^a_{-}  e^b_{i}  = 0.
\end{equation}

Given some point $p \in \check{M}$ sufficiently near the reference geodesic $\check{z}(u)$, identify a $u$ coordinate associated with $p$ by solving the equation
\begin{equation}
	e^a_{-}(u) \check{\nabla}_a \check{\sigma}(\check{z}(u),p) = 0.
	\label{FermiNormal}
\end{equation}
Here, $\check{\sigma}(p,p')$ denotes Synge's function in the spacetime $(\check{M}, \check{g}_{ab})$. Once $u = u( p )$ has been fixed using \eqref{FermiNormal}, the remaining three coordinates of $p$ are determined by defining the tetrad components of the ``separation vector'' $- \check{\nabla}^a \check{\sigma} (\check{z}(u (p) ), p)$ to be the coordinates $\check{v}(p)$ and $\check{\bx}(p)$:
\begin{equation}
	-\check{\nabla}^a \check{\sigma}  = \check{v} e^a_{-}  + \check{x}^i e^a_i .
\end{equation}
Inverting this relation,
\begin{subequations}
\begin{align}
	\check{v} (p) := e^a_{+}(u( p )) \check{\nabla}_a \check{\sigma}( \check{z}(u( p )), p ) ,
	\\
	\check{x}^i (\check{p}) := - \delta^{ij} e^a_{j}(u( p )) \check{\nabla}_a \check{\sigma}( \check{z}(u( p )), p ) .
\end{align}
\end{subequations}
Together, these equations and \eqref{FermiNormal} define a Fermi-like coordinate system $(u,\check{v}, \check{\bx})$ near the reference geodesic $\check{z}(u)$. Given any $u' \in \mathbb{R}$, the point $\check{z}(u')$ has coordinates $u=u'$ and $\check{v} = \check{\bx} = 0$ in this chart. 

The Penrose limit involves a 1-parameter family of transformations on the components $\check{g}_{\check{\mu} \check{\nu}}$ of the metric in the coordinates $(u,\check{v}, \check{\bx})$. Consider, in particular, the substitutions
\begin{equation}
	u \rightarrow u, \quad \check{v} \rightarrow v := \lambda^{-2} \check{v}, \quad \check{\bx} \rightarrow \bx := \lambda^{-1} \check{\bx}
	\label{PenroseScale}
\end{equation}	
for any $\lambda > 0$. In the limit $\lambda \rightarrow 0$, this transformation can be interpreted as ``zooming up'' on the reference geodesic $\check{z}(u)$ and then boosting along it by a similar factor. All components $\check{g}_{\mu\nu}$ of the metric in the coordinate system $(u,v,\bx)$ vanish as $\lambda \rightarrow 0$. Expanding the line element in powers of $\lambda$, the first non-vanishing term is proportional to $\lambda^2$ \cite{BlauPenrose}:
\begin{align}
	\rmd \check{s}^2  = \lambda^2 \big[- 2 \rmd u \rmd v - \check{R}_{+i+j}(u) x^i x^j \rmd u^2 + |\rmd \bx|^2 \big] 
\nonumber
\\
	~ + O(\lambda^3).
	\label{PenroseExp}
\end{align}
Here, $\check{R}_{+i+j}(u)$ denotes the appropriate tetrad components of the Riemann tensor on the reference geodesic:
\begin{equation}
	\check{R}_{+i+j}(u) := \check{R}_{abcd}(\check{z}(u)) e^a_{+}(u) e^b_{i}(u) e^c_{+}(u) e^d_{j}(u) .
\end{equation}
Comparing \eqref{PlaneWaveGen} and \eqref{PenroseExp}, it is clear that
\begin{equation}
	g_{\mu\nu} := \lim_{\lambda \rightarrow 0} \lambda^{-2} \check{g}_{\mu\nu}
	\label{PenroseMetric}
\end{equation}
is -- regardless of the original geometry -- the metric of a plane wave spacetime in Brinkmann coordinates with the amplitude and polarization profile
\begin{equation}
	H_{ij} (u) = - \check{R}_{+i+j}(u).
	\label{PenroseH}
\end{equation}
In this sense, the geometry near any\footnote{If the reference geodesic intersects a singularity and therefore cannot be extended to infinitely large values of its affine parameter, one can still perform a Penrose limit. The only difference is that the resulting plane wave spacetime is slightly different from the type described in Sects. \ref{Sect:ppGeometry} and \ref{Sect:Geometry}. In such cases, the coordinate $u$ would no longer take all values in $\mathbb{R}$ and $H_{ij}(u)$ may be unbounded for finite $u$.} null geodesic is equivalent to that of an appropriate  plane wave spacetime.

\begin{figure}
	\includegraphics[width=.95\linewidth]{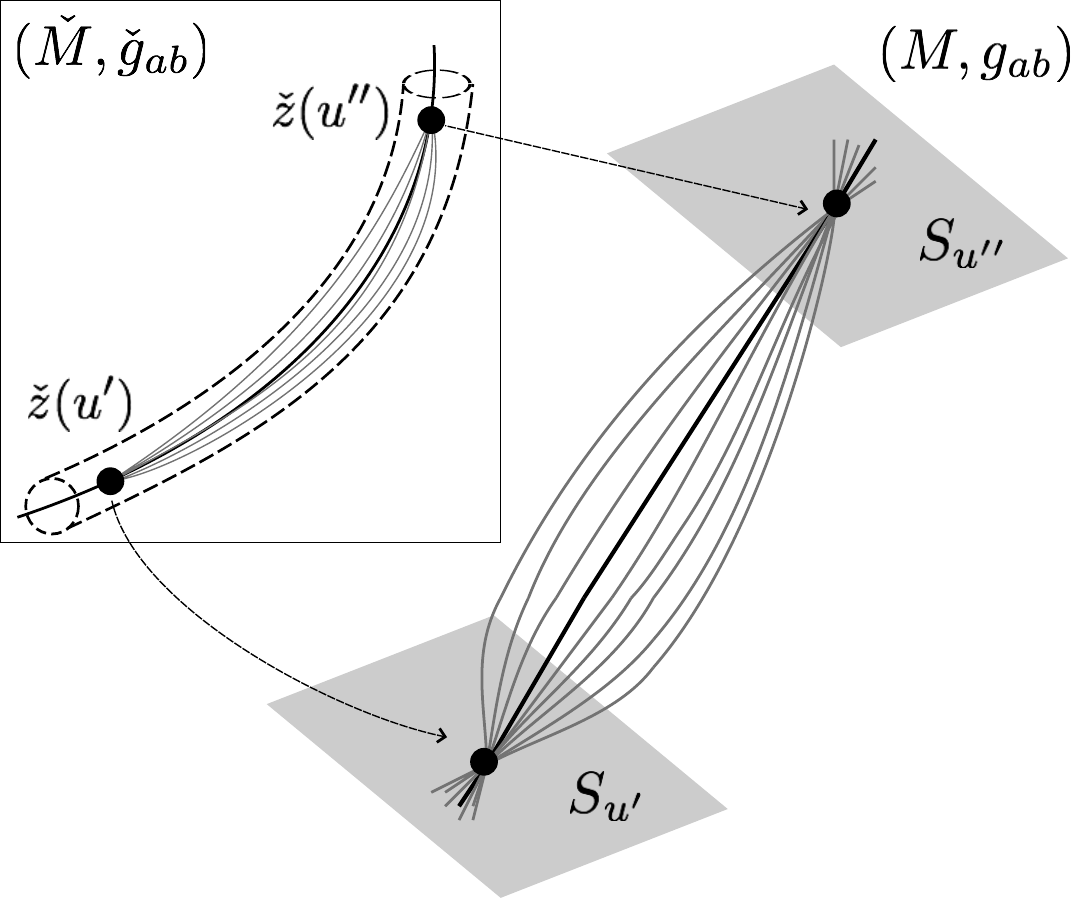}
	\caption{Schematic illustrating that Penrose limits map conjugate points to conjugate points. The inset shows the reference null geodesic in the spacetime $(\check{M}, \check{g}_{ab})$ along with a number of nearby geodesics. The points $\check{z}(u')$ and $\check{z}(u'')$ are conjugate on the reference geodesic. They are mapped to the conjugate hyperplanes $S_{u'}$ and $S_{u''}$ in the associated plane wave spacetime.}
	\label{Fig:PenroseConj}
\end{figure}

\begin{figure}
	\includegraphics[width=.95\linewidth]{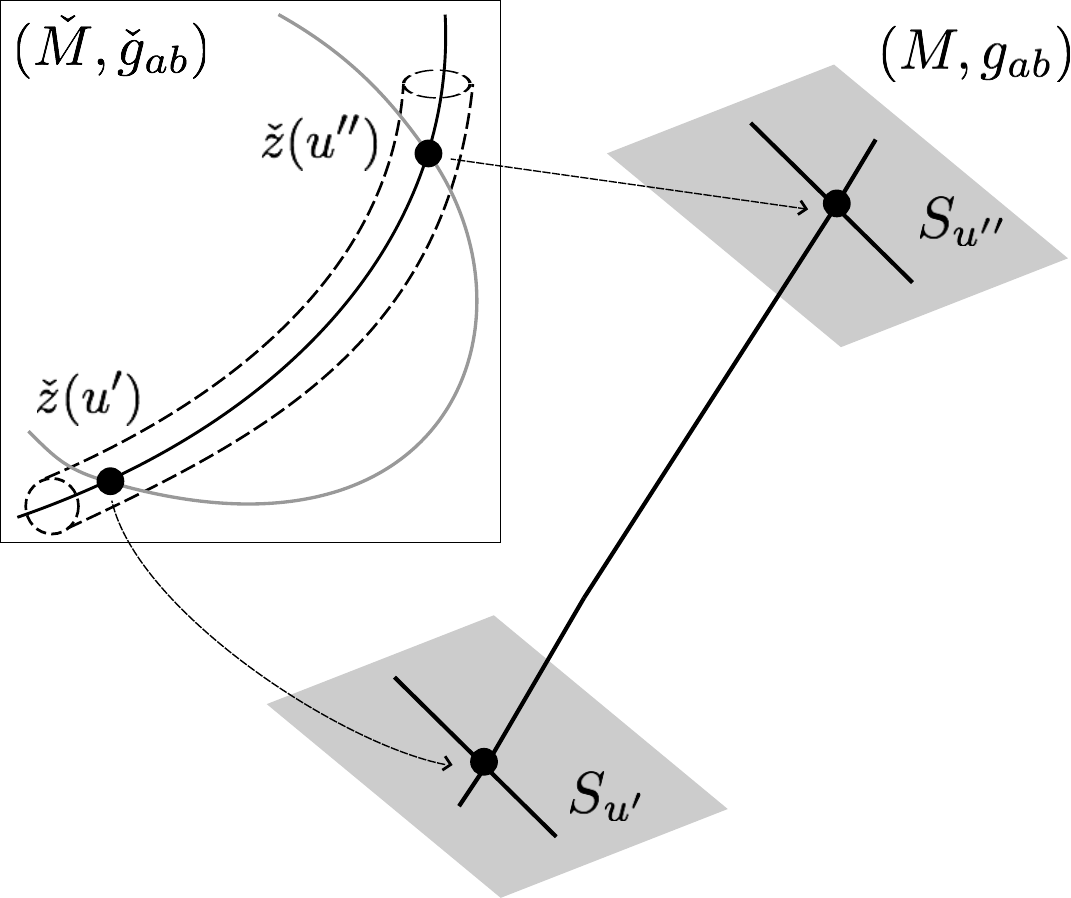}
	\caption{The effect of a Penrose limit on a null curve which intersects the reference geodesic at the points $\check{z}(u')$ and $\check{z}(u'')$. Such a curve is mapped to null geodesics on the hyperplanes $S_{u'}$ and $S_{u''}$. Note that one continuous curve in the original spacetime $(\check{M}, \check{g}_{ab})$ is mapped into two disconnected curves in the associated plane wave spacetime $(M,g_{ab})$.}
	\label{Fig:PenroseNonConj}
\end{figure}

For any choice of reference geodesic, the Penrose limit preserves various properties of the original spacetime $(\check{M}, \check{g}_{ab})$ in the associated plane wave spacetime $(M, g_{ab})$ \cite{GerochHereditary, BlauHereditary, BlauNotes}. For example, conformally-flat spacetimes are always mapped to conformally-flat plane waves. Similarly, vacuum (Ricci-flat) spacetimes are always mapped to vacuum plane waves. In general, the number of linearly independent Killing fields cannot decrease after taking a Penrose limit.

For every $u \in \mathbb{R}$, the Penrose limit maps the point $\check{z}(u') \in \check{M}$ on the reference geodesic into a point with Brinkmann coordinates $u=u'$ and $v=\bx=0$ in the associated plane wave spacetime. This implies that the reference curve -- which is a null geodesic in $(\check{M},\check{g}_{ab})$ -- is mapped into a null geodesic in $(M,g_{ab})$. 

Crucially, the conjugate point structure of $\check{z}(u)$ is identical in both the original and plane wave spacetimes. If $\check{z}(u')$ and $\check{z}(u'')$ are two points that are conjugate along the reference geodesic in the original spacetime $(\check{M}, \check{g}_{ab})$, the hyperplanes $S_{u'}$ and $S_{u''}$ are conjugate in the associated plane wave spacetime $(M,g_{ab})$. The multiplicities of the conjugate pairs $( \check{z}(u'), \check{z}(u''))$ and $(S_{u'}, S_{u''})$ are identical. See Fig. \ref{Fig:PenroseConj}.

It is also important to note the effect of a Penrose limit on a smooth curve in the original spacetime which intersects the reference geodesic at, say, $\check{z}(u')$. It is straightforward to show from \eqref{PenroseScale} that all such trajectories (which are not infinitesimal deformations of the reference geodesic) are mapped to the $u=u'$ hyperplane $S_{u'}$. They are geodesics with respect to the plane wave metric $g_{ab}$, and are therefore straight lines in the coordinates $(v,\bx)$ which pass through $v = \bx = 0$.  Any curves which are null or timelike at the intersection point $\check{z}(u')$ (and some that are spacelike there) are mapped to the null geodesic with Brinkmann coordinates $u=u'$ and $\bx=0$. See Fig. \ref{Fig:PenroseNonConj}.

\subsection{Singular structure of generic scalar Green functions}
\label{Sect:GenericGreen}

We now consider a Green function $\check{G}(p, p')$ associated with the scalar wave equation 
\begin{equation}
	\check{L} \check{\Phi} = - 4 \pi \check{\rho}
	\label{WaveEqScalarGen}
\end{equation}
in an arbitrary spacetime $(\check{M}, \check{g}_{ab})$. As in \eqref{GenericWaveEqn}, the principal part of the linear differential operator $\check{L}$ is to be given by $\check{g}^{ab} \check{\nabla}_a \check{\nabla}_b$. Unlike in the plane wave field equation \eqref{WaveEqScalar}, we allow for additional terms involving at most one derivative (so fields with, e.g., mass or non-minimal coupling to the curvature may be considered). Any Green function associated with \eqref{WaveEqScalarGen} is required to satisfy
\begin{equation}
	\langle \check{G}(p,p'), \check{L}^\dag \check{\varphi}(p) \rangle = - 4 \pi \check{\varphi}(p')
	\label{WaveEqScalarGenGreen}
\end{equation}
for all test functions $\check{\varphi} \in C^\infty_0(\check{M})$. Here, $\check{L}^\dag$ denotes the adjoint of $\check{L}$. 

Penrose limits can be thought of as zooming in on a particular null geodesic $\check{z}(u)$. It follows that a plane wave Green function $G(p,p')$ could only be expected to describe the action of a generic Green function $\check{G}(p,p')$ near $\check{z}(u)$. A more precise statement of this form is that we would like to consider the action of $\check{G}(p,p')$ on test functions $\varphi(u,v,\bx) \in C^\infty_0(\mathbb{R}^4)$ that are fixed in the scaled coordinates $(u,v,\bx)$ related to the Fermi-like coordinates $(u, \check{v}, \check{\bx})$ via \eqref{PenroseScale}. Given some $\varphi$, define a 1-parameter family of test functions $\check{\varphi}_\lambda$ such that
\begin{equation}
	\check{\varphi}_\lambda (u, \check{v}, \check{\bx} )  := \varphi ( u, \lambda^{-2} \check{v}, \lambda^{-1} \check{\bx} )
	\label{ScaledTest}
\end{equation}
for all $\lambda >0$. Test functions of this type always remain near $\check{z}(u)$ when $\lambda$ is sufficiently small. 

The action of any smooth function -- call it $\check{\mathcal{V}}(p)$ -- on a test function $\check{\varphi}_\lambda$ of the form \eqref{ScaledTest} is given by
\begin{align}
	\langle \check{\mathcal{V}} , \check{\varphi}_\lambda \rangle &= \int \! \rmd u \rmd \check{v} \rmd^2 \check{\bx} \, \check{\mathcal{V}}(u, \check{v}, \check{\bx}) \check{\varphi}_\lambda ( u, \check{v}, \check{\bx} )
	\nonumber
	\\
	&= \lambda^4 \int \! \rmd u \rmd v \rmd^2 \bx \, \check{\mathcal{V}} (u, \lambda^2 v, \lambda \bx) \varphi(u,v,\bx).
	\label{SmoothScaling}
\end{align}
This clearly scales like $\lambda^4$ as $\lambda \rightarrow 0$. One would therefore expect any tail terms in $\check{G}(p,p')$ to scale like $\lambda^4$ when acting on test functions $\check{\varphi}_\lambda$.

Portions of $\langle \check{G}(p,p'), \check{\varphi}_\lambda(p) \rangle$ depending on parts of $\check{G}(p,p')$ which are singular on $\check{z}(u)$ decrease more slowly than $\lambda^4$ in the Penrose limit $\lambda \rightarrow 0$. Consider, for example, Synge's function $\check{\sigma}(p,p')$ for pairs of points that are sufficiently close that the standard definition of this object remains well-defined. Then the definition \eqref{SigDef} and the Penrose limit metric $g_{ab}$ given by \eqref{PenroseMetric} suggest that
\begin{equation}
	\sigma \sim \lambda^{-2} \check{\sigma}.
\end{equation}
Hence,
\begin{equation}
	\delta( \check{\sigma} ) \sim \lambda^{-2} \delta(\sigma), \quad \mathrm{pv} \left( \frac{1}{\check{\sigma}} \right) \sim \lambda^{-2} \mathrm{pv} \left( \frac{ 1 }{ \sigma } \right) .
	\label{SingScale}
\end{equation}
These are the most singular terms that one would expect to find in $\check{G}(p,p')$. It is therefore reasonable to expect that $\langle \check{G}, \check{\varphi}_\lambda \rangle$ scales like $\lambda^4 \lambda^{-2} =\lambda^2$ as $\lambda \rightarrow 0$. 

We now use this heuristic argument as motivation to \textit{define} the ``leading order singular portion'' of a generic scalar Green function $\check{G}(p,p')$. For any $p' = \check{z}(u') \in \check{M}$ lying on the reference geodesic and any test function $\check{\varphi}_\lambda$ that is, as described above, fixed in the scaled coordinates $(u,v,\bx)$, define a linear operator $G(p,p')$ by
\begin{equation}
	\langle G(p,p'), \varphi(p) \rangle := \lim_{\lambda \rightarrow 0} \lambda^{-2} \langle \check{G}(p,p'), \check{\varphi}_\lambda (p) \rangle. 
	\label{PlaneGreenDef}
\end{equation}
The Hadamard form \eqref{Hadamard} and the estimates \eqref{SingScale} guarantee that this limit exists at least for test functions whose supports lie sufficiently close to $p'$. We assume, however, that the limit exists for \textit{all} test functions of the form \eqref{ScaledTest}.

The notation in \eqref{PlaneGreenDef} suggests that $G(p,p')$ is a Green function in an appropriate plane wave spacetime. To establish that this is indeed the case, consider families of test functions generated by $g^{ab} \nabla_a \nabla_b \varphi$, where $g^{ab}$ is the inverse of the Penrose limit metric \eqref{PenroseMetric} and $\nabla_a$ is the associated covariant derivative operator. Substitution into \eqref{PlaneGreenDef} and use of \eqref{WaveEqScalarGenGreen} shows that
\begin{align}
	\langle G, g^{ab} \nabla_a \nabla_b \varphi \rangle = \lim_{\lambda \rightarrow 0} \langle \check{G}, \check{L}^\dag \check{\varphi}_\lambda \rangle = - 4 \pi \varphi (p')
\end{align}
for all $\varphi \in C^\infty_0 (\mathbb{R}^4)$. Comparison with \eqref{BoxGWeak} shows that the operator $G(p,p')$ is a Green function for a scalar field propagating on a plane wave spacetime with metric \eqref{PenroseMetric}. It follows that appropriate components of generic Green functions behave like plane wave Green functions near null geodesics. Properties of plane wave Green functions derived in Sect. \ref{Sect:Green} may therefore be used to understand some aspects of scalar Green functions in more general spacetimes. 

To summarize, fix a background spacetime $(\check{M}, \check{g}_{ab})$ in which a scalar field $\check{\Phi}$ propagates according to \eqref{WaveEqScalarGen}. Fix a point $p' \in \check{M}$ corresponding to the location of some small disturbance in $\check{\Phi}$. The effect of such a disturbance may now be followed in a neighborhood of some affinely-parameterized null geodesic $\check{z}(u)$ which passes through $p' = \check{z}(u')$. Perform a Penrose limit using $\check{z}(u)$ and the metric $\check{g}_{ab}$. Such a limit requires a choice of tetrad $\{ e_{\pm}^a(u) , e^a_i (u) \}$ along the reference geodesic. This is to be constructed using the prescription described in Sect. \ref{Sect:PenroseDef}. Defining a Fermi-like coordinate system $(u,\check{v}, \check{\bx})$ and performing the scaling \eqref{PenroseScale} produces [via \eqref{PenroseMetric}] a plane wave metric in Brinkmann coordinates with the wave profile \eqref{PenroseH}.

The (say) retarded Green function $\check{G}_\mathrm{ret}(p,p')$ associated with \eqref{WaveEqScalarGenGreen} is related to the retarded plane wave Green function $G_\mathrm{ret}(p,p')$ constructed in Sect. \ref{Sect:GlobalSoln}. For any test function $\varphi \in C^\infty_0(\mathbb{R}^4)$, \eqref{ScaledTest} may be rewritten as
\begin{equation}
	\lim_{\lambda \rightarrow 0} \lambda^{-2} \langle \check{G}_\mathrm{ret} , \check{\varphi}_\lambda \rangle = \langle G_\mathrm{ret}, \varphi \rangle + \langle \Gamma, \varphi \rangle.
	\label{PenroseGreen}
\end{equation}
Here, $\Gamma(p,p')$ is an appropriate solution to the homogeneous wave equation
\begin{equation}
	\langle \Gamma, g^{ab} \nabla_a \nabla_b \varphi \rangle = 0
\end{equation}
associated with the plane wave spacetime. $\Gamma(\cdot,p')$ vanishes in the Penrose limit of any normal neighborhood of $p'$ [as computed in the original spacetime $(\check{M}, \check{g}_{ab})$], but need not vanish globally. We return to this point shortly. 

For the moment, consider only the first term on the right-hand side of \eqref{PenroseGreen}. It is clear from the discussion in Sect. \ref{Sect:GlobalSoln} that, fixing $p'$, $G_\mathrm{ret}(\cdot,p')$ is proportional either to $\delta(\sigma)$ or $\mathrm{pv}(1/\sigma)$. It can switch between these two possibilities and also switch sign. If there is nothing in $\Gamma(\cdot,p')$ that remains singular on curves where $\sigma(\cdot, p') = 0$, such terms provide a precise sense in which generic retarded Green functions $\check{G}_\mathrm{ret}(p,p')$ have singularities that ``look like'' either $\delta(\sigma)$ or $\mathrm{pv}(1/\sigma)$ near null geodesics [where $\sigma(\cdot, p')=0$]. It is simple to determine which of these forms is appropriate by considering the points conjugate to $p'$ along $\check{z}(u)$. These may be found by first using \eqref{PenroseH} to construct $\mathbf{H}(u)$ from $\check{R}_{abc}{}^{d}(\check{z}(u))$. $\mathbf{H}(u)$ can then be used to compute the matrix $\bB(u,u')$ defined by \eqref{JacobiSpatial} and \eqref{ABboundary}. A point $\check{z}(\tau_n)$ is conjugate to $\check{z}(u') = p'$ if and only if $\det \bB(\tau_n, u') = 0$. The multiplicity of such a conjugate pair is equal to the nullity of $\bB(\tau_n,u')$.

The discrete set of points conjugate to $\check{z}(u')$ on $\check{z}(u)$ generically divide the reference geodesic into a number of regions corresponding to the $\mathcal{N}_n(u')$ defined at the beginning of Sect. \ref{Sect:Geometry} (recall Fig. \ref{Fig:Normal}). Rules \eqref{Rule1}-\eqref{Rule4} may be used to find the leading order singular structure of $\check{G}_\mathrm{ret}(\cdot, p')$ in each of these regions using only the multiplicities of intervening conjugate points. 

This argument takes into account only the contribution to $\check{G}_\mathrm{ret}(p,p')$ from the first term on the right-hand side of \eqref{PenroseGreen}. In general, the second term on the right-hand side of this equation may also be important. It could be required if, as described in the introduction, null geodesics emanating from $p'$ later intersect $\check{z}(u)$ at a point that is not conjugate to $p'$. Such phenomena introduce new singularities in a neighborhood of the reference geodesic whose locations cannot be predicted using only the limited geometric information preserved in the Penrose limit. As illustrated in Fig. \ref{Fig:PenroseNonConj}, Penrose limits map any null geodesic in the full spacetime which intersects the reference geodesic at, say, $\check{z}(u')$ into a null geodesic in the plane wave spacetime confined to the hyperplane $S_{u'}$. One might therefore expect to take into account the singularities transported by such geodesics using an appropriate $\Gamma(p,p')$ in \eqref{PenroseGreen} that is singular when $u=u'$, $v \in \mathbb{R}$, and $\bx = 0$. If there is a surface full of null geodesics that transversely intersect the reference geodesic, an appropriate $\Gamma(p,p')$ might be singular throughout $S_{u'}$. It is not, however, clear precisely what form $\Gamma(p,p')$ should take.

As a very simple model for this phenomenon, consider the field
\begin{equation}
	\Phi(p) = \delta(\sigma(p,p')) + \alpha \delta (\sigma(p,p''))
\end{equation}
in flat spacetime with $\alpha$ an arbitrary constant and $\sigma(p', p'') \neq 0$. For $p$ different from $p'$ and $p''$, this satisfies the homogeneous equation $\nabla^a \nabla_a \Phi = 0$. It might be viewed as approximating a Green function in curved spacetime near some small segment of a null geodesic emanating from $p'$. The term proportional to $\alpha$ schematically represents the effect of a transversely intersecting null geodesic not associated with a conjugate point. 

Consider a null geodesic starting at $p'$ and construct a Fermi-like coordinate system $(u,\check{v},\check{\bx})$ as described above. Adjusting the origin of the $u$ coordinate appropriately, $\Phi(p)$ has the explicit form
\begin{align}
	\Phi & (u, \check{v}, \check{\bx} ) = \delta \Big( - u \check{v} + \frac{1}{2} | \check{\bx} |^2 \Big) 
	\nonumber
	\\
	&~ + \alpha \delta \Big( -(u-u'') (\check{v}- \check{v}'') + \frac{1}{2} | \check{\bx} - \check{\bx}''|^2 \Big).
\end{align}
If the light cone of $p''$ intersects the reference geodesic somewhere, $\check{v}'' \neq 0$. Scaling the coordinates as in \eqref{PenroseScale} then results in
\begin{align}
	\Phi = & ~ \lambda^{-2} \delta \Big( - u v + \frac{1}{2} | \bx |^2 \Big) 
\nonumber
\\
	&	~ + \alpha \delta \Big( (u-u'')  \check{v}'' + \frac{1}{2} | \check{\bx}''|^2 \Big).
\end{align}
It is clear that the second line of this equation becomes negligible as $\lambda \rightarrow 0$. This suggests -- but does not prove -- that transverse intersections of the light cone not associated with conjugate points do not survive the Penrose limit at all [i.e., $\Gamma = 0$ in \eqref{PenroseGreen}]. It is possible that a different result might arise if, e.g., the intersection point were conjugate along some of the connecting geodesics, but not along the reference geodesic.

We can only conclude that there might exist cases where $\Gamma \neq 0$ in \eqref{PenroseGreen}. If so, the singular support of $\Gamma$ necessarily extends to $|v| \rightarrow \infty$. Such singularities appear quite different from those associated with $G_\mathrm{ret}(p,p')$. There is a sense in which they are ``frozen'' at specific affine times on the reference geodesic.

\subsubsection*{Examples}

We now discuss some consequences of the above results. General statements are made regarding Green functions associated with conformally-flat spacetimes and an important class of vacuum spacetimes. Some more specific examples are also mentioned briefly. 

The simplest general statement following from the argument of Sect. \ref{Sect:GenericGreen} concerns scalar Green functions in spacetimes whose metrics are conformally flat. As noted above, all Penrose limits of conformally-flat spacetimes are conformally-flat plane waves. Furthermore, all conformally-flat plane waves have metrics $g_{ab}$ with the form \eqref{PlaneWaveConfFlat}. It is evident from this together with \eqref{PlaneWaveGen} and \eqref{JacobiSpatial}-\eqref{ZeroDet} that all conjugate points in such spacetimes have multiplicity 2. The plane wave Green function $G_\mathrm{ret}(p,p')$ is therefore given by \eqref{GretDeg}. Via \eqref{PenroseGreen}, a similar structure also appears in the Green function $\check{G}_\mathrm{ret}(p,p')$ associated with the full spacetime $(\check{M}, \check{g}_{ab})$. This provides the sense in which the 2-fold singular structure \eqref{SingStruct2} is present in retarded scalar Green functions associated with all conformally-flat spacetimes.

For null geodesics passing through a vacuum (Ricci-flat) region of some spacetime $(\check{M},\check{g}_{ab})$, the associated Penrose limit is a vacuum plane wave with the metric \eqref{PlaneWaveRiccFlat}. The wave profile in such a case satisfies
\begin{equation}
	\mathrm{Tr} \, \mathbf{H}(u) = \delta^{ij} \check{R}_{+i+j} (\check{z}(u))= 0.
\end{equation}
If
\begin{equation}
	\check{R}_{+i+j} (\check{z}(u)) = h(u) \bar{H}_{ij}
	\label{hRiemann}
\end{equation}
for some constant matrix $\bar{\mathbf{H}}$, the resulting plane wave is said to be linearly polarized. An appropriate rotation of the spatial components $e^a_i(u)$ of the tetrad used to perform the Penrose limit can then be used to set
\begin{equation}
	\bar{\mathbf{H}} = \pm \left(
	\begin{matrix}
	1 & 0 \\ 
	0 & -1
	\end{matrix}
	\right) .
\end{equation}
It follows that the $h(u)$ appearing in \eqref{hRiemann} can be identified (up to a sign) with the $h_{+}(u)$ in \eqref{PlaneWaveRiccFlat}. If $h(u)$ is either entirely non-negative or entirely non-positive, it is clear from \eqref{JacobiSpatial}-\eqref{ZeroDet} that any conjugate points which may exist must have multiplicity 1. $G_\mathrm{ret}(p,p')$ then has the form \eqref{Green4Fold}. It follows from \eqref{PenroseGreen} that in the vacuum case where the Riemann tensor along the reference geodesic has the form \eqref{hRiemann} and the $h(u)$ appearing in that equation does not pass through zero (but may sometimes equal zero), $\check{G}_\mathrm{ret}(p,p')$ contains the repeating 4-fold pattern of singular structures \eqref{SingStruct4}. 

Many of the explicit computations of four-dimensional Green functions found in the literature fall into one of the two classes of spacetimes just described. In the conformally-flat case, scalar Green functions have been computed in both the Einstein static universe and Bertotti-Robinson spacetimes \cite{MarcPrivate}. As expected from our argument using Penrose limits, the singular structures of retarded Green functions associated with both of these spacetimes have been found to have a 2-fold pattern with the form \eqref{SingStruct2}.

Another important example in the literature is the retarded scalar Green function associated with Schwarzschild spacetime. All Penrose limits of Schwarzschild\footnote{Some null geodesics of Schwarzschild intersect the central singularity. The associated Penrose limits are then singular plane waves. Before this point, however, all discussion above remains valid.}  have the form \eqref{hRiemann} with $h(u) \geq 0$ \cite{BlauNotes,QED2}. We therefore predict that retarded Green functions in Schwarzschild possess the 4-fold singular structure \eqref{SingStruct4}. This is indeed what was observed in the explicit computations carried out in \cite{CausticsSchw}.

More generally, we may consider scalar Green functions associated with all Kerr spacetimes. Penrose limits of Kerr (and all other Petrov type D spacetimes) are discussed in \cite{QED2}. It is easily shown from this that Penrose limits of Kerr result in wave profiles with the form \eqref{hRiemann}. For a reference geodesic with specific angular momentum $l_z$ about the symmetry axis and Carter constant $q$, it is shown in \cite{QED2} that
\begin{equation}
	h(u) = \frac{ 3 M [ (a-l_z)^2 + q ] }{ [ r^2(u) + a^2 \cos^2 \theta(u)]^{5/2} }.
\end{equation}
Here, $M$ and $a M$ are the mass and angular momentum associated with the Kerr background. $r(u)$ and $\theta(u)$ are the Boyer-Lindquist coordinates of the reference geodesic at the affine time $u$. It is clear that $h(u)$ cannot change sign, so we predict that retarded scalar Green functions in Kerr spacetime contain the 4-fold singular structure \eqref{SingStruct4}.

\subsection{Tensor Green functions}
\label{Sect:TensorGeneralize}

Thus far, we have considered only Green functions associated with the propagation of scalar fields. It is straightforward to partially extend our results to also allow for fields with nonzero tensor rank. In particular, we now show that the leading order singular structure of tensor Green functions in arbitrary spacetimes can be understood using appropriate plane wave Green functions. No attempt is made, however, to also derive the form of those plane wave Green functions as we have done in the scalar case. 

As an example, consider a field $\check{A}_a$ propagating on a spacetime $(\check{M}, \check{g}_{ab})$ and satisfying
\begin{equation}
	\check{L} \check{A}_a = - 4 \pi \rho_a.
	\label{TensorWave}
\end{equation}
Here $\check{L}$ is any second-order linear differential operator whose principal part is equal to the d'Alembertian $\check{g}^{ab} \check{\nabla}_a \check{\nabla}_b$. The wave equation \eqref{TensorWave} is naturally associated with Green functions $\check{G}_{a}{}^{a'}(p,p')$ satisfying
\begin{equation}
	\langle \check{G}_{a}{}^{a'} (p,p'), \check{L}^\dag \check{\varphi}^a(p) \rangle = - 4 \pi \check{\varphi}^{a'}(p')
	\label{TensorWaveGreen}
\end{equation}
for all smooth vector fields $\check{\varphi}^a(p)$ with compact support. 

Now choose a point $p' \in \check{M}$ and consider the behavior of $G_{a}{}^{a'}(\cdot, p')$ near a null geodesic $\check{z}(u)$ passing through $p' = \check{z}(u')$. As above, we choose a tetrad and construct a Fermi-like coordinate system $(u, \check{v}, \check{\bx})$ in a neighborhood of $\check{z}(u)$. It is also useful to consider the scaled coordinates $(u,v,\bx)$ defined by \eqref{PenroseScale}. 

We now seek an analog of \eqref{PlaneGreenDef}. This requires choosing an appropriate family of test functions similar to \eqref{ScaledTest}. Given an arbitrary test function $\varphi^\mu (u, v, \bx)$ in the plane wave spacetime which results from the Penrose limit of $(\check{M}, \check{g}_{ab})$ and $\check{z}(u)$, reverse the coordinate transformation \eqref{PenroseScale} to obtain
\begin{subequations}
\begin{align}
	\check{\varphi}_\lambda^u (u, \check{v}, \check{\bx} ) := \varphi^u (u, \lambda^{-2} \check{v}, \lambda^{-1} \check{\bx} )
	\\
	\check{\varphi}_\lambda^{\check{v}} (u, \check{v}, \check{\bx} ) := \lambda^2 \varphi^v (u, \lambda^{-2} \check{v}, \lambda^{-1} \check{\bx} )
	\\
	\check{\varphi}_\lambda^{\check{i}} (u, \check{v}, \check{\bx} ) := \lambda \varphi^i (u, \lambda^{-2} \check{v}, \lambda^{-1} \check{\bx} )
\end{align}
\end{subequations}
in the unscaled coordinates $(u, \check{v}, \check{\bx})$. Similarly, choose some covector $v_{\mu'}$ which remains fixed (for all $\lambda>0$) in the scaled coordinates $(u, v, \bx)$. Then define
\begin{equation}
	\langle G_{\mu}{}^{\mu'},  \varphi^\mu v_{\mu'} \rangle := \lim_{\lambda \rightarrow 0} \lambda^{-2} \langle \check{G}_{\check{\mu}}{}^{\check{\mu}'} ,  \check{\varphi}_\lambda^{\check{\mu}} v_{\check{\mu}'} \rangle.
\end{equation}
Considering test functions of the form $g^{\nu \rho} \nabla_\nu \nabla_\rho \varphi^\mu$, where $g_{\mu\nu}$ denotes the Penrose limit metric \eqref{PenroseMetric}, it is straightforward to show using \eqref{TensorWaveGreen} that 
\begin{equation}
	\langle G_{\mu}{}^{\mu'} , v_{\mu'} g^{\nu \rho} \nabla_\nu \nabla_\rho \varphi^\mu \rangle = - 4 \pi \varphi^{\mu'}(p') v_{\mu'}.
\end{equation}
It follows that the operator $G_{\mu}{}^{\mu'}(p,p')$ is, as the notation suggests, a Green function associated with the plane wave spacetime obtained by taking a Penrose limit with $(\check{M}, \check{g}_{ab})$ and $\check{z}(u)$. 

This argument carries through essentially without change for Green functions associated with all higher-rank tensor fields. We have thus established that there is a sense in which the leading order singular behavior of all tensor wave equations can be understood by considering appropriate plane wave Green functions.

\section{Discussion}

This paper discusses the transport of disturbances in scalar fields propagating on curved spacetimes. In particular, we study how light cone caustics affect the character of singularities appearing in the relevant Green functions. This problem is addressed in two steps. First, explicit Green functions are obtained for massless scalar fields propagating on all non-singular four-dimensional plane wave spacetimes. We then show in Sect. \ref{Sect:Penrose} that Penrose limits provide a sense in which certain aspects of these solutions are universal: The leading order singular structure of scalar Green functions associated with essentially all four-dimensional spacetimes can be described by appropriate plane wave Green functions. 

The plane wave Green functions we obtain are summarized in Sect. \ref{Sect:GlobalSoln}. They are globally defined and fully explicit [up to the calculation of the $2 \times 2$ matrices $\bA$ and $\bB$ defined by \eqref{JacobiSpatial} and \eqref{ABboundary}]. Almost everywhere, plane wave Green functions are found to have a ``Hadamard-like'' component. Using $\sigma$ and $\Delta$ to denote Synge's function and the van Vleck determinant (which are well-defined almost everywhere in plane wave spacetimes), we find that there exist Green functions that switch between the forms $\pm \sqrt{|\Delta|} \delta(\sigma)$ and $\pm \sqrt{|\Delta|} \mathrm{pv} (1/\pi \sigma)$ after each pass through a conjugate hyperplane. 

As described in Sect. \ref{Sect:GenericGreen}, there is a sense in which (say) retarded Green functions $\check{G}_\mathrm{ret}(\cdot, p')$ satisfying \eqref{WaveEqScalarGenGreen} in generic spacetimes contain similar Hadamard-like terms near any future-directed null geodesic emanating from $p'$. Fixing some point $p$ on such a null geodesic (that is not conjugate to $p'$), precisely which Hadamard form is appropriate depends only on the pattern of multiplicities of all points conjugate to $p'$ that lie between $p'$ and $p$. Following rules \eqref{Rule1}-\eqref{Rule4}, crossing a non-degenerate (multiplicity 1) conjugate point is found to change a Green function involving $\sqrt{|\Delta|} \delta(\sigma)$ into one involving $\sqrt{|\Delta|} \mathrm{pv} (1/\pi \sigma)$. Conversely, non-degenerate conjugate points transform Green functions proportional to $\sqrt{|\Delta|} \mathrm{pv} (1/\pi \sigma)$ into ones proportional to $- \sqrt{|\Delta|} \delta(\sigma)$. One pass through a conjugate point with multiplicity two is seen to have the same effect as two passes through conjugate points with multiplicity one. This merely reverses signs: $\sqrt{|\Delta|} \delta(\sigma) \rightarrow - \sqrt{|\Delta|} \delta(\sigma)$ or $\sqrt{|\Delta|} \mathrm{pv} (1/\pi \sigma) \rightarrow - \sqrt{|\Delta|} \mathrm{pv} (1/\pi \sigma)$. 

In this way, we have derived and made significantly more precise Ori's comments \cite{Ori} regarding changes in the singularity structure of Green functions due to light cone caustics. The result is a simple universal rule that is -- unlike most results regarding caustics -- naturally stated in terms of distributions on the spacetime manifold (as opposed to statements involving Fourier transforms). 

It is interesting to note that the object $\sigma$ appearing in the leading order singular structure of a generic Green function $\check{G}_\mathrm{ret}$ is not the world function $\check{\sigma}$ associated with the spacetime $(\check{M}, \check{g}_{ab})$. $\check{\sigma}$ is typically ill-defined when its arguments are widely separated. $\sigma$ is, instead, the world function of an appropriate plane wave spacetime obtained via a Penrose limit. This \textit{is} well-defined almost everywhere. Similar comments also apply to the van Vleck determinant $\Delta$, which effectively measures the ``strength'' of the leading order singular terms appearing in a generic retarded Green function. Explicit forms for both $\sigma$ and $\Delta$ are easily computed in arbitrary spacetimes using the results of Sects. \ref{Sect:Bitensors} and \ref{Sect:PenroseDef}.

The rules we derive for changes in a Green function's singular structure have a simple heuristic interpretation. One might think of degenerate conjugate points as events where bundles of light rays are perfectly focused in every direction. Sharp solutions involving $\delta(\sigma)$ might therefore be expected to remain sharp after passing through a degenerate conjugate point. Similarly, more diffuse solutions like $\mathrm{pv} (1/\sigma)$ might be expect to remain diffuse in such an encounter. Conjugate points with multiplicity 1 are different. They focus null geodesics in only one transverse direction. It is therefore reasonable to expect sharp solutions like $\delta(\sigma)$ to be ``blurred out'' by such structures. Somewhat less intuitive is that the nature of this blurring is always such that another pass through a non-degenerate conjugate point ``resharpens'' the field back into a form involving $\delta(\sigma)$.

An important special case of this work concerns the behavior of retarded Green functions associated with scalar fields in the Kerr spacetime. All conjugate points appearing on null geodesics of Kerr are non-degenerate. Scalar Green functions in Kerr therefore change singularity structure according to the 4-fold pattern \eqref{SingStruct4}. This result includes as a special case the 4-fold behavior observed by Dolan and Ottewill \cite{CausticsSchw} in Schwarzschild Green functions. 

The problem of wave propagation in curved spacetime is a very general one with many applications. Our results may therefore be useful in a number of fields. One possible application concerns the computation of self-forces: What is the force exerted by a small object on itself in a curved spacetime? One may assume that the total field is the retarded solution and find the force that this exerts on a given body. In generic spacetimes, the result depends on the object's past history at least via the tail term $\mathcal{V}(p,p')$ appearing in \eqref{GretGeneral}. It has, however, been less clear precisely how light cone caustics in the distant past contribute to an object's self-field (see, e.g., \cite{CausticsNariai, CausticsSchw,Theo} for some related discussion). More generally, it is important to understand ``how much'' of a charge's past history influences its current self-field. Our results may be useful in answering this question.

Other applications could exist even in systems where the spacetime curvature is negligible.  Wave equations on curved spacetimes are mathematically equivalent to various physically different problems in flat spacetime. For example, one might use the same equations to describe the propagation of acoustic waves in a moving fluid or electromagnetic waves in certain classes of permeable materials \cite{AnalogGrav, Analog2}. It would be interesting to translate the results of this paper to more readily apply to problems such as these. The physical meaning of the Penrose limit would be particularly interesting to understand in some of these ``analog gravity'' systems.

There are two additional ways in which this work could be extended. Most obviously, it would be extremely useful to generalize our results to apply to wave equations involving tensor fields with nonzero rank. The singularity structure of disturbances in, e.g., electromagnetic fields and metric perturbations could then be understood in a relatively simple way. We carry out one portion of this task in Sect. \ref{Sect:TensorGeneralize}, where Penrose limits are used to relate generic tensor Green functions to appropriate plane wave Green functions. Although there does not appear to be any significant obstacle to doing so, we have not made any attempt to compute plane wave Green functions for higher-rank tensor fields in this paper. A complete discussion of the leading order singularity structure of tensor Green functions must therefore wait for later work.

It might also be interesting to extend our work to higher numbers of dimensions. In the four-dimensional case considered here, the rules describing how Green functions transition between different singular structures suggest that passing through a conjugate point with multiplicity 2 is equivalent to two passes through conjugate points with multiplicity 1. While it appears likely that such a rule generalizes for larger multiplicities, it would be interesting to verify this directly. Is one pass through a conjugate point with multiplicity $n\geq 1$ in a spacetime with dimension $d \geq n+2$ equivalent to $n$ passes through conjugate points with multiplicity 1? Questions related to higher dimensions are perhaps not only of mathematical interest. Higher-dimensional plane wave spacetimes and Penrose limits have found extensive use in string theory and related subjects \cite{String1, String2, String3, BlauNotes}.

\appendix

\section{Properties of $\bA(u,u')$ and $\bB(u,u')$}
\label{Sect:ABProperties}

The $2 \times 2$ matrices $\bA(u,u')$ and $\bB(u,u')$ defined by \eqref{JacobiSpatial} and \eqref{ABboundary} play a central role in the geometry of plane wave spacetimes. This appendix briefly reviews some of their most important properties.

First note that any two solutions $\mathbf{C}(u)$ and $\mathbf{D}(u)$ to the modified oscillator equation \eqref{EDef} satisfy
\begin{equation}
\partial_u (\mathbf{C}^\intercal \partial_u \mathbf{D}-\partial_u \mathbf{C}^\intercal \mathbf{D} ) = 0.
\label{DAbel}
\end{equation}
This is a simple application of Abel's identity. Using the boundary conditions \eqref{ABboundary} with the identifications $\mathbf{C}(u) \rightarrow \bA(u,u')$ and $\mathbf{D}(u) \rightarrow \bB(u,u')$, it follows that
\begin{equation}
\bA^\intercal \partial_u \bB -\partial_u \bA^\intercal \bB = \bm{\delta}.
\label{Abel}
\end{equation}

In general, $\bA$ and $\bB$ are not symmetric matrices. Simple combinations of them are, however, symmetric. Applying \eqref{DAbel} with $\mathbf{C} , \mathbf{D} \rightarrow \bA$ demonstrates that $\bA^\intercal \partial_u \bA$ is one such example:
\begin{equation}
\bA^\intercal \partial_u \bA = (\bA^\intercal \partial_u \bA )^\intercal.
\label{SymMatrix}
\end{equation}
Right-multiplying this equation with $\bA^{-1}$ and left-multiplying with $(\bA^{-1})^\intercal$ shows that $\partial_u \bA \bA^{-1}$ and its inverse are also symmetric (wherever they exist). All of these comments continue to hold with the replacement $\bA \rightarrow \bB$.

Using these observations together with \eqref{Abel} further shows that almost everywhere,
\begin{equation}
\bB \bA^\intercal = ( \partial_u \bB \bB^{-1} - \partial_u \bA \bA^{-1} )^{-1}.
\label{BASym}
\end{equation}
The right-hand side of this equation is symmetric, so $\bB \bA^\intercal$ must be symmetric as well. 

A similar argument shows that the symmetry condition \eqref{SymE} on the matrix $\mathbf{E}(U)$ involved in the transformation between Rosen and Brinkmann coordinates is automatically satisfied for all $U$ if it satisfied for any $U$. It also follows that $\dot{\mathbf{E}} \mathbf{E}^{-1}$ is symmetric wherever it exists.

$\bA$ and $\bB$ can be computed by directly solving the differential equations \eqref{JacobiSpatial} with the boundary conditions \eqref{ABboundary}. Alternatively, both matrices may be found by integrating any single $\mathbf{E}(u)$ satisfying \eqref{EDef} and \eqref{SymE}. Use of \eqref{RosenH} and \eqref{DAbel} with $\mathbf{C} \rightarrow \mathbf{E}$ and $\mathbf{D} \rightarrow \bB$ demonstrates that 
\begin{equation}
\bB(u,u') =  \int_{u'}^u \rmd u'' \mathbf{E}(u) \bm{\mathcal{H}}^{-1}(u'') \mathbf{E}^\intercal (u')
\label{BFormula}
\end{equation}
as long as $\bm{\mathcal{H}}^{-1}(\cdot)$ is defined throughout the interval $(u,u')$. The same assumption also leads to a formula for $\bA$:
\begin{align}
\bA(u,u') = \bm{\delta} +  \int_{u'}^u \rmd u'' \mathbf{E}(u) \bm{\mathcal{H}}^{-1}(u'')
\nonumber
\\
~ \times [ \dot{\mathbf{E}}(u'')  - \dot{\mathbf{E}}(u')  ]^\intercal . 
\end{align}
Similar equations may be derived for $u$ and $u'$ arbitrarily separated by considering a number of different $\mathbf{E}$ matrices which are invertible in a suitable set of overlapping intervals. One interesting consequence of \eqref{BFormula} is that
\begin{equation}
\bB(u,u') = - \bB^\intercal(u',u).
\end{equation}

\section{Distributional character of $\mathcal{G}^\sharp_{n^\pm}$ and $\mathcal{G}^\flat_{n^\pm}$}
\label{Sect:Distributions}

It is not \textit{a priori} clear that the functionals $\mathcal{G}^\sharp_{n^\pm}(p,p')$ and $\mathcal{G}^\flat_{n^\pm}(p,p')$ used in the ansatz \eqref{GAnsatz3} for the scalar Green function are well-defined. From their definitions \eqref{IdeltaDef} and \eqref{IpvDef}, these objects take as input an ``observation point'' $p' = (u',v',\bx') \in \mathbb{R}^4$ and any test function $\varphi_n(p) = \varphi_n(u,v,\bx)$ that is smooth and with compact support in the set $\mathcal{T}_n(p')$ defined in Sect. \ref{Sect:GreenOnConj}.

By definition, a linear form $\langle G^\sharp_{n^\pm} (p,p'), \varphi_n(p) \rangle$ is a distribution if and only if, for every fixed $p'$ and every compact subset $\omega \subset \mathcal{T}_n(u')$, there exist semi-norm estimates of the form
\begin{equation}
	| \langle \mathcal{G}^\sharp_{n^\pm}, \varphi_{\omega} \rangle | \leq C_{\omega} \sum_{|\alpha| \leq N_{ \omega }} \sup | \partial_\alpha \varphi_{\omega} |
	\label{SemiNormGen}
\end{equation}
for all $\varphi_{\omega}$ with $\mathrm{supp} \, \varphi_{\omega} \subset \omega$ \cite{FriedlanderDistributions}. $\alpha$ denotes a multi-index and $|\alpha|$ its order. The non-negative numbers $C_{\omega}$ and $N_{\omega}$ depend only on the region $\omega$, and not on any details of the test function. Estimates of this type are straightforward if $\omega$ does not pass through the hyperplane $S_{\tau_n(u')}$ conjugate to $S_{u'}$. More generally, one must consider separately the possible multiplicities the conjugate pair $(S_{\tau_n(u')}, S_{u'})$. 

It is simplest to derive estimates like \eqref{SemiNormGen} by choosing finite (nonzero) numbers $u_\omega$, $x_\omega$, and $v_\omega$ such that the compact rectangular region defined by
\begin{subequations}
	\label{SRect}
	\begin{align}
	|u-\tau_n(u')|< &~ u_\omega ,  \quad |v-v'| < v_\omega,
	\\
	&|\bx-\hat{\bA}_n \bx'| < x_\omega ,
	\end{align}
\end{subequations}
entirely encloses $\omega \subset \mathcal{T}_n(u')$. As in Sect. \ref{Sect:CausticGeo}, $\hat{\bA}_n(u') := \bA\big(\tau_n(u'), u' \big)$. It is also convenient to introduce the $\infty$-norm $\| \cdot \|_{\omega}$ in $\omega$: 
\begin{equation}
\| \cdot \|_{\omega} := \sup_{\omega} | \cdot |.
\end{equation}

\subsection{Degenerate conjugate points}

Suppose that $\tau_n(u')$ is associated with conjugate points of multiplicity $2$. Choose an arbitrary compact region $\omega \subset \mathcal{T}_n(u')$ and finite numbers $u_\omega$, $x_\omega$, and $v_\omega$ that define a rectangular region \eqref{SRect} which encloses $\omega$. 

Consider the definition \eqref{IdeltaDef} for $\langle \mathcal{G}^\sharp_{n^\pm}, \varphi_{\omega} \rangle$ when $\mathrm{supp} \, \varphi_{\omega} \subset \omega$. This consists of an integral over the coordinates $u$ and $\bx$. It is clear that the integrand vanishes whenever $|\bx-\hat{\bA}_n \bx'| > x_\omega$. If, however, $\varphi_{\omega}$ has support sufficiently close to $S_{\tau_n}$, a sharper bound may be placed on the maximum value of $|\bx-\hat{\bA}_n \bx'|$ that needs to be considered by using $v_\omega$. This is because the function $\chi(u,u'; \bx, \bx')$ defined by \eqref{VDef} has the form \eqref{VExpandDeg} in this region. It follows that the integrand vanishes when
\begin{equation}
	\left| \chi_n - \frac{1}{2} \frac{ | \bx - \hat{\bA}_n \bx' |^2 }{ \tau_n -u }   \right| > v_\omega.
\end{equation}
This is implied by the stronger condition
\begin{equation}
	| \bx - \hat{\bA}_n \bx' |^2 > 2 |\tau_n-u| \left( v_\omega + \| \chi_n \|_{\omega} \right).
	\label{xRestrict}
\end{equation}
It follows that the spatial support of the integrand in \eqref{IdeltaDef} shrinks at least as fast as $\sqrt{|\tau_n-u|}$ as $u \rightarrow \tau_n$.

Without loss of generality, we may choose $v_\omega$ to be sufficiently large that, e.g.,
\begin{equation}
	v_\omega > \| \chi_n \|_{\omega}.
	\label{SvBound}
\end{equation}
The spatial integral in \eqref{IdeltaDef} may therefore be limited to 
\begin{equation}
	| \bar{\bx} | < 2 \sqrt{ v_\omega },
\end{equation}
where $\bar{\bx}$ is defined by \eqref{xBarDef}. Using $\bar{\bx}$ as an integration variable, \eqref{IdeltaDef} changes to
\begin{align}
	\langle \mathcal{G}^\sharp_{n^\pm} , \varphi_{\omega} \rangle := \pm \lim_{\epsilon \rightarrow 0^+} \int_{\tau_n \pm \epsilon}^{\tau_n \pm u_\omega} \! \rmd u \int_{ |\bar{\bx}| < 2 \sqrt{ v_\omega } } \! \rmd^2 \bar{\bx} 
	\nonumber
	\\
	~ \left( \frac{\sqrt{|(\tau_n-u)^2 \Delta|}}{|u-u'|} \right)  \varphi_{\omega}(u,v'+\chi,\bx).
\end{align}
It follows from \eqref{VanVleckDeg} that the integrand in this equation is everywhere bounded. Hence,
\begin{align}
	| \langle \mathcal{G}^\sharp_{n^\pm} , \varphi_\omega \rangle | \leq 4 \pi u_\omega v_\omega \left\| \frac{\sqrt{(\tau_n-u)^2 |\Delta|}}{|u-u'|} \right\|_\omega 
	\nonumber
	\\
	~ \times \sup | \varphi_\omega | < \infty.
	\label{SemiNormSharpDeg}
\end{align}

Although $\Delta$ is, in general, unbounded in $\omega$, this argument shows that $\langle \mathcal{G}^\sharp_{n^\pm} , \varphi_\omega \rangle$ is always finite. All integrals in its definition converge. Furthermore, they converge absolutely. The order of integration does not matter in \eqref{IdeltaDef}. This establishes that the $\mathcal{G}^\sharp_{n^\pm}$ are well-defined linear functionals. They are also distributions. Eq. \eqref{SemiNormSharpDeg} provides a semi-norm estimate of the form \eqref{SemiNormGen} with $N_\omega = 0$ and
\begin{equation}
	C_\omega = 4 \pi u_\omega v_\omega \left\| \frac{\sqrt{(\tau_n-u)^2 |\Delta|}}{|u-u'|} \right\|_\omega.
\end{equation}
$N_\omega$ and $C_\omega$ depend, as required, only on $\omega$ (and not on the test function $\varphi_\omega$).

Establishing similar bounds for the functionals $\mathcal{G}^\flat_{n^\pm}$ is more complicated. Although it is again useful to change the integration variable $\bx$ to $\bar{\bx}$ in the definition \eqref{IpvDef}, this leaves a result where neither the integrand nor the integration volume are bounded. 

The first simplification arises by using  \eqref{VExpandDeg} and \eqref{SvBound} to deduce that the integrand in \eqref{IpvDef} vanishes when, e.g., 
\begin{equation}
	|\bar{\bx}|>0, \qquad \Sigma > \frac{1}{2} | \bar{\bx} |^2 + 2 v_\omega,
\end{equation}
or
\begin{equation}
	|\bar{\bx}|>2 \sqrt{ v_\omega } , \qquad \Sigma <\frac{1}{2} | \bar{\bx} |^2 - 2 v_\omega.
\end{equation}
Additionally, it is clear from \eqref{xBarDef} and \eqref{SRect} that the integrand also vanishes when
\begin{equation}
	| \bar{\bx} | > \frac{ x_\omega }{ \sqrt{ |\tau_n-u| } }. 
\end{equation}
We shall assume that $x_\omega$ has been chosen to be sufficiently large that
\begin{equation}
	x_\omega > \sqrt{6 u_\omega v_\omega},
	\label{xOmega}
\end{equation}
so
\begin{equation}
	\frac{ x_\omega }{ \sqrt{ |\tau_n-u|} } > \sqrt{6 v_\omega}.
	\label{SxSv}
\end{equation}
The factor of $6$ here is a fairly arbitrary number. Any other choice greater than $4$ could also be used. 

These considerations suggest that \eqref{IpvDef} should be split into two parts:
\begin{widetext}
	\begin{align}
		\langle \mathcal{G}^\flat_{n^\pm} , \varphi_\omega \rangle = \mp \lim_{\epsilon \rightarrow 0^+ }  \int_{\tau_n \pm \epsilon}^{\tau_n \pm u_\omega } \rmd u \left( \frac{ \sqrt{ (\tau_n -u )^2 |\Delta| } }{ u-u' } \right) \left[ \int_{|\bar{\bx}| <  \sqrt{6 v_\omega } } \! \! \rmd^2 \bar{\bx}  \int_0^{5 v_\omega } \! \! \rmd \Sigma   \int_{-1}^1 \rmd \nu \, \partial_v \varphi_\omega (u,v' +\chi + \nu \Sigma, \bx )  \right.
		\nonumber
		\\
		\left. ~ \mp \int_{ |\bar{\bx}| > \sqrt{ 6 v_\omega } }^{ |\bar{\bx}| < x_\omega / \sqrt{|\tau_n-u|}  } \!  \! \! \rmd^2 \bar{\bx}  \int_{ \frac{1}{2} | \bar{\bx} |^2 - 2 v_\omega  }^{ \frac{1}{2} | \bar{\bx} |^2 + 2 v_\omega  } \! \rmd \Sigma  \left( \frac{ \varphi_\omega (u, v' + \chi  \mp \Sigma, \bx)}{\Sigma} \right) \right] ,
	\end{align}
\end{widetext}
Both integrands here are manifestly bounded. The integral on the top line is also carried out over a finite volume. It is therefore trivial to provide a bound for it. The integral on the lower line must be bounded somewhat more carefully. The final result is that
\begin{align}
	&| \langle \mathcal{G}^\flat_{n^\pm} , \varphi_\omega \rangle |  \leq 
	4 \pi v_\omega \left\| \frac{\sqrt{(u-\tau_n)^2 |\Delta|}}{u-u'} \right\|_\omega 
	\nonumber
	\\
	& \qquad ~ \times  ( 15 u_\omega v_\omega \sup | \partial_v \varphi_\omega| + \zeta_\omega \sup | \varphi_\omega | ) ,
	 \label{SemiNormDegFlat}
\end{align}
where
\begin{equation}
	\zeta_\omega := 2 \int_0^{u_\omega} \rmd \bar{u} \int_{ \sqrt{3/2} }^{ x_\omega /  \sqrt{ 4 \bar{u} v_\omega } } \rmd \bar{r} \, \bar{r} \ln \left( \frac{ \bar{r}^2 + 1 }{ \bar{r}^2- 1 } \right).
	\label{pvconst}
\end{equation}
Noting that
\begin{equation}
	0 < \bar{r} \ln \left( \frac{ \bar{r}^2 + 1 }{ \bar{r}^2- 1 } \right) <  2
\end{equation}
in the relevant range, a straightforward integration shows that $\zeta_\omega < \infty$. It follows that the linear functional $\langle \mathcal{G}^\flat_{n^\pm}, \varphi_\omega \rangle$ is always finite. It is bounded by the semi-norm estimate \eqref{SemiNormDegFlat}, so $\mathcal{G}^\flat_{n^\pm}$ is a distribution.

\subsection{Non-degenerate conjugate points}

Suppose now that $\tau_n(u')$ is associated with conjugate points of multiplicity 1 and choose a compact region $\omega \subset \mathcal{T}_n(u')$ in the same manner as in the previous case.  Using \eqref{VExpandSimp}, the integrand of \eqref{IdeltaDef} is easily seen to vanish when
\begin{align}
	\left|\chi_n - \frac{1}{2} \frac{ [ \hat{ \mathbf{q}}_n  \cdot ( \bx - \hat{\bA}_n \bx' ) ] ^2 }{ \tau_n - u } \right|>v_\omega.
\end{align}
Introducing the variables $(\tilde{x}^1,\tilde{x}^2)$ defined by \eqref{XTilde}, this inequality is implied by the stronger condition 
\begin{equation}
	(\tilde{x}^1)^2 > 2 ( v_\omega + \| \chi_n \|_\omega).
\end{equation}
Also note that the integrand of \eqref{IdeltaDef} vanishes when 
\begin{equation}
	|\tilde{x}^2|> x_\omega.
\end{equation} 

We assume that the parameters $u_\omega$, $v_\omega$, and $x_\omega$ are chosen such that \eqref{SvBound} holds. It then follows that \eqref{IdeltaDef} can be rewritten as 
\begin{align}
	\langle \mathcal{G}^\sharp_{n^\pm} , \varphi_\omega \rangle := \pm \lim_{\epsilon \rightarrow 0^+} \int_{\tau_n \pm \epsilon}^{\tau_n \pm u_\omega} \! \! \! \rmd u \int_{ |\tilde{x}^1|< 2 \sqrt{v_\omega} } \! \! \! \rmd \tilde{x}^1 \int_{ |\tilde{x}^2|< x_\omega} \! \! \! \rmd \tilde{x}^2
\nonumber
\\
	~ \left( \frac{\sqrt{|(\tau_n-u) \Delta|}}{|u-u'|} \right)  \varphi_\omega (u,v'+\chi,\bx).
\end{align}
It follows from \eqref{VanVleckSimp} that the integrand in this equation is everywhere bounded. This means that
\begin{align}
| \langle \mathcal{G}^\sharp_{n^\pm} , \varphi_\omega \rangle | \leq  8 u_\omega x_\omega \sqrt{v_\omega} \left\| \frac{\sqrt{|(\tau_n-u) \Delta|}}{|u-u'|} \right\|_\omega 
\nonumber
\\
~ \times \sup | \varphi_\omega | < \infty.
\label{SemiNormSharpSimp}
\end{align}
Again, we see that all integrals in the definition of $\mathcal{G}^\sharp_{n^\pm}$ converge. Eq. \eqref{SemiNormSharpSimp} provides a semi-norm estimate of the form \eqref{SemiNormGen}, so this operator is, as claimed, a distribution near non-degenerate conjugate hyperplanes. 


We now establish similar bounds for the functionals $\mathcal{G}^\flat_{n^\pm}$. Using  \eqref{VExpandSimp} and \eqref{SvBound} we find that the integrand in \eqref{IpvDef} vanishes when
\begin{align}
	|\tilde{x}^1|>0, \qquad \Sigma > \frac{1}{2} |\tilde{x}^1|^2 + 2  v_\omega,
\end{align}
or
\begin{align}
	|\tilde{x}^1|>2 \sqrt{ v_\omega } ,\qquad \Sigma <\frac{1}{2} | \tilde{x}^1 |^2 - 2  v_\omega.
\end{align}
This integrand also vanishes when
\begin{align}
	| \tilde{x}^1 | > \frac{  x_\omega }{ \sqrt{ |\tau_n-u| } } \qquad \mathrm{or} \quad 	| \tilde{x}^2 | > x_\omega .
\end{align}

We assume again that the \eqref{xOmega} holds, so \eqref{SxSv} is true. Eq. \eqref{IpvDef} can then be rewritten as
\begin{widetext}
	\begin{align}
	\langle \mathcal{G}^\flat_{n^\pm} , \varphi_\omega \rangle = \mp \lim_{\epsilon \rightarrow 0^+ }  \int_{\tau_n \pm \epsilon}^{\tau_n \pm u_\omega } \! \! \! \rmd u \left( \frac{ \sqrt{ |(\tau_n -u )\Delta| } }{ u-u' } \right) \int_{|\tilde{x}^2| <  x_\omega   } \! \! \! \rmd \tilde{x}^2 \Bigg[  \int_{|\tilde{x}^1| <  \sqrt{6 v_\omega } } \! \! \rmd \tilde{x}^1 \int_0^{5 v_\omega } \! \! \rmd \Sigma   \int_{-1}^1 \! \! \rmd \nu \, \partial_v \varphi_\omega (u,v' + \chi + \nu \Sigma, \bx ) 
	\nonumber
	\\
	 ~ \mp  \int_{ |\tilde{x}^1| > \sqrt{ 6 v_\omega } }^{ |\tilde{x}^1| < x_\omega / \sqrt{|\tau_n-u|}  } \!  \! \! \rmd \tilde{x}^1 \int_{ \frac{1}{2} |\tilde{x}^1 |^2 - 2 v_\omega  }^{ \frac{1}{2} | \tilde{x}^1 |^2 + 2 v_\omega  } \! \! \! \rmd \Sigma  \left( \frac{ \varphi_\omega (u, v' + \chi  \mp \Sigma, \bx)}{\Sigma} \right) \Bigg] .
	\end{align}
\end{widetext}
This implies the bound
\begin{align}
	& | \langle \mathcal{G}^\flat_{n^\pm} , \varphi_\omega \rangle |  \leq 
	8 x_\omega \sqrt{v_\omega} \left\| \frac{\sqrt{|(u-\tau_n)\Delta|}}{u-u'} \right\|_\omega 
\nonumber
\\
	& \quad ~ \times  ( 5\sqrt{6} u_\omega v_\omega \sup | \partial_v \varphi_\omega| + \Upsilon_\omega \sup | \varphi_\omega | ),
\label{SemiNormSimpFlat}
\end{align}
where 
\begin{equation}
	\Upsilon_\omega := \int_0^{u_\omega} \rmd \bar{u} \int_{ \sqrt{3/2} }^{ x_\omega /  \sqrt{ 4 \bar{u} v_\omega } } \rmd \bar{r} \ln \left( \frac{ \bar{r}^2 + 1 }{ \bar{r}^2- 1 } \right).
\end{equation}
The integrand in this equation is bounded from above by $2$ (and from below by $0$), so $\Upsilon_\omega< \infty$. It follows that the linear functional $\mathcal{G}^\flat_{n^\pm}$ is a distribution near non-degenerate conjugate hyperplanes.

Together, the results of this appendix establish that for any nonzero integer $n$ such that $\tau_n(u') \in T(u')$, $\mathcal{G}^\sharp_{n^\pm}(p,p')$ and $\mathcal{G}^\flat_{n^\pm}(p,p')$ are well-defined distributions throughout $\mathcal{T}_n(u')$.

\acknowledgements

This work was supported in part by NSF grant PHY08-54807 to the University of Chicago. The authors thank Barry Wardell, Aaryn Tonita, Marc Casals, Robert Wald, and Volker Perlick for useful discussions and comments.

\end{document}